\documentclass[a4paper,11pt]{article}
\pdfoutput=1 

\usepackage{jheppub} 
\usepackage[utf8]{inputenc}
\usepackage[normalem]{ulem}
\usepackage{setspace}
\usepackage{subfigure}
\usepackage{hyperref}
\definecolor{refkey}{gray}{0.45}
\definecolor{labelkey}{RGB}{155,48,48}
\definecolor{UI_blue}{RGB}{32, 64, 151}
\definecolor{UI_red}{RGB}{187, 62, 24}
\definecolor{UI_blue2}{RGB}{0, 84, 147}
\definecolor{UI_red2}{RGB}{159, 32, 66}
\definecolor{UI_gray}{RGB}{169, 169, 169}
\definecolor{UI_sepia}{RGB}{112, 66, 20}
\definecolor{UI_bittersweet}{RGB}{254, 111, 94}
\definecolor{UI_emerald}{RGB}{80, 200, 120}
\definecolor{UI_olivegreen}{RGB}{181, 179, 92}
\definecolor{UI_cadetblue}{RGB}{95, 158, 160}
\definecolor{UI_fuchsia}{RGB}{255, 0, 255}
\definecolor{UI_midnightblue}{RGB}{25, 25, 112}
\definecolor{UI_royalblue}{RGB}{0,35, 102}
\definecolor{UI_periwinkle}{RGB}{204, 204, 255}
\definecolor{UI_redorange}{RGB}{255, 83, 73}
\definecolor{UI_brickred}{RGB}{203,65,84}	
\definecolor{UI_forestgreen}{RGB}{34, 139, 34}
\definecolor{UI_tan}{RGB}{210,180,140}	
\definecolor{UI_burlywood}{RGB}{222,184,135}
\definecolor{UI_burlywood}{RGB}{192,64,0}
\definecolor{UI_darkorchid}{RGB}{153,50,204}

\usepackage{mathtools}
\usepackage{physics}
\usepackage{simplewick}
\usepackage[obeyFinal]{todonotes}

\def\ba{\begin{align}}\def\ea{\end{align}}
\def\beq{\begin{eqnarray}}\def\eeq{\end{eqnarray}}
\def\be{\begin{equation}}\def\ee{\end{equation}}
\def\ben{\begin{equation}}
	\def\een{\end{equation}}
\def\bea{\begin{eqnarray}}
	\def\eea{\end{eqnarray}}

\def\hPi{{\hat{\Pi}}}

\def\hP{\hat{P}}
\def\hX{\hat{X}}

\def\cO{{\cal{O}}}

\def\trho{{\tilde{\rho}}}
\def\fr{\frac}

\def\eq#1{(\ref{#1})}

\def\beq{\begin{eqnarray}}\def\eeq{\end{eqnarray}}
\def\be{\begin{equation}}\def\ee{\end{equation}}
\def\ba{\begin{align}}\def\ea{\end{align}}
\def\beq{\begin{eqnarray}}\def\eeq{\end{eqnarray}}
\def\be{\begin{equation}}\def\ee{\end{equation}}
\def\ben{\begin{equation}}
	\def\een{\end{equation}}
\def\bea{\begin{eqnarray}}
	\def\eea{\end{eqnarray}}

\def\vm{{\vec{m}}}

\def\hPi{{\hat{\Pi}}}

\def\hP{\hat{P}}
\def\hX{\hat{X}}

\def\cO{{\cal{O}}}

\def\trho{{\tilde{\rho}}}
\def\fr{\frac}

\def\eq#1{(\ref{#1})}

\title{Entanglement Entropy in Internal Spaces and Ryu-Takayanagi Surfaces}
	\preprint{\parbox{3cm}{TIFR/TH/22-45}}
\author{Sumit R. Das$^1$,}
\author{Anurag Kaushal$^{2,3}$,}
\author{Gautam Mandal$^2$,}
\author{Kanhu Kishore Nanda$^2$,}
\author{Mohamed Hany Radwan$^1$,}
\author{Sandip P. Trivedi$^2$}

\affiliation{$^1$Department of Physics and Astronomy, University of Kentucky, Lexington, KY 40506, U.S.A.}
\affiliation{$^2$Department of Theoretical Physics, Tata Institute of Fundamental Research, Mumbai 400005, INDIA}
\affiliation{$^3$International Centre for Theoretical Sciences,
	Tata Institute of Fundamental Research, Shivakote, Bengaluru 560089, INDIA.  }

\emailAdd{sumit.das@uky.edu}\emailAdd{anuragkaushal314@gmail.com}
\emailAdd{mandal@theory.tifr.res.in}
\emailAdd{kanhu.nanda@tifr.res.in}
\emailAdd{M.Radwan@uky.edu}\emailAdd{sandip@theory.tifr.res.in}

\abstract{We study minimum  area surfaces associated with a region, $R$, of an internal space. For example, for a warped product involving an  asymptotically $AdS$ space and an internal space $K$, the region $R$ lies in  $K$ and the surface ends on $\partial R$. We find that the result of Graham and Karch can be avoided  in the presence of warping, and such surfaces can sometimes exist for a general region $R$.  When such a warped product  geometry arises in the IR from a higher dimensional asymptotic AdS, we argue that the area of the surface can be related to  the entropy arising from entanglement of internal degrees of freedom of the boundary theory. We study  several examples, including  warped or direct products involving $AdS_2$, or higher dimensional $AdS$ spaces, with the internal space,   $K=R^m, S^m$; $Dp$ brane geometries and their near horizon limits; and several geometries with a UV cut-off. We find that such RT surfaces often exist and can be useful probes of the system, revealing information about finite length  correlations, thermodynamics and  entanglement. We also make  some preliminary  observations about the role such surfaces can play in bulk reconstruction,
 and their relation to  subalgebras of observables  in the boundary theory.}

\begin{document}
	
\begin{flushright}
\end{flushright}
	
\maketitle
\flushbottom

\section{Introduction}
\label{sec:intro}
	
Starting from  the seminal work of Ryu and Takayanagi, extremal or RT surfaces have played a key role in shaping our understanding of holography \cite{Ryu:2006bv} and \cite{ryu2006aspects}. The RT proposal was extended to more general situations by \cite{hubeny2007covariant}, a proof for the proposal was given in \cite{lewkowycz2013generalized} and extensions to include matter contributions were discussed in \cite{faulkner2013quantum,jafferis2016relative}. These developments lead to a better understanding of bulk reconstruction and its connection to quantum error correction, \cite{harlow2017ryu,harlow2018tasi}. For a more complete set of references of these developments see \cite{harlow2018tasi,rangamani2017holographic} and references therein.	
	 
In concrete realisations of $AdS$  holography the  space-time description typically contains an additional internal space, which can be thought of as a geometrisation of the global symmetries in the field theory dual. For example, in $AdS_5\times S^5$ the internal $S^5$ encodes the $SO(6)$ R-symmetry of the ${\cal N}=4$ SYM theory. These internal spaces do not play a role in the conventional RT surfaces mentioned above, which wrap all the internal directions and are associated with a region $R$ of the spatial boundary of  the $AdS$ space. The interpretation of the RT surface in the boundary hologram is that it calculates the entanglement entropy of region $R$. 
	 
The presence of the extra internal space on the bulk side however prompts the question as to  whether one should consider another type of extremal surface in the bulk. In the simplest of cases such a surface would wrap all the spatial directions on which the field theory lives and instead be associated with a region $R$ of the internal space, ending at the boundary of spacetime on the co-dimension $1$ subspace of the internal space, $\partial R$. Such a surface could perhaps compute a suitably defined measure of  entanglement entropy for internal degrees of freedom in the field theory when they  are divided into two parts, corresponding to the region $R$ and its complement, in the bulk internal space. And it could hopefully play a role in understanding bulk reconstruction, say for a bulk subregion which  includes only some part of the internal space. We will often  refer to such RT surfaces as ``internal RT surfaces" below, to distinguish them from their more conventional counterparts. 
	 
There has been some recent discussion in the literature in the context of field theories, on what such a notion of entanglement entropy could be, associated with dividing the internal degrees of freedom rather than the space in which the field theory lives.  This notion has been called  target space entanglement, since it refers to target space degrees of freedom, rather than the base space on which the field theory lives, and we will discuss it further shortly. 

Before proceeding, we note that the issue we have raised above, about RT surfaces associated with internal space subregions, is particularly sharp  in  a version of gauge-gravity duality where the $AdS$ space-times is  replaced by the near horizon geometry of $D0$ branes \cite{Itzhaki:1998dd}. The duality is well motivated by  reasoning entirely analogous to \cite{maldacena1999large,aharony2000large},  with the boundary theory, in this case, being    gauged  quantum mechanics obtained by dimensionally reducing the ${\cal N}=4$ SYM theory.  
All the bulk directions in this case correspond to target space degrees in the Quantum mechanics and these degrees must somehow contain  information about the bulk, and its various sub-regions.  The behaviour of Internal RT surfaces would be an important guide in asking how this happens.

\subsection{Target Space Entanglement:}
Internal RT surfaces, which are associated with a region of the internal space, as we will explain below, will be the main focus of this paper. But before going on to discuss them let us pause here to review briefly some of the definitions which have been suggested for target space entanglement entropy in the literature. 
One such definition  was given in \cite{das2020bulk,Mazenc:2019ety,das2021gauge,hampapura2021target} and is associated with a subalgebra of gauge invariant operators resulting from a constraint in the target space. Consider for example the D0 brane quantum mechanics referred to above whose bosonic sector has nine $N \times N$ matrices, $\hX^I, I = 1 \cdots 9$. 
A set of gauge invariant operators is of the form \ben
\cO = {\rm Tr} \left[ \hX^I \hX^J \hPi_K \cdots \right]
\label{0-1}
\een
where $\hPi_K$ are the conjugate momenta to $\hX^K$.
Consider a hermitian matrix $f(X^I)$ which is a function of the matrices $X^I$. The target space constraint is defined in terms of   a matrix valued projector
\ben
\hP = \int_A dy~ \delta [y{\bf  I } - f(\hX^I)]
\label{0-2}
\een
where $A$ denotes an interval on the real line, $[y_1,y_2]$. ${\bf I }$ above denotes the identity $N\times N$ matrix  and the delta function  is best defined in Fourier space as 
\be
\label{defdel}
\delta [y{\bf  I }- f(\hX^I)]=\int dk~
e^{i k  (y {\bf I}-f(\hX^I) )}
\ee

Then a projected version of an operator of the form (\ref{0-1}) is 
\ben
\cO_P = {\rm Tr} \left[\hP \hX^I \hP \hX^J \hP \hPi_K \cdots \hP\right]
\label{0-3}
\een
These projected operators form a sub-algebra associated with the constraint $f(\hX^I) \in [y_1,y_2]$. The projector selects the eigenvalues of $f(\hX^I)$ which lie in the interval $A$, as can be seen easily in a gauge were $f(\hX^I)$ is diagonal. The full Hilbert space becomes a sum over sectors labelled by the number $n$ of eigenvalues which lie in the interval $A$. For each sector, there is an un-normalized reduced density matrix $\trho_n$ which evaluates the expectation values of the projected operators in this sub-algebra and ${\rm Tr} \trho_n$ is the probability for $n$ eigenvalues to be in $A$. The von Neumann entropy of the full reduced density matrix $\rho = \oplus \trho_n$ is an entanglement entropy. This is entanglement in the target space of the quantum mechanics. Consider for example a function $f(X^I) = X^1$. The constraint then restricts the eigenvalues of $X^1$ to
lie in the interval $A$ and the subalgebra corresponds to integrating out the eigenvalues which lie outside $A$. In addition, in a sector labelled by $n$, the matrix elements of the other matrices which lie outside a $n \times n$ block are also integrated out. In this model, the eigenvalues of the matrices denote locations of D0 branes while the off-diagonal elements are open strings joining different branes. Therefore, this target space entanglement provides a notion of entanglement in the bulk as perceived by D0 branes. 

For two dimensional non-critical bosonic string theory, the dual is the gauged quantum mechanics of a single hermitian matrix (see, e.g. \cite{Klebanov:1991qa,Jevicki:1993qn,Ginsparg:1993is} for a review). As is well known, this theory becomes equivalent to a theory of $N$ free fermions in an external potential living on the space of eigenvalues. The target space entanglement described above is then the base space entanglement of the second quantized theory of fermions \cite{Das:1995vj}, \cite{Das:1995jw, Hartnoll:2015fca}. Upto a fuzziness of the order of the string scale, the density of fermions is identified with the single dynamical mode of string theory : this precise notion of target space entanglement provides an approximate notion of entanglement in the two dimensional bulk space-time.  Significantly, the scale which makes the entanglement entropy finite is the fermi level which is the inverse of the coupling constant of the collective field theory of the density. This indicates that from the string theory point of view the finiteness is a non-perturbative phenomenon. This can be explictly seen in a model of fermions in a box in the absence of any external potential \footnote{This is good enough since the UV finiteness of a small interval should be insensitive to the details of the potential.} \cite{Das:2022nxo}. In this case, the leading contribution to the entanglement entropy is divergent to any finite order in a perturbation expansion in inverse powers of the fermi momentum. However the perturbation series can be summed and the re-summed result is finite.

The discussion above can be generalized to field theories in higher dimensions in several ways see \cite{das2020bulk, das2021gauge} for more details. 

There is another definition  of target space entanglement which is associated with internal symmetries in the boundary theory,  suggested in \cite{karch2015holographic} see also \cite{Anous:2019rqb}. This  is as follows: 
Take  the case where the internal space is $K = S^n$ which is a geometrization of a $SO(n+1)$ R-symmetry of the field theory which has scalar fields $X^I, I = 1 \cdots (n+1)$. Gauge invariant operators are then labelled by $SO(n+1)$ quantum numbers and are of the form of symmetric traceless products of the $X^I$\footnote{We are using the usual correspondence between spherical harmonics and polynomials made out of the coordinates $x^I$.} ,
\ben
\cO_{l,\vm} (\xi) = {\rm Tr} \left[ X^{(I}(\xi) X^J (\xi) \cdots X^{K)}(\xi)-{\rm trace} \right]
\label{0-6}
\een
where $\xi$ denotes a base space coordinate.
One can then fold these operators with $S^n$ spherical harmonics $Y_{l,\vm}(\theta_i)$ and form 
\ben
\cO(\xi,\theta_i) = \sum_{l,\vm} Y_{l,\vm}(\theta_i) \cO_{l,\vm} (\xi) 
\label{0-7}
\een
where $\theta_i, i = 1 \cdots n$ are angles on $S^n$. This is an operator which is localized in the base space of the field theory and in $S^n$ which is part of the target space. One can then consider the operators  obtained by taking $\xi$ to include the whole base space and with $\theta_i$ corresponding to a subregion $A$ of the $S^n$, and obtain a subalgebra of observables  by taking sums and products of operators of this type. 

This is also a target space entanglement, but different from the one which is obtained by imposing a projector of the type (\ref{0-2}).
The projector (\ref{0-2}) is applied on the individual matrix fields as in (\ref{0-3}) {\em before taking a trace} and the projected operator is made gauge invariant by taking a trace at the end. In contrast, the restriction on the angles $\theta_i$ in (\ref{0-7}) is performed on operators which are already gauge invariant.  As argued above, the first type of target space entanglement is naturally associated with measurements made on D branes in the bulk. On the other hand, the sub-algebra of operators obtained by restricting the angles in (\ref{0-7}) is closer in spirit to a restriction on supergravity modes which source these operators.

It was  conjectured \cite{karch2015holographic}  that an RT surface anchored on the boundary of $A$ and smeared along the $AdS$ space directions evaluates the von Neumann entropy for the sub-algebra of the second type we have discussed, eq.(\ref{0-6}) (\ref{0-7}). Such a subalgebra continues to make sense for instances of holography where the bulk space is not a product or asymptotically $AdS$, e.g. D0 brane holography, as discussed in \cite{Anous:2019rqb}.

%

For other ideas of entanglement in internal space see \cite{Gautam:2022akq}, and a related notion of entwinement see \cite{Balasubramanian:2014sra,Balasubramanian:2018ajb}.
One comment is worth making. This paper deals with entanglement of internal RT surfaces, but consider for a moment a conventional RT surface associated with a region of the base space in the boundary. Here it is clear that the subalgebra
whose entropy is being calculated by the RT surface is akin to the second one above, eq.(\ref{0-6}) (\ref{0-7}) and obtained by restricting to gauge invariant operators with support in the base space region of interest. Given the locus in the bulk of the RT surface we can also find another subalgebra of the first type we discussed above, by taking the function $f$ in eq.(\ref{0-2}), to correspond to this locus. It is unlikely that the RT surface will calculate the entanglement of this subalgebra as well. 

\subsection{A Road Map for What Follows}
The conjecture that the  second kind of target space entanglement we discussed above is  related to the area of  RT surfaces  associated with the internal spaces is quite natural.
However  so far there has been no very definite substantiation of this claim. 
One reason for this is a result due to  Graham and Karch \cite{graham2014minimal} (which we will refer to as the GK theorem below). This theorem pertains to 
extremal surfaces in direct  product spacetimes of the form, $Y^{AdS}_{n+2} \times K_m$, where $Y^{AdS}_{n+2}$ is an asymptotically $AdS$ space time and $K_m$ is an internal  compact space. The result, which we also discuss in section \ref{compact},  is that an RT surface, associated with a region $R$ of $K_m$, i.e. ending  at   asymptotic infinity on the boundary of $R$, $\partial R$,  
can exist only in very special circumstances, namely when $\partial R$ itself is a minimal surface of $R$ 
\footnote{Note that here, following the standard nomenclature, by a minimal surface of $R$, we mean a surface whose area does not change upto first order in small perturbations. This is to be contrasted with what is meant by  RT surfaces being  minimal. The  RT surface  minimises its   area subject to the boundary conditions at asymptotic infinity, or at the UV cut-off, being held fixed} .
This makes the study of such internal space  RT surfaces seem less interesting than their conventional counterparts. 
In the discussion below we will often refer to RT surfaces associated with a region $R$ of the internal space as being ``anchored" at the boundary in the internal space, or simply as being  internal RT surfaces.

Now, if we do not take the boundary of $Y^{AdS}_{n+2} \times K_m$ to be at asymptotic infinity, but instead at a finite cut-off, this restriction is lifted but the significance of the resulting RT surfaces is not clear then. Such surfaces were studied in  \cite{mollabashi2014entanglement} for the case of $AdS_5\times S^5$ with a UV cut-off and   it was proposed that the area of this surface evaluates the entanglement entropy between $SU(M) \times SU(N-M)$ subsectors of the $SU(N)$ gauge theory \footnote{A related discussion in the context of entanglement between two interacting scalar fields appears in 
\cite{MohammadiMozaffar:2015clv}}. 
Such surfaces  were also studied in \cite{karch2015holographic} by studying the Coulomb branch and it was suggested that the second definition above, eq.(\ref{0-6}) (\ref{0-7}), using the R symmetry group,  could provide a precise gauge invariant definition in
the boundary for the entropy evaluated by area of such an RT surface.


In one situation RT surfaces associated with general subregions $R$  of the internal space must clearly exist. 
This is the case when the  geometry can be extended further towards the UV, and the internal space directions  in the IR geometry are found to lie along  the directions of an asymptotic $AdS_{n+2}$ space. The  RT surface ending on $\partial R$, i.e. anchored in the internal space in the IR,   when extended to the full UV geometry, is then just a conventional RT surface, anchored at the boundary of the $AdS_{n+2}$ in the usual manner. And in the boundary dual to the $AdS_{n+2}$, the surface calculates the conventional entanglement entropy of   the boundary theory. 

A very concrete example, discussed in section \ref{ads2}, is provided by an  extremal RN black brane/hole where the IR theory is a warped product $AdS_2 ``\times" K_n$, with $K_n$ being $R^n$ or $S^n$. 
This example clearly shows that the GK theorem can be evaded. The key reason turns out to be the fact that   the IR geometry is not a direct product space, but instead a warped product spacetime. 
Before proceeding, we note that in what follows,  for brevity, we will often use the symbol $``\times"$ to denote a warped, rather than a direct product.


Motivated by this example we  consider several other cases in this paper, some containing  warping or  a UV cut-off  \footnote{Some of the  geometries we considered are not known, at least  to us,  to arise as solutions of gravity theories. We studied them here to illustrate  the various types of possibilities which can arise.}. In section \ref{adsn} we study higher dimensional 
cases where the warped product is of the form $AdS_{n+2}``\times " R^m$, with $n>0$.  We find that when the warping reaches a finite, non-vanishing,  limit in the deep IR, the behaviour is quite different from the $AdS_2``\times" S^n$, $AdS_2``\times" R^n$,  cases. Most importantly, in these product spaces the dependence of the entanglement entropy on the size of the subregion staurates once the latter exceeds a critical size. This indicates that whenever these product spaces appear as IR geometries of a higher dimensional asymptotically $AdS_{m+n+2}$, the UV field theory must have a  gap, with correlations having finite extension along the directions which will become the internal dimensions.
At low temperatures, $T$,  and for sufficiently big regions in $R^m$, the difference between the  finite temperature and zero temperature areas is independent of the  cutoff and scales like $T^n$.
In section \ref{compact} the asymptotic behaviour of  RT surfaces in general  warped backgrounds is discussed  and some explicit examples including asymptotically flat cases and $Dp$ brane geometries are discussed. 
In section \ref{uvcutoff} geometries with a UV cut off are discussed. Our investigation reveals that internal RT surfaces can exhibit a rich set of behaviours, sometimes,  going deep into the interior, and in other cases being close to the UV cut-off at the boundary. There can also be interesting cases, like in conventional RT surfaces, of finite length correlations in the system.

We conclude  the paper with a discussion  in section \ref{disc}. Here we also comment on the possible relation of internal RT surfaces with various subalgebras, including the ones discussed above, and also what we can hope to learn about bulk reconstruction using such RT surfaces, in the future. Important details are in appendices \ref{fta}, \ref{globaladsn}, \ref{Moredetana} and \ref{CFT}.

		\section{$ AdS_2$ in the Infrared}
	\label{ads2}
	In this section we consider situations which give rise to an $AdS_2$ geometry in the IR. This could arise for example in an extremal RN brane which is asymptotic to $AdS_4$ or to $AdS_n, n>2$ in general; or it could arise from an extremal RN solution in asymptotically flat space. For concreteness we focus here in the case of four dimensional spacetimes with an $AdS_2$ in the infra red. We expect our main conclusions  to extend to dimensions greater than $4$ as well. 
	Below, we first consider the case of an extremal RN brane in $AdS_4$ and in the following subsection the near-horizon geometry of a more general warped spacetime and examine extremal surfaces in these cases. 
	
	\subsection{$AdS_4\rightarrow AdS_2$ }
	\label{flowads4to2}
	We begin by  considering  an extremal Reissner-Nordstorm (RN) brane in the asymptotic $AdS_4$ with the boundary theory being CFT$_3$. The bulk metric is given by,
	\begin{equation}
		ds^2 = -f(r) dt^2 + \frac{dr^2}{f(r)} + r^2 (dx^2 + dy^2) 
		\label{metstart}
	\end{equation}
	with, 
	\begin{equation}
		r^2 f(r) = (r-r_h)^2 {(r^2 + 2 r r_h + 3 r_h^2)\over R_4^2},
	\end{equation}
	where $r_h$ is the location of the horizon. In large $r$ limit we have,
	\begin{equation}
		f(r) \rightarrow \frac{r^2}{R_4 ^2},
	\end{equation}
	while in the near horizon region,
	\begin{equation}
		f(r) \rightarrow \frac{(r-r_h)^2}{R_2 ^2}.
	\end{equation}
	Here $R_4$ is the AdS radius and,
	\begin{equation}
		\label{defr2}
		R_2 =\frac{R_4}{\sqrt{6}}
	\end{equation}
	is the radius of $AdS_2$.
	
	In the  field theory dual to the $AdS_4$ system this geometry corresponds to a state with chemical potential 
	\be
	\label{chemp}
	\mu = \sqrt{3} { r_h\over R_4^2}.
	\ee
	
	We want to find the entanglement entropy of the infinite strip,
	\begin{equation}
		x \in \left(\frac{-l}{2}, \frac{l}{2}\right), y \in (0, L)
	\end{equation}
	at the boundary. We will eventually take $L$ to infinity. To compute the entanglement entropy we first compute the minimal area following the RT prescription. We take a constant slice in time, then the area is given by,
	\be
	\label{area0}
	A=\int_{x=-l/2}^{x=l/2} dx r L \sqrt{ \frac{r'^2}{f(r)} + r^2 }
	\ee
	with $r' = \frac{dr}{dx}$.
	
	Note that the extremal area surface will be symmetric with $r(x)=r(-x)$, and $x$ will be single valued along the surface going from $(-l/2,l/2)$. The Euler Lagrange equation can be derived by taking $x$ to be the independent  variable and extremising with respect to $r(x)$. However we do not need this second order equation itself. Since the action is independent of $x$ the conjugate momentum, more correctly thought of as the Hamiltonian, will be conserved. 
	This is given by 
	\begin{equation}
	\label{defH}
    	H = P_r r' - \mathcal{L} = - \frac{r^3}{\sqrt{ \frac{r'^2}{f(r)} + r^2 }}
    \end{equation}
	leading to, 
	\be
	\label{simpl}
	(r^6-H^2r^2)={H^2\over f(r)} r'^2
	\ee	
	
	Before proceeding we note that while it is not strictly correct to regard $r$ as the independent variable, since it is not single valued along the surface, one can obtain the same conserved charge $H$ anyway by loosely regarding $x$ as the dependent variable and $r$ as the independent one; the conjugate momentum for $x$ then is given by $-H$.

At the turning point, $r'=0$, from eq.(\ref{simpl}) we then learn that 
	\begin{equation}
		H^2 = r_0 ^4.
	\end{equation}
	Hence, denoting ${\dot{x}}=dx/dr=1/r'$ we learn that 
	\begin{equation}
		\Dot{x} = \sqrt{\frac{r_0 ^4}{f(r) (r^6 - r_0 ^4 r^2) }}. \label{deltax}
	\end{equation}
	Integrating we have,
	\begin{equation}
		\Delta x = \int_{r_0} ^{r_{UV}} dr \sqrt{\frac{r_0 ^4}{f(r) (r^6 - r_0 ^4 r^2) }}. \label{deltaxint}
	\end{equation}

	In analysing the resulting solution and its area it will be convenient to consider  half the surface 
	along which $x\in (0,l/2)$ and $r\in (r_0, r_{UV})$ where $r_0$ is the turning point and $r_{UV}$ is the UV cut-off for $r$. This has area 
	\be
	\label{area1}
	A=L \int_{r_0}^{r_{UV}} dr {r^3\over \sqrt{f(r) (r^4-r_0^4)}}
	\ee

	For a large value of strip length  $l$, we expect the minimal area to go into the near horizon area with  $r_0$ being close to $r_h$, i.e. ${(r_0-r_h)\over r_h} \ll 1$.  We would like to examine the change in $x$, as the surface traverses the region far from the horizon, where the geometry is asymptotically $AdS_4$ and in the near-horizon region, which is approximately 
	$AdS_2$. We separate the two regions at the radial location $r_B$ where $r_B$ is of $\order{r_h}$, and  ${(r_B-r_h)\over (r_0-r_h)}\gg 1$.
	

	
	
	
	\textbf{Near the horizon:}
	\newline
     In this region the change in $x$, which we denote by $\Delta x$ is given by
	\begin{align}
		\Delta x &= \int_{r_0} ^{r_B} \sqrt{\frac{r_0 ^4}{f(r) (r^6 - r_0 ^4 r^2) }} dr  \nonumber\\
		& = \int_{r_0} ^{r_B} \sqrt{R_2 ^2 \frac{r_0 ^4}{(r-r_h)^2 (r^6 - r_0 ^4 r^2) }} dr  = \int_{r_0} ^{r_B} \frac{R_2 r_0 ^2}{ r (r-r_h) \sqrt{(r^4 - r_0 ^4)}} dr \nonumber \\
		& \simeq \int_{r_0} ^{r_B} \frac{R_2 r_0 ^2}{ r_0 (r-r_h) \sqrt{4 r_0 ^3 (r - r_0)}} dr  = \int_{r_0} ^{r_B} \frac{R_2}{2 r_0 ^{\frac{1}{2}}(r-r_h) \sqrt{(r - r_0)}} dr \label{eqx11} \\
		& \simeq \int_{r_0} ^{r_B} \frac{R_2}{2 r_0 ^{\frac{1}{2}} (r-r_0 + r_0 - r_h) \sqrt{(r - r_0)}} dr = \int_{0} ^{z_B} \frac{R_2}{r_0 ^{\frac{1}{2}}} \frac{1}{ (z^2 + \epsilon^2)} dz = \frac{R_2}{\epsilon r_0 ^{\frac{1}{2}}} \arctan{\left(\frac{z_B}{\epsilon}\right)},\label{eqx12}
	\end{align}
	with,
	\be
	\label{defeps}
	\epsilon=\sqrt{r_0-r_h}, z= \sqrt{r-r_0}, z_B = \sqrt{r_B-r_0}.
	\ee
	We see that by taking $\epsilon\rightarrow 0$, i.e. $r_0\rightarrow r_h$,  $\Delta x$ can be made as big as we want, with 
	\begin{equation}
		\label{tp2}
		\Delta x \propto \frac{R_2}{\sqrt{r_0 (r_0 - r_h)}}.
	\end{equation}

  \textbf{Far from the horizon:}

    The near horizon approximation we just did actually gives an upper bound on the value of the original integral, eq.\eqref{deltaxint}. This is due to the fact that the integrand is always positive and the near horizon approximation is always bigger than the original integrand even if we extend it to the large $r$ region. This implies that the remaining part of the integral has a contribution

    \begin{equation}
          \Delta x _{FH} = \int_{r_B}^{r_{UV}} \sqrt{\frac{r_0 ^4}{f(r) (r^6 - r_0 ^4 r^2)}} < \frac{R_2}{\epsilon r_0 ^{\frac{1}{2}}} \Big(\arctan{\left(\frac{z_{UV}}{\epsilon}\right)}-\arctan{\left(\frac{z_B}{\epsilon}\right)\Big)}
    \end{equation}

    Where 

    \be
    z_{UV}=\sqrt{r_{UV}-r_0}
    \ee
    
    Invoking the condition that ${(r_B-r_h)\over (r_0-r_h)}\gg 1$, we can use the asymptotic expansion of the $arctan$ function and taking $r_{UV} \rightarrow \infty$ we get
   
    \be
	\Delta x_{FH} < \frac{R_2}{\sqrt{r_0 (r_B-r_0)}} .
	\ee
    
     Using the fact that $r_B, r_0 \sim \order{r_h}$ we get from the above equation,
    \begin{equation}
    	\Delta x_{FH} \sim \frac{R_4}{r_h} \label{farho}.
    \end{equation}
    And so in contrast with the contribution coming from the near horizon region we see that there is a finite bound on the contribution to $\Delta x$ coming from this region.

	
	
	Let us now turn to evaluating the area, $A$, of the extremal surface. We have in mind taking the strip length $l$ to be large enough, so that we get an extremal surface that goes close to the horizon.
	\newline
	 \textbf{Far from the horizon:}
	
 We start by evaluating the $r > r_B$ region contribution to the area this time. Using a somewhat different strategy we Taylor expand the integrand in powers of ${(r_0-r_h)\over r_h}$ so we have
 
 \begin{align*}
		\Delta A & = \int_{r_B} ^{r_{UV}} L \frac{r^3 dr}{\sqrt{f(r) (r^4 - r_0 ^4)}} \\
		& \sim \int_{r_B} ^{r_{UV}}  L \Bigg[{\frac {{r}^{3}}{ \sqrt{f(r)  \left( {r}^{4}-{{r_h}}^{4} \right) }}}+2\frac {{r}^{3}{{ r_h}}^{4}}{\sqrt{f(r)  \left({r}^{4}-{{r_h}}^{4}\right)} \left({r}^{4}-{{r_h}}^{4} \right) }\Big(\frac{{r_0}-{r_h} }{r_h}\Big) +\mathcal{O}\Big(\Big(\frac{{r_0}-{r_h} }{r_h}\Big)^2\Big)\Bigg]dr \end{align*}

 After integration, the first term produces a diverging contribution when $r_{UV} \rightarrow \infty$, but this contribution does not depend on the location of the turning point and as such is universal for these surfaces and we can just subtract it. 
 All other higher order terms give finite contributions and clearly vanish in the limit $r_0 \rightarrow r_h$. This clearly shows that in this limit we should focus on the contribution coming from the near horizon region.
 
    \textbf{Near the horizon:}
	This region gives, 
	\begin{align*}
		\Delta A & = \int_{r_0} ^{r_B} L \frac{r^3 dr}{\sqrt{f(r) (r^4 - r_0 ^4)}} \\
		& \simeq {R_2\over 2}  \int_{r_0} ^{r_B} L r_0 ^{\frac{3}{2}} \frac{dr}{(r-r_h) \sqrt{r-r_0}} .
	\end{align*}
	It is easy to see that  this integral is of the same form as  \eqref{eqx11}. Denoting  $L$ by  $\Delta y$ and noting that  $r_0 \simeq  r_h$ we can write the area of the surface in the near-horizon region as,
	\begin{equation}
		\label{fpads2}
		\Delta A \simeq \Delta x \Delta y \ r_h^2.
	\end{equation}
	We see that the finite part of the entanglement which is independent of the UV cut-off $\delta$ comes from the near horizon region. It is extensive in the area $\Delta x \Delta y$, like a thermodynamic entropy. We note from the metric, eq.(\ref{metstart}) that in our conventions $\Delta x, \Delta y$ are dimensionless.
	Working with re-scaled lengths $\Delta x  \Delta y\rightarrow R_4^2 \Delta x \Delta y\equiv V$ then  it is easy to see  that the contribution this area makes to the entanglement is  
	\be
	\label{conen}
	{\Delta A \over 4 G_N}\simeq {R_4^2\over 4 G_N} V \mu^2 \sim N^2 V \mu^2,
	\ee
	where we have used eq.(\ref{chemp})
	This makes it clear that  it is the chemical potential which plays the role of the ``effective temperature" here. 
	
	Let us note that an alternate derivation of eq.(\ref{fpads2}), along with the first subleading correction is discussed in more detail in Appendix \ref{CFT}.
	
	It is worth understanding more quantitatively how big $\Delta x$ must be for the surface to go deep inside the near-horizon region. 
	If the surface goes deep inside, the turning point must be very close to the horizon, giving rise to the condition. 
	\be
	\label{condd}
	{r_0-r_h\over r_h}  \ll 1.
	\ee
	From  eq.(\ref{tp2})  this leads to,
	\be
	\label{conds}
	{R_4^2\over r_0 r_h (\Delta x)^2} \ll  1 \implies R_4 \Delta x \gg {R_4^2\over r_h}\sim {1\over \mu}
	\ee
	where in the  we have substituted $r_0\simeq r_h$, $R_2\sim R_4$  and used the relation eq.(\ref{chemp}). Noting that in our units $R_4 \Delta x$ has dimensions of length we see that the length of the strip  we are considering in the UV boundary theory must be much bigger than unity in units of the chemical potential for the RT surface to penetrate deep inside. Since the theory dual to extremal RN geometry has only one scale $\mu$, this is to be expected. 
	
	Let us end this subsection with some comments. 
	First,  while we considered a particular geometry on the boundary, corresponding to a strip of extent $l$, our conclusions are much more general. It is clear that for a general region on the boundary, once its size is much bigger than the inverse chemical potential, the corresponding extremal surface will go deep into the near-horizon region, and the finite part (independent of the UV cut-off ) of the entanglement entropy will then be extensive in the volume of the region in units 
	of the chemical potential. 
	
	Second, since we see that the finite part of the entanglement, independent of the UV cut-off arises from the near-horizon region it is worth restating our results 
	above in terms of the $AdS_2\times R^2$ region itself.  The result will then be valid much more generally, independent of the 
	UV asymptotics which give rise to the $AdS_2$ region. 	From the point of view of a 1-dimensional dual to the $AdS_2$ we would be computing entanglement in target space. The $AdS_2$ has a radius $R_2$.  For our present purposes we take the boundary to be at radial location $r_B$ satisfying the condition 
	\be
	\label{condb}
	{r_B-r_h\over r_h}<\sim  \order{1}.
	\ee
	Now consider a region  ${\cal R}$ of the transverse $R^2$ at $r=r_B$ and an  extremal surface which at $r=r_B$ is ``pegged" on the boundary of ${\cal R}$. For the case considered above $R_4\sim R_2$, eq.(\ref{defr2}). Thus the condition eq.(\ref{conds})  can be equally well stated as 
	\be
	\label{statcd}
	\Delta x \gg {R_2\over r_h},
	\ee
	and we learn that for ${\cal R}$ having a linear extent meeting  eq.(\ref{statcd}) the extremal surface will go very close to the horizon.
	The resulting area will then scale with the volume  of the region ${\cal R}$, $\Delta x \Delta y $,  leading to an entropy
	\be
	\label{enta}
	S={1\over 4 G_N} r_h^2 \Delta x \Delta y.
	\ee
	
	Third, so far we have been considering planar black branes. The case of a  black hole, where we start with $AdS_4$ in global coordinates in the UV and flow to $AdS_2\times S^2$ in the IR, will be discussed in section \ref{other}.
	Finally, these considerations can be extended to near-extremal RN branes easily as is discussed in Appendix \ref{fta}. 			
	
	\subsection{More General Black Branes with  $AdS_2$ in the IR}
	\label{mgads2}
	We saw above that when the region on the $AdS_4$ boundary had a large enough spatial extent, the dominant contribution to the RT surface, after a UV subtraction,  comes from the near-horizon region. Keeping this in mind it is worth understanding the key features of the calculation above from a near-horizon point of view. In this section therefore we  consider a more general situation where the near horizon region is of the form
	\begin{eqnarray}
		\label{nh}
		ds^2 & = & -f(r) dt^2 + {dr^2\over f(r)} + \Phi^2 (dx^2+dy^2),\label{nh1}\\
		f(r) & = & {(r-r_h)^2\over R_2^2},\label{nh2}
	\end{eqnarray}
	i.e. with a near horizon $AdS_2$ of radius $R_2$ and the volume of the internal space being given by the dilaton $\Phi^2$. We take the dilaton to vary with $r$; as we will see this variation is actually important in obtaining the minimal areas surface. 
	Note that this near-horizon region could have arisen  from an $AdS_4$ or an $AdS_{d+1}$ geometry in the UV, or even from a geometry with very different asymptotic behaviour. 
	
	We consider a boundary of the $AdS_2$ spacetime which is located at $r=r_B$ and take a region ${\cal R}$ on the boundary of extent $\Delta x$ and $\Delta y$ in the $x,y$ directions respectively. For simplicity we again take ${\cal R}$  to be a strip with $\Delta y = L$ being very big, i.e. $L\rightarrow  \infty$. We are interested in the analogue of the RT surface ending at $r_B$ along the boundary of  ${\cal R}$.

	Repeating the analysis above, and omitting some of the steps, we get that the conserved quantity, which  is analogous to $H$ in  eq.(\ref{defH}) but  denoted  as $P_x$ below, is given by 
	\be
	\label{valP}
	P_x={ \Phi^3 {\dot{x}} \over \sqrt{{1\over f} + \Phi^2 {\dot x}^2 }}.
	\ee

	If the dilaton takes value $\Phi_0$ at turning point, we have 
	\be
	\label{valpx}
	P_x=\Phi_0^2,
	\ee
	and the total extent traversed along the $x$ direction, $\Delta x$, in going from the boundary of the $AdS_2$ region to the turning point is 
	\be
	\label{rtra}
	\Delta x = \int_{r_0}^{r_B} dr {\sqrt{\Phi_0^4\over f(r) \Phi^2 ( \Phi^4-\Phi_0^4 )}},
	\ee
	where $r_B,r_0$ are the radial locations of the boundary of the $AdS_2$ region and of the turning point corresponding to $\Phi_0^2$, respectively. The formula for area is,
	\begin{equation}
		A = L \int_{r_0}^{r_B} dr \frac{\Phi^3}{\sqrt{f(r) ( \Phi^4-\Phi_0^4 )}}. \label{areagen}
	\end{equation}
	
	Now drawing from the discussion of the previous section let us take $\Delta x$ to be sufficiently big, we expect then that  
	the dominant contribution to $\Delta x$ will come from the region close to the turning point, which in turn will be close to the horizon. 
	
	In this region we can approximate the dilaton as
	\be
	\label{appr}
	\Phi^4-\Phi_0^4\simeq4 \Phi_0^3  (\Phi-\Phi_0),
	\ee
	where $\Phi_0$ is the value of the dilaton at the horizon (the attractor value). 
	Note truncating the expansion and only keeping the first term in the Taylor series is justified only if 
	\be
	\label{condsgen}
	{\Phi-\Phi_0\over \Phi_0}\ll 1.
	\ee
	We will see shortly that  this condition is self-consistently true. 
	Inserting  eq.(\ref{appr}) in eq.(\ref{rtra}) we get 
	\be
	\label{lo}
	\Delta x \sim  \int_{r_0}^{r_B} dr{1\over \sqrt{\Phi_0}(r-r_h) \sqrt{\Phi(r)-\Phi_0}},
	\ee
	where we have substituted for $f(r)$ from  eq.(\ref{nh2}). 
	Eq.(\ref{lo})  is analogous to eq.(\ref{eqx12}) above (with $\Phi(r)=r$) and determines the turning point as a function of $\Delta x$
	We see that the nature of the variation of $\Phi$ in the near-horizon region is important in determining where the turning point is located.

	Let us now consider a situation where both the horizon value of the dilaton, $\Phi_h$ and its first derivative with respect to $r$, $\Phi'_h$,
	do not vanish.  This is what happens in the case considered in the previous section where we took $\Phi=r$. 
	In such cases the analysis of the previous section immediately leads to 
	\be
	\label{conca}
	\Delta x \propto {1\over \sqrt{\Phi_0} \sqrt{\Phi_0-\Phi_h}}.
	\ee
	And from eq.(\ref{areagen}) we see  that the area to good approximation, after an $r_B$ dependent subtraction,  goes like
	\be
	\label{vala}
	A\sim \Delta x \Delta y \Phi_h^2.
	\ee
	From eq.(\ref{conca}) it follows that for sufficiently large $\Delta x$ the condition eq.(\ref{condsgen}) is also valid. 
	
	One comment is worth making here. Notice that in going from eq.(\ref{rtra}) to eq.(\ref{lo}) we replaced the metric component $f(r)$ by its leading behaviour $(r-r_h)^2$, however we had to keep the radial variation of the dilaton from its horizon value. It is easy to see that the deviation of $f(r)$ from its leading behaviour only makes a subleading contribution
	for RT surfaces, when $\Delta x$ is sufficiently big, so that $r_0$, the turning point, meets the condition, eq.(\ref{condd}) (more generally the area $\Delta x \Delta y$  has to be   sufficiently big). 
	The entanglement calculation therefore reveals something important about the near $AdS_2$ region, also seen in low- energy scattering, etc. Namely, that only the deviation of the dilaton from its horizon value is important, and not of the components of the metric along the
	$r,t$ directions, although they are formally of the same fractional order $\order{r-r_h\over r_h}$. This fact is responsible for why, quite universally, the near-$AdS_2$ region in near-extremal black holes and branes is described by JT gravity and the Schwarzian action, \cite{nayak2018dynamics, moitra2019extremal,maldacena2016remarks,iliesiu2021statistical,Heydeman:2020hhw}.

	The advantage of our more general discussion here  is that we can readily consider other situations as well. 
	For example, consider a case where the attractor value of the dilaton, $\Phi_h$,  does not vanish, and   in the vicinity of the horizon it takes the form
	\be
	\label{formdila}
	\Phi = \Phi_h + \alpha (r-r_h)^{p},
	\ee
	where $p$ is a  general  positive number. We also take $\alpha>0$, so that the dilaton decreases as it approaches the horizon. 
	
	It is easy to see that eq.(\ref{lo}) is still valid in this case as long as  eq.(\ref{condsgen}) is true. The integral can then be done in terms of the variable
	\be
	\label{varz}
	z=\sqrt{\Phi-\Phi_0}
	\ee which leads to an expression very analogous to eq.(\ref{eqx12}),
	\be
	\label{lt}
	\Delta x\sim {1\over \sqrt{\Phi_0}} \int {d z \over z^2 + \epsilon^2},
	\ee
	where $\epsilon^2=\Phi_0-\Phi_h$.
	As a result $\Delta x$ is given, even in the more general case meeting eq.(\ref{formdila}),  by eq.(\ref{conca}) and the Area  is given by eq.(\ref{vala}). We also see that eq.(\ref{condsgen}) is valid for sufficiently large $\Delta x$.

	
	We have not discussed the  the ``far region" between the $AdS_2$ boundary and the boundary of the UV region in this section. 
	As long as this far region,  after a suitable UV subtraction,  does not make a significant contribution to the finite part of the RT surface's area,    we then see  in considerable generality that the dominant behaviour of the area will be given by eq.(\ref{vala}) for sufficiently large values of   $\Delta x, \Delta y$. 
	
	Let us finally consider one more cases where the horizon  value  $\Phi_h$ vanishes. We take a metric of the form		 
	\be
	ds^2   =  -g(r) dt^2 + {dr^2 \over f(r)} + \Phi^2 (dx^2+dy^2),\label{nhmeta}
	\ee
	\be
	\Phi  =  (r-r_h)^\delta. \label{nhmetb}
	\ee
	Here $g(r)$ vanishes at $r=r_h$ but we will not have to be specific about  its exact behaviour. In such cases	from eq.(\ref{rtra}) we get that,
	\begin{align}
		\Delta x & \sim \int_{r_0}^{r_B} dr {\sqrt{\Phi_0^4\over (r-r_h)^2 \Phi^2 ( \Phi^4-\Phi_0^4 )}} \nonumber \\
		& \sim \int_{1}^{\xi_B} \frac{1}{\Phi_0}\frac{d \xi}{\xi^{1+\delta} \sqrt{\xi^{4 \delta}-1}},   \label{delta}
	\end{align}
	where we have taken,
	\be
	\label{defxia}
	\xi = \frac{r-r_h}{r_0 - r_h}.
	\ee
	The integral in \eqref{delta} is well behaved and will  give a finite numerical value leading, for sufficiently big $\Delta x$, to the result 

	\be
	\label{resint}
	\Delta x \sim { \kappa \over \Phi_0}.
	\ee
	with,
	\begin{equation}
		\kappa = \frac{\sqrt{\pi} \Gamma(\frac{3}{4})}{\delta \Gamma(\frac{1}{4})}
	\end{equation}
	As a result we see that in this case too, when $\Delta x$ increases, $\Phi_0$ must decrease as $r_0\rightarrow r_h$, so that the turning point comes closer to the horizon. 
	In fact, the relation, eq.(\ref{resint}) agrees with eq.(\ref{conca}) after setting $\Phi_h=0$. And actually it is easy to see that the result eq.(\ref{resint}) follows just from a simple scaling relation. The spatial part of the metric in eq.(\ref{nhmeta})  (where we do not include the $ dt^2$ term since it is not   relevant  for determining the minimum area surface)  is given by
	\be
	\label{spatp}
	ds^2={dr^2\over (r-r_h)^2} + (r-r_h)^{2 \delta} (dx^2+dy^2),
	\ee
	and therefore invariant under $(r-r_h)\rightarrow \lambda (r-r_h)$,  $ (x,y) \rightarrow {1\over \lambda^{\delta}} (x,y) $. 
	The relation eq.(\ref{resint}) follows from this.
\par Finally let us  turn to evaluating the area in this case. From  eq.(\ref{areagen}) it is given by 
	\begin{align}
		A & \sim L \int_{r_0}^{r_B} dr  \frac{\Phi^3}{\sqrt{(r-r_h)^2 (\Phi^4 - \Phi_0 ^4)}} \nonumber \\
		& \sim L \Phi_0 \int_{1}^{\xi_B} d\xi \frac{\xi^{(3\delta - 1)}}{\sqrt{\xi^{4 \delta}-1}} \sim - \kappa L \Phi_0,
	\end{align}
	where we have used the variable $\xi$ defined in eq.(\ref{defxia}) and in the second line, subtracted an $r_B$ dependent term which makes the resulting integral well behaved and gives a finite numerical factor.
From eq.(\ref{resint}), and denoting $L=\Delta y$  we see then that in this case 
	\be
	\label{relaa}
	A \sim - \kappa^2 {\Delta y \over \Delta x}.
	\ee
 	This result can be equivalently written as 
 \be
 \label{equr}
 A\sim - \Delta y \Delta x \Phi_0^2,
 \ee
 which is more analogous to what was obtained above, eq.(\ref{vala}). From eq.(\ref{equr}) we also see that the result for $A$ is consistent with the scaling symmetry discussed above. 
	
	
	Let us end by noting that  if $g(r)$ the $g_{tt}$ component, eq.(\ref{nhmeta}) vanishes as 
	$g(r)\sim (r-r_h)^{2\alpha}$ then the near horizon geometry above is of Lifshitz type, with scaling symmetry,
	$(r-r_h)\rightarrow \lambda (r-r_h), t\rightarrow t/\lambda^\alpha$, $(x,y)\rightarrow  {1\over \lambda^\delta} (x,y)$.


	\section{Higher Dimensional $AdS$ spaces } 
	\label{adsn}
	Here we consider metrics of the form  $AdS_{n+2}$ $``\cross"$ $R^m$, with $n>0$,  
	\begin{equation}
		ds^2 = R^2 r^2 (-dt^2 + \sum_{i} (dx^i)^2) + R^2 \frac{dr^2}{r^2}  + R^2 \Phi(r) ^2  \sum_{\mu} (dy^{\mu})^2. \label{productmet}
	\end{equation}
	The indices $i$ range from 1 to $n$ while $\mu = 1,2,....m$. $R$ is the radius of $AdS_{n+2}$ and  $\Phi$ is the dilaton that depends in general on $r$. Note that this radial dependence of  $\Phi$, results in the spacetime being   a warped product, which is why, consistent with the   notation introduced at the beginning, we used the  symbol $``\cross"$  above.

	Let us also note that a geometry of this type can sometimes arise  in the IR starting from an $AdS_M$ space where $M>n+2$, in the UV. At the end of  subsection \ref{consdil} we will in fact consider one such example of a $AdS_3 ``\cross" R^2$  dimensional spacetime which is asymptotic in the UV to $AdS_5$. 
	
	We take the boundary of this spacetime to be at a large value of $r$ which we denote by $r_{UV}$.
	In a manner analogous to the previous section we are interested here  in a region on the boundary at constant time, say $t=0$ which corresponds to a strip of length $l$ along $y_1$ 
	\begin{equation}
		0 \leq y_1 \leq 2\Delta y_1.
	\end{equation}	The region extends fully along all the remaining $y^2, \cdots y^m$ directions which we take  to have a volume $V_{m-1}$. In the bulk we are then interested in an extremal surface at $t=0$ which ends at $r_{UV}$  on the boundary of this strip at $y_1=0, y_1= 2 \Delta y_1$ and wraps all the remaining $y^\mu$ directions.   
	
	The area of this surface is given by 
	\begin{equation}
		A = R^{(m+n)} V_{n+m-1} \int r^{n} \Phi ^{m-1} \sqrt{\frac{dr^2}{r^2} + \Phi ^2  (dy_1)^2}.   \label{areaeq}
	\end{equation}
	The factor in front is the volume of the rest $n+m-1$ directions. 
	The conjugate momenta to $y_1$, analogous to $H$, eq.(\ref{defH}), is,
	\begin{equation}
		P_y = r^n \frac{\Phi ^{m+1} \dot{y_1}}{\sqrt{\frac{1}{r^2} + \Phi ^2 \Dot{y_1}^2}}.
	\end{equation}
	Rearranging the above equation we have,
	\begin{equation}
		\Dot{y_1} = \frac{P_y}{r \Phi \sqrt{r^{2n} \Phi ^{2 m} -P_y^2 }}.\label{turneq}
	\end{equation}
	The turning point is given by,
	\begin{equation}
		P_y = r_0 ^n \Phi_0 ^m.
	\end{equation}
	Thus we can write,
	\begin{equation}
		\Delta y_1 = \int_{r_0} ^{r_{UV}} \frac{P_y}{r \Phi \sqrt{r^{2n} \Phi ^{2 m} -P_y^2 }} dr. \label{yeq}
	\end{equation}
	Note that $\Delta y_1$ is only half the interval of the strip. 
	Then \eqref{yeq} becomes,
	\begin{eqnarray}
		\Delta y_1 &= & \int_{r_0} ^{r_{UV}} \frac{ r_0 ^n \Phi_0 ^m}{r \Phi \sqrt{r^{2n} \Phi ^{2 m} -r_0 ^{2 n} \Phi_0 ^{2m}}} dr\nonumber\\
		&=& \frac{1}{\Phi_0} \int_{1} ^{\xi_{UV}} \frac{d \xi}{\xi \Tilde{\Phi} (\xi) \sqrt{\xi^{2n} (\Tilde{\Phi} (\xi)) ^{2 m} -1}}. \label{exply}
	\end{eqnarray}
	Here we have taken $\xi = \frac{r}{r_0},\xi_{UV} = \frac{r_{UV}}{r_0}, \Tilde{\Phi} = \frac{\Phi}{\Phi_0}$.

		\subsection{Constant Dilaton:}
	\label{consdil}
	
	Let us now consider the special case where $\Phi(r)$ a constant independent of $r$. 
	Note that in this case we have a direct product space $AdS_{n+2}\times R^m$. One of the reasons for studying this example  is that the internal space $R^m$ is non-compact. Therefore the considerations of the Graham-Karch theorem, mentioned at the beginning of this note, do not directly apply.

	We denote the constant value of the dilaton as $\alpha$ below. 
	In this case ${\tilde \Phi}=\Phi/\Phi_0=1$  and from eq.(\ref{exply}) we get that

	\begin{equation}
		\Delta y_1 = \int_{r_0}^{r_{UV}} \frac{r_0 ^n }{\alpha r \sqrt{r^{2n} - r_0 ^{2n}}} d r = \frac{1}{\alpha} \int_{1}^{\xi_{UV}} \frac{d \xi}{\xi \sqrt{\xi ^{2n} -1}} .\label{constphi}
	\end{equation} 
	Here $\xi = \frac{r}{r_0}$ and  $\xi_{UV} = \frac{r_{UV}}{r_0}$. 
	
	The equation \eqref{constphi} can be exactly integrated and the answer is,
	\begin{equation}
		n \alpha \Delta y_1 = \arcsec(\xi_{UV} ^n) \implies r_0 ^n = r_{UV} ^n \cos( n \alpha \Delta y_1). \label{turningpointconstphi}
	\end{equation}
	For $r_0\ll r_{UV}$, i.e., $\xi_{UV} \gg 1$, $ n\alpha \Delta y_1 \rightarrow \pi/2$. This behaviour can be understood directly from the integral eq.(\ref{constphi}). 
	Rewriting the  integration as,
	\begin{equation}
		\Delta y_1 = \frac{1}{\alpha} \int_{1}^{\infty} \frac{d \xi}{\xi\sqrt{\xi ^{2n} -1}} - \frac{1}{\alpha} \int_{\xi_{UV}}^{\infty} \frac{d \xi}{\xi \sqrt{\xi^{2n} -1}},
	\end{equation}
	one finds that the first term can be exactly integrated. And since we assume that $\xi_{UV} \gg 1$ we can expand the second term in the large $\xi$ limit and only keep the leading term. Then we get,
	\begin{equation}
		\Delta y_1 = \frac{\pi}{2 n \alpha} - \frac{r_0 ^n}{n \alpha r_{UV}^n}, \label{resulty}
	\end{equation}
	which gives the behaviour for small $\frac{r_0}{r_{UV}}$. 
	
	More generally, the shape of the surface $r(y_1)$ is given by
	\begin{equation}
		r^n = r_{UV}^n \cos(n\alpha \Delta y_1)\sec(n\alpha(y_1- \Delta y_1))
		\label{shape}
	\end{equation}
	Using eq.\eqref{yeq} we get for the area,
	\begin{align}
		A &= 2 R^{(m+n)} V_{n+m-1} \alpha ^{m-1} \int_{r_0}^{r_{UV}} \frac{r ^{2n} d r }{r \sqrt{r ^{2n} -r_0 ^{2n}}}  \nonumber \\
		& =  2 R^{(m+n)}V_{n+m-1} \alpha ^{m-1} \frac{r_0 ^n}{n } \sqrt{\xi_{UV} ^{2n} -1} = 2 R^{(m+n)} \frac{V_{n+m-1}}{n} \alpha ^{m-1} \sqrt{r_{UV} ^{2n} - r_0 ^{2n}} \nonumber \\
		&=  2 R^{(m+n)} \frac{V_{n+m-1}}{n} \alpha ^{m-1} r_{UV} ^n \sin( n \alpha \Delta y_1). \label{constphiarea1}
	\end{align}
	We see that $A$  is divergent, as expected,  when $r_{UV} \rightarrow \infty$. Comparing it with the area of surface which just ``hangs" near the boundary, i.e. at $r=r_{UV}$,
	\begin{equation}
		A_{2} = 2 R^{(m+n)}  \frac{V_{n+m-1}}{n} \alpha ^{m-1} r_{UV} ^n (n \alpha \Delta y_1), \label{constphiarea2}
	\end{equation}
	we see that $A_2$ is always greater than $A$. 
	
	There is a third surface also which we should also keep in mind. 
	This surface has $\dot{y}=0$ and has two parts to it with $y$ taking the two values at the end of the strip. One branch is at $y_1=0$ and the other at $y_1= 2 \Delta y_1$. 	The area of these two parts put together is 
	\be
	\label{atwop}
	A_3=2 R^{(m+n)} {V_{n+m-1}\over n}\alpha^{m-1} r_{UV}^n.
	\ee
	For $ \Delta y_1 < \frac{\pi}{2 n \alpha}$, it is easy to see that the smallest area is given by $A$ in eq.(\ref{constphiarea1}). 
	For $ \Delta y_1 > \frac{\pi}{2 n \alpha}$ the surface corresponding to eq.(\ref{constphiarea1}) ceases to exist and the minimal area  is given by 
	$A_3$ eq.(\ref{atwop}). 
	Notice that both $A$ and $A_3$ scale like the UV cutoff $r_{UV}^n$ and thus are strongly UV dependent.

	The exchange of dominance  we have found indicates that there are quantum correlations whose behaviour changes significantly for length scales along the transverse $R^m$ directions exceeding (in suitable units) the value 
	\be
	\label{lcrit}
	(\Delta y)_{\rm crit}={\pi \over 2n \alpha}
	\ee
	This behaviour is similar to what was observed for base space entanglement in  $Dp$ brane geometries  due to a    phase transition analogous to the confinement/deconfinement  transition\cite{klebanov2008entanglement}. 
	
	To get a better understanding of this behaviour it is useful to consider a situation where the $AdS_{n+1}\times R^m$ arises in the IR starting from a higher dimensional $AdS$ space. 
	One such example is provided by the flow studied in d'Hoker and Kraus \cite{d2009magnetic,DHoker:2010xwl},  where a magnetic field  $B$ is turned on in the field theory dual to $AdS_5$  resulting in the IR in an $AdS_3\times R^2$ geometry\footnote{Strictly speaking in this case the dilaton in the IR cannot be taken to be constant, and will vary along the radial direction, but the qualitative features do not change, as is also discussed in section \ref{vardilsec} .}. We will take the magnetic field component $F_{xy}$ to be non-zero, and the $R^2$ in the IR then lies along the $x,y$ directions. 
	
	In this case the UV scale for the $AdS_3\times R^2$ geometry $r_{UV}$ is set by the magnetic field $B$ with 
	\be
	\label{valrm}
	r_{UV} \sim \sqrt{B}R^2,
	\ee
	where $R$, the radius of the $AdS_3$, is of the same order as the radius of the $AdS_5$. 
	The horizon  value of the dilaton is also  determined by $B$ and goes like 
	\be
	\label{valdih}
	\Phi^2=  {B\over \sqrt{3} R^2}.
	\ee
	Thus the condition $\Delta y ={\pi \over 2n \alpha}$ becomes in terms of a re-scaled  coordinate $y \rightarrow R y $, which has dimensions of length on the $AdS_5$ boundary, the condition
	\be
	\label{condxx}
	\Delta y \sqrt{B}\sim O(1).
	\ee
	with the coefficient on the RHS depending on the precise relation between $R, R_5$ etc. 
	We see in this example then that  the length scale which characterises the exchange of dominance of the RT surfaces  is set by the magnetic field.
	The system, in the presence of the magnetic field acquires quantum correlations along the $x,y$ directions  of length scale $L\sim 1/\sqrt{B}$. For regions in the $x-y$ plane  directions which are much bigger than this scale the entanglement changes in behaviour and starts  scaling with the perimeter of this region with the corresponding surface being given in the $AdS_3\times R^2$ region by   $A_3$ which is independent of $\Delta y_1$.
	For regions of smaller size but close to the transition, so that the surface enters the IR $AdS_3$ region, the corresponding entanglement is given by 
	$A$, eq.(\ref{constphiarea1}), instead of $A_3$. Note also that in this case $n=1$ and the scaling of both $A,A_3$ as $V_n r_{UV} ^n$, noted above,   means that the quantum correlations  arising due to the magnetic  field are extensive in the 3rd direction, which lies along the $AdS_3$, with a scale set by $B$. 
	
	Let us end this subsection with a few comments. 
	First, it is  worth comparing the behaviour we have found above with what we saw for the $AdS_2$ case (with $n=0$). In the  results we have obtained above depends  the  UV cut-off appears in a multiplicative manner. In contrast for the $AdS_2$ the UV dependence is an additive factor and once it is removed the renormalised entropy has an interesting finite part in the  independent of the UV cutoff. And there was no  exchange of dominance between extremal surfaces with the UV finite part eq.(\ref{fpads2}) being  extensive in the volume of the transverse space, in units of the chemical potential. 
	
	Second, our consideration in this section tell us that the features we saw above for the flow from $AdS_5\rightarrow AdS_3\times R^2$ should be true more generally for flows from higher dimensional $AdS$ spaces to $AdS_{n+2}, n>0$. In the $AdS_5$  example we considered the magnetic  field, $B$, sets the scale for both the horizon value of the dilaton $\alpha$  and the UV cut-off $r_{UV}$.  However it could well be that more generally these two scales are different. In such cases the entanglement entropy in the transverse directions will be characterised by the scale set by  the horizon value of the dilaton,  and this scale will decide the exchange of dominance between the two extremal surfaces.  The resulting entanglement will always scale extensively with the volume of the $AdS_{n+2}$ with a scale $r_{UV}$ set by the energy  in the UV at which departures  from the  $AdS_{n+2}$ occur. 
	Third, one can consider for this case with a constant dilaton, a finite temperature deformation where the $AdS_{n+2}$ is replaced by a black brane in $AdS_{n+2}$ space. This is discussed in the appendix in \ref{ftadsn}.
Here it is shown that the difference between the finite temperature and zero temperature areas is independent of the UV cutoff, for a sufficiently big region,  and scales as $T^n$ (where $T$ is the temperature) in the low temperature limit.
	
	Finally, one can also consider replacing the poincare patch $AdS_{n+2}$ with hyperbolic space in global coordinates. This is discussed in the appendix in \ref{globaladsn}.

	\subsection{Non Constant Dilaton}
	\label{vardilsec}
	After dealing with constant dilaton in the last subsection, we will now discuss dilatons that depend on $r$. Our considerations are quite general, we only assume that the dilaton is monotonic, growing as one goes away from the horizon to larger values of $r$.  
	
	To know how $\Delta y_1$ changes we use \eqref{exply}. We reproduce the equation here for convenience.
	\begin{equation}
		\Delta y_1 = \frac{1}{\Phi_0} \int_{1} ^{\xi_{UV}} \frac{d \xi}{\xi \Tilde{\Phi} (\xi) \sqrt{\xi^{2n} (\Tilde{\Phi} (\xi)) ^{2 m} -1}}.  \label{exply2}
	\end{equation}
	Here we have taken $\xi = \frac{r}{r_0},\xi_{UV} = \frac{r_{UV}}{r_0}, \Phi_0=\Phi(r_0)$, and $ \Tilde{\Phi} = \frac{\Phi}{\Phi_0}$,  $r_0$ is the turning point of the RT surface.

	We saw in the constant dilaton case that as $r_0\rightarrow 0$, and  approached the horizon, $ \Delta y_1$ approached it's maximum. To examine the behaviour of eq.(\ref{exply2}) we first note that  as $r_0 \rightarrow 0$, $\xi_{UV} \rightarrow \infty$ for any fixed $r_{UV}$. So, as $r_0 \rightarrow 0$
	\be
	\label{bdgen}
	\Tilde{\Phi} (\xi)= \frac{\Phi(r)}{\Phi(r_0)}=\frac{\Phi(r_0 \xi) } {\Phi(r_0)}\rightarrow 1.
	\ee
	The last limit is obtained by taking $\xi$ fixed and $r_0\rightarrow 0$, and is appropriate for evaluating the integral in eq.(\ref{exply2}), since the range for $\xi$ becomes $ [1,\infty)$, independent of $r_0$,  in the limit when $r_0\rightarrow 0$. 
	
	Note that in the above manipulation we have assumed that $\Phi(r=0)$  is nonzero and also taken $\Phi$ to be a smooth function near $r=0$. We take 
	\be
	\label{defga}
	\Phi(r=0)=\alpha
	\ee
	below,
	then we have from eq.(\ref{exply2}) that,
	\begin{equation}
		\Delta y_1 = \frac{1}{\alpha} \int_{1} ^{\infty} \frac{d \xi}{\xi  \sqrt{\xi^{2n} -1}}  = \frac{\pi}{2 n \alpha}. \label{maxdely}
	\end{equation}
	So we see that even with a varying dilaton the maximum value of $ \Delta y_1$ remains the same as eq.(\ref{lcrit}). In the Fig \ref{dilatondely} we plot \eqref{exply2} for the case $\Phi=\alpha + \beta r^\delta$, and show that the behvaiour of $\Delta y_1$  confirms the result \eqref{maxdely}. 
	
{\begin{figure}
		\centering
		\subfigure[]{\includegraphics[width=0.45\textwidth]{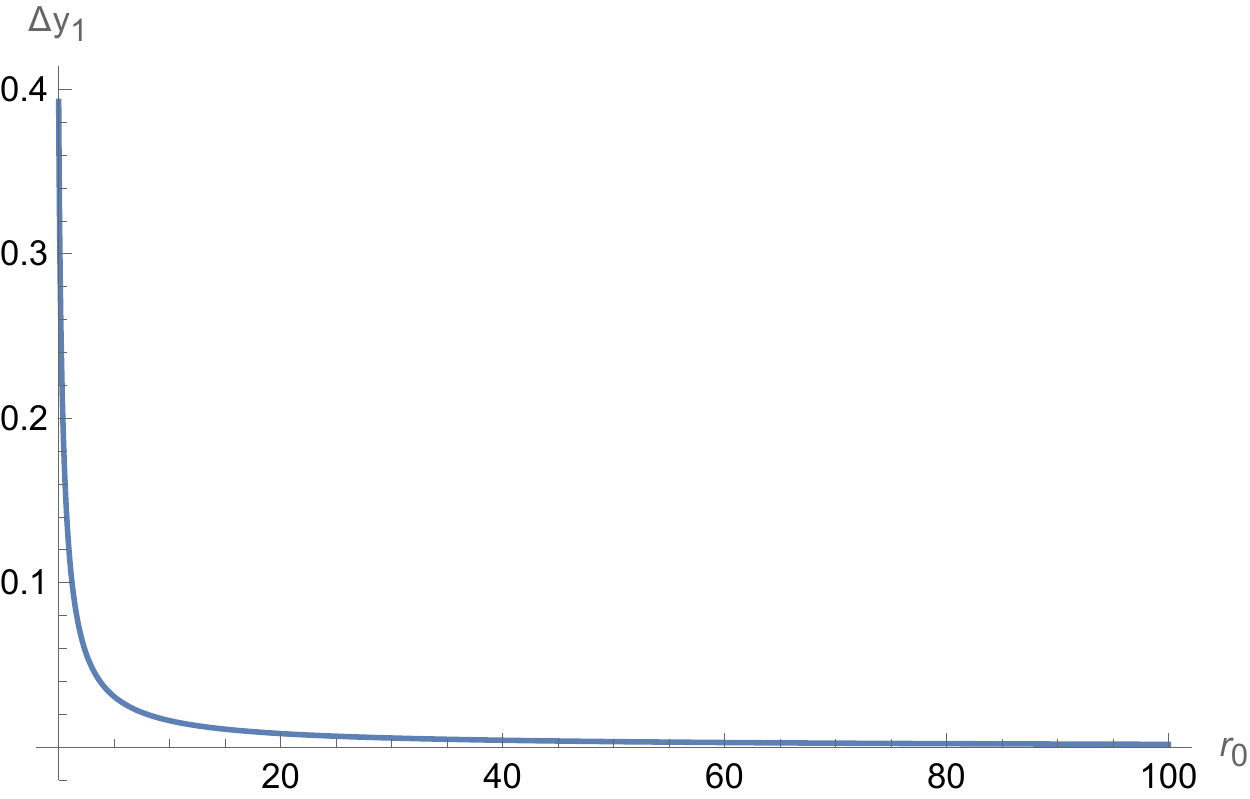}}\hfill
		\subfigure[]{\includegraphics[width=0.45\textwidth]{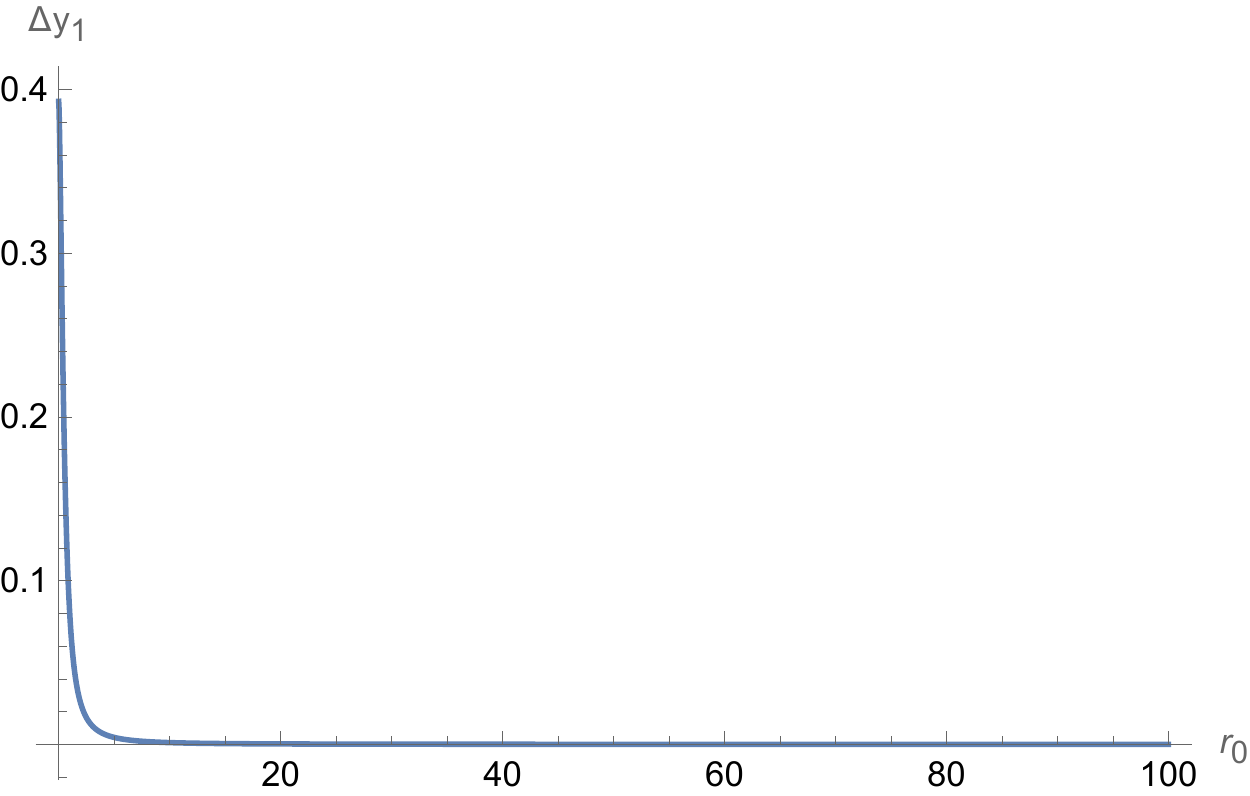}} 
		\caption{$r_{UV} =10^4, n=m=\alpha=\beta=2$. (a) $\delta=1$, (b) $\delta=2$. The maximum value in both cases is the same and agrees with analytical result \eqref{maxdely}.}
		\label{dilatondely}
\end{figure}}
			

Let us now turn to the area of the extremal surface. 
From  \eqref{exply} and  \eqref{areaeq} we get, for the surface that has a turning point at $r_0$,
\begin{equation}
A_1 = 2 R^{(m+n)} V_{n+m-1} \int_{r_0} ^{r_{UV}} dr \frac{r^{2 n -1} \Phi ^{2m-1}}{\sqrt{r^{2 n} \Phi ^{2m} - r_0 ^{2 n} \Phi_0 ^{2m}}}. \label{A1eq}
\end{equation}
This area should be compared with two other surfaces. 
			
The surface which ``hangs" at the boundary has the area,
\begin{equation}
A_2 = 2 R^{(m+n)} V_{n+m-1} r_{UV} ^n \Phi_{UV} ^m \Delta y_1 \label{A2eq}.
\end{equation}
Finally there is a third surface, analogous to eq.(\ref{atwop}), which consists of two disconnected pieces with $\dot {y_1}=0$. It has area,
\begin{equation}
				A_3 =2 R^{(m+n)} V_{n+m-1}  \int_{0} ^{r_{UV}} r^{n-1} \Phi ^{m-1} dr \label{A3eq}.
\end{equation}
	To proceed we use the fact that $\Phi$ is a monotonic function, which increases away from the horizon, as $r$ increases. It is then easy to see that in the integral in eq.(\ref{A3eq}), $\alpha<\Phi<\Phi_{UV}$ and therefore, 
	\be
	\label{baxx}
2 R^{(m+n)} V_{n+m-1}{r_{UV}^n\over n} \alpha^{m-1} <A_3<	2 R^{(m+n)} V_{n+m-1}{r_{UV}^n\over n} \Phi_{UV}^{m-1}.
\ee

Now consider what happens when $\Delta y_1$ is bigger that $(\Delta y)_{\rm crit}$, eq.(\ref{lcrit}). 
It is then easy to see from the upper limit in eq.(\ref{baxx})   that $A_3$ is less than  $A_2$. This proves that $A_3$ is the RT surface, of lowest area, in this case. 
	
	To illustrate the behaviour when all three surfaces exist, i.e. $\Delta y< (\Delta y)_{\rm crit}$,  		
as  a concrete example we take $\Phi=2+2 r$.  And in the Fig \ref{Areaplot} we have plotted for this case, $\frac{A_1}{A_3}$,  $\frac{A_1}{A_2}$ with $R=1$. As we can see $A_1<A_2$ and $A_1<A_3$. Hence $A_1$ has the smallest area
when this surface exists.
		
{\begin{figure}
\centering
\subfigure[]{\includegraphics[width=0.45\textwidth]{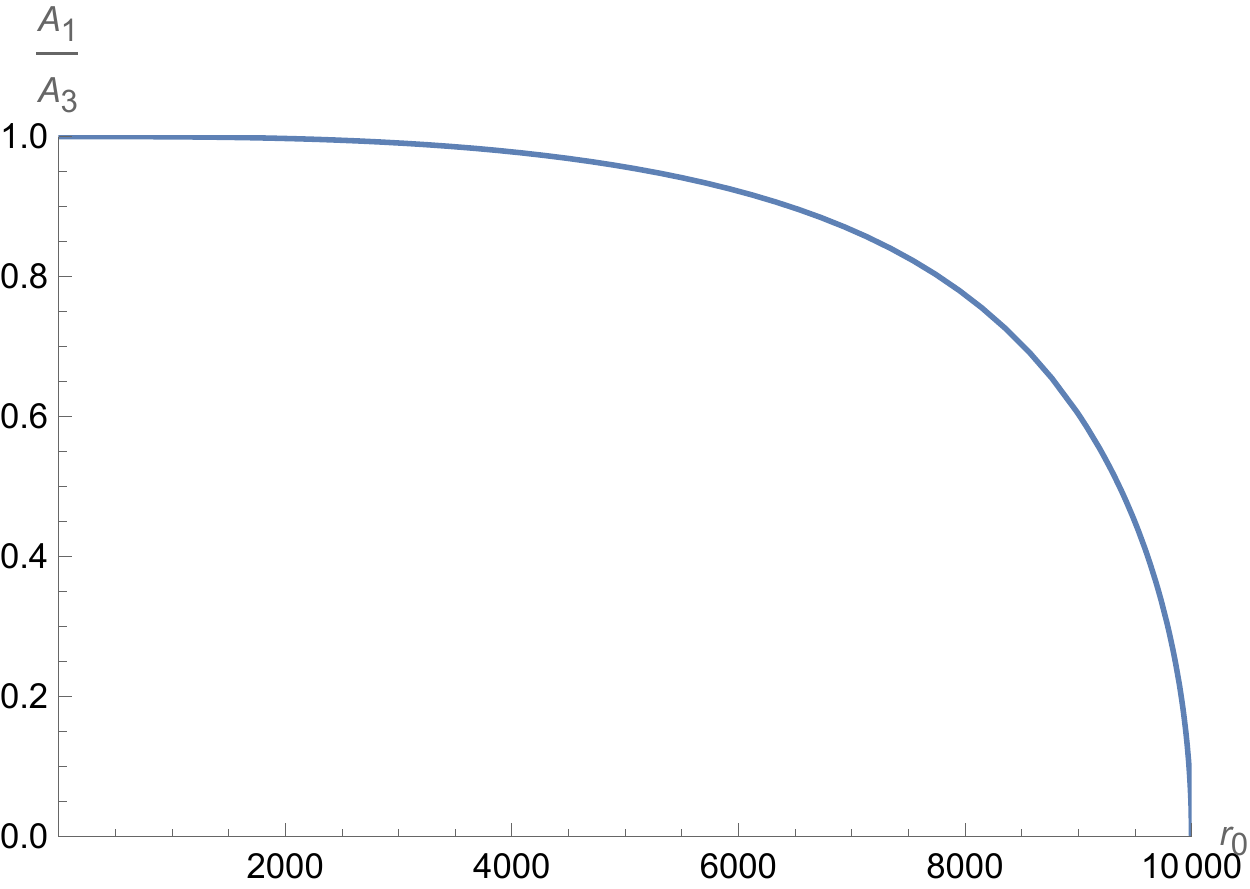}}\hfill
\subfigure[]{\includegraphics[width=0.45\textwidth]{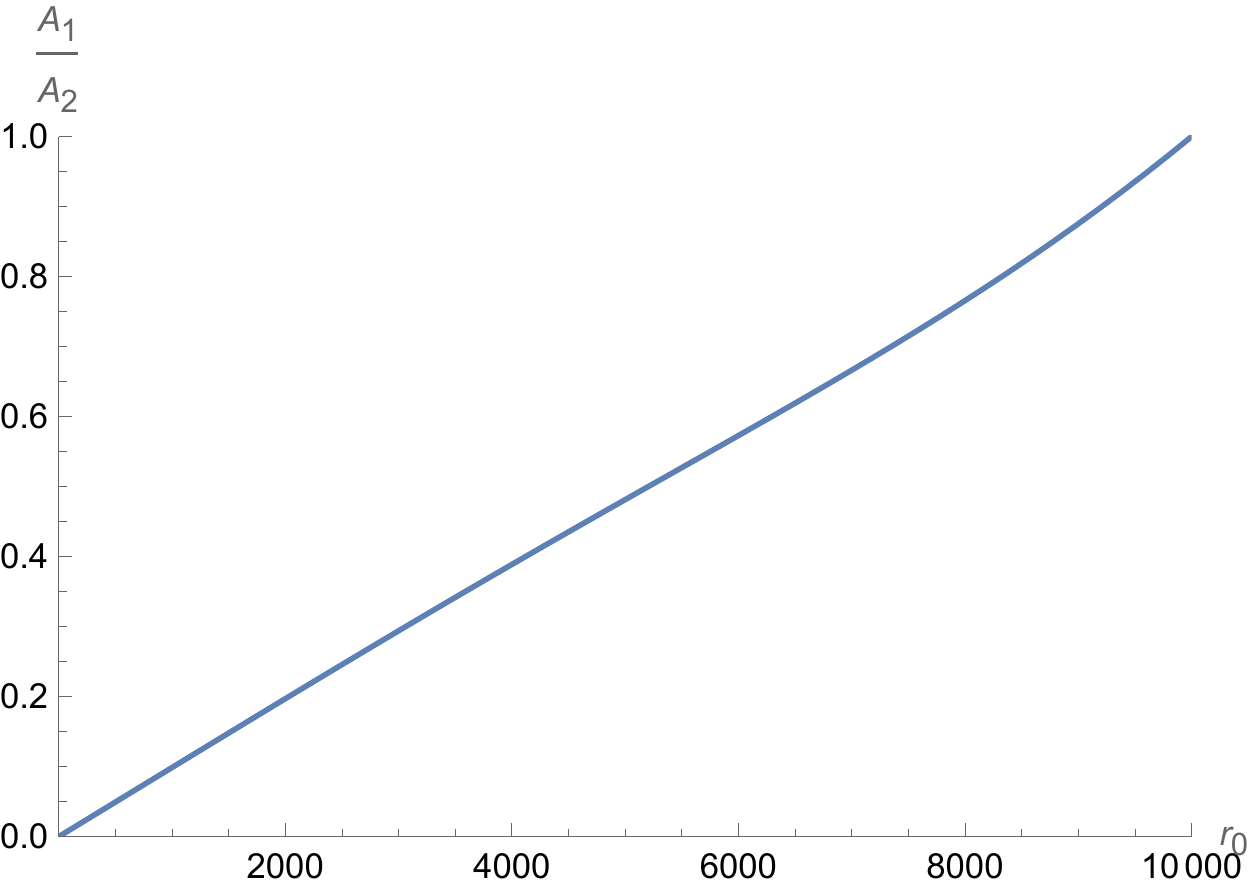}} 
\caption{$r_{UV}=10^4, n=m=\alpha=\beta=2, \delta=1$.}
\label{Areaplot}
\end{figure}}
We have also varied the parameters $\alpha, \delta$ in some range and find that $A_1$ continues to be the minimum area surface in this range. More generally, for varying dilaton profiles it is not so straightforward  to analyse which of the three surfaces has the smallest area. 
	
	Finally, let us note that when $\Delta y>(\Delta y)_{\rm crit}$ from the lower limit in eq.(\ref{baxx}) we get
	\be
	\label{calx}
	2 R^{(m+n)} V_{n+m-1}{r_{UV}^n\over n} \alpha^{m-1} <A_3
	\ee
	This shows that $A_3$ grows at least as fast as 
	\be
	\label{a3g}
	A_3	 \sim r_{UV}^n		V_{n+m-1}	\alpha^{m-1}.
	\ee
	
\subsubsection{Dilaton Vanishing at the Horizon} 
We end by considering the case where the horizon value of the dilaton, $\alpha$,  vanishes. 
As an example, we can  consider the case where the  dilaton is  of the form, $\Phi = \beta r^{\delta}$, with $\beta,\delta >0$. Then from \eqref{exply} we get,
\begin{equation}
\Delta y_1 = \frac{1}{\Phi_0} \int_{1} ^{\infty} \frac{d \xi}{\xi ^{1+ \delta} \sqrt{\xi^{2n + 2m \delta} -1}} - \frac{1}{\Phi_0} \int_{\xi_{UV}} ^{\infty} \frac{d \xi}{\xi ^{1+ \delta} \sqrt{\xi^{2n + 2m \delta} -1}}
\end{equation}
where $\xi_{UV} = \frac{r_{UV}}{r_0}$. The first integral can be exactly calculated and the second integral can be expanded in the limit $\xi_{UV} \gg 1$.  Keeping only the first term in the expansion of second integral we get,
\begin{equation}
\Delta y_1 = \frac{1}{\Phi_0}  \frac{\sqrt{\pi}}{\delta} \frac{\Gamma(\frac{1}{2} + \frac{\delta}{2 (m \delta + n)})}{\Gamma(\frac{\delta}{2 (m \delta + n)})} - \frac{1}{\Phi_0}\frac{1}{n + (m+1) \delta} \left(\frac{r_0}{r_{UV}}\right)^{n+(m+1)\delta}. \label{case3y}
\end{equation}
Since $\Phi_0\rightarrow 0$ as the turning point $r_0\rightarrow 0$, we see that now the maximum value for $\Delta y_1$ is infinite. This  agrees with the limit $\alpha\rightarrow 0$ in eq.(\ref{maxdely}).

As above, to find the smallest area surface we need to compare three surfaces. $A_1$ which is the surface with the turning point at $r_0$ \eqref{A1eq}, $A_2$ given in  \eqref{A2eq}, and the third surface along which $y_1$ is a constant with   area,
\begin{equation}
A_3 =2 R^{(m+n)} V_{n+m-1}  \int_{0} ^{r_{UV}} r^{n-1} \Phi ^{m-1} dr = 2 R^{(m+n)} V_{n+m-1} \beta^{m-1} \frac{1}{n + (m-1) \delta} r_{UV} ^{n+(m-1) \delta}.
\end{equation}
We have not carried out a general comparison among these three. 
It is easy to see that in some range of parameters $A_1$ is the smallest area. E.g., taking $\Phi = 2 r^2$, we have plotted 
$\frac{A_1}{A_3}$ and $\frac{A_1}{A_2}$ in Fig \ref{Areaplot2} (with $r_0$ in units of $R$).  We see that $A_1$ is indeed the smallest area. 
					

{\begin{figure}
\centering
\subfigure[]{\includegraphics[width=0.45\textwidth]{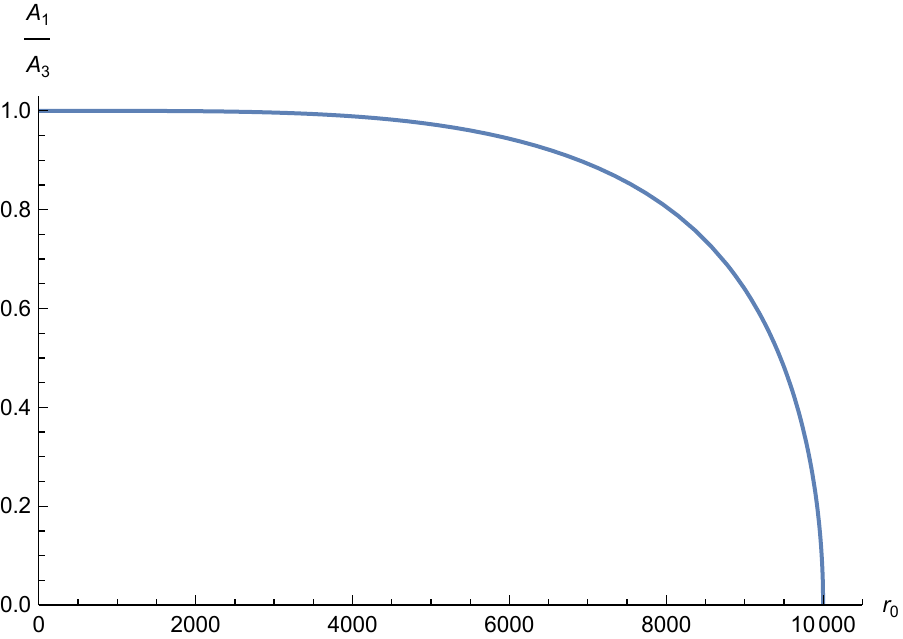}}\hfill
\subfigure[]{\includegraphics[width=0.45\textwidth]{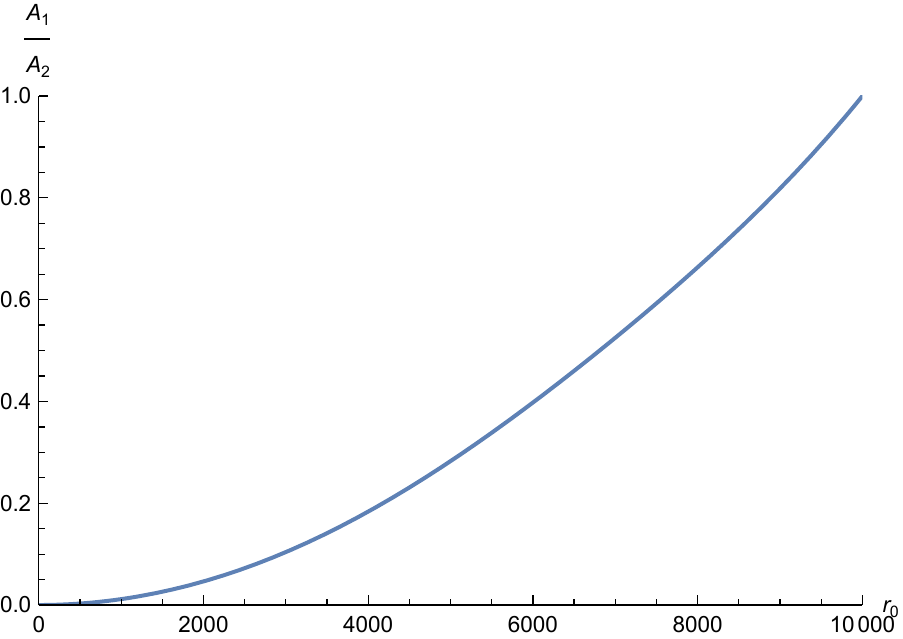}} 
\caption{$r_{UV}=10^4, n=m=\delta=\beta=2$.}
\label{Areaplot2}
\end{figure}} 
							
We end with one comment.  
 If $\Phi$ vanishes as a power law $\Phi\rightarrow r^\gamma$ when $r\rightarrow 0$, the spacetime in the near-horizon region is of a Lifshitz type, with eq.(\ref{productmet}) becoming
\be
\label{nhpm}
ds^2=R^2 r^2 (-dt^2+\sum_i(dx^i)^2) + R^2{dr^2\over r^2} + R^2 r^{2\gamma} \sum_\mu (dy^\mu)^2.
\ee
This metric  is invariant under $r\rightarrow \lambda r$, $(t,x^i)\rightarrow (\lambda)^{-1} (t,x^i)$, $y^\mu\rightarrow \lambda^{-\gamma}y^\mu$.
Thus we see that  $y^\mu$ directions are also part of the ``base space" directions in the Lifshitz IR theory in this case. 
And  the entanglement entropy in the IR theory which would be computed by the RT surface we are considering, would in fact be a kind of base space entanglement in the IR theory. 
							
\section{Compact Transverse Spaces}
\label{compact}
In this section we will examine extremal surfaces when the internal space is compact. We start in section \ref{GKtheorem} by first reviewing the Graham-Karch theorem for the simple case of a direct product $AdS_{n+2}\times S^m$ space and consider surfaces which at the asymptotic boundary of $AdS_{n+2}$, end on a sphere $S^{m-1}$. This $S^{m-1}$ itself bounds a ``spherical cap" of the internal $S^m$. Next, in section \ref{warp} we show how the GK theorem can be evaded once one includes warping and replaces the direct product of $AdS_{n+2}$ and $S^m$ by a warped product. We work out the necessary conditions, imposed by the asymptotic analysis, for the GK theorem to be evaded in this case. In section \ref{examples} we work out some explicit examples showing that once the necessary conditions are met,  extremal surfaces which do not end on minimal sub-manifolds of the internal $S^m$ do exist. Our analysis in section \ref{warp} is in fact quite general and applies also to other cases which are  warped products not containing  asymptotic hyperbolic space as a factor. In section \ref{other} we then apply our results to asymptotically flat spacetimes, and also to the $Dp$ brane geometries,  and show that the necessary conditions to evade the GK theorem are in fact  not met in these cases. 
							
\subsection{Direct Product Spaces and The Graham-Karch theorem}
\label{GKtheorem}
Let us begin by first discussing some of the key results obtained by Graham and Karch (GK), \cite{graham2014minimal}.
							
We consider a space of the form $X\times T$ where $X$ is an asymptotically hyperbolic space, $AdS_{n+2}$, and $T$ is a compact space. 
It is important in the following discussion of this subsection that the full space is a direct product of these two factors, rather than a warped product. 
							
							
We will be interested  in  a minimal area sub-manifold in $X\times T$  which is itself asymptotic (in the limit when  one goes to the boundary of $X$)  to a product sub-manifold in both the $X$ and the $T$ factors. An important result of GK   then says  that the asymptotic form of the submanifold in the $T$ factor wrapped by this surface is itself a minimal submanifold of $T$. By minimal submanifold of $T$ we mean that its area does not change at first order under small fluctuations of the submanifold, thus it can actually be a minimum, maximum or saddle of the area functional. 
We will refer to this result as the GK theorem below. 
							
\begin{figure}[h]
	\begin{center}
		\begin{tikzpicture}
			\draw[thick, blue] (0,0) circle [radius=2cm];
			
			\filldraw[color=black, fill=gray!50, very thick] (-1.732,1) arc [start angle=-190, end angle=10, x radius=17.58mm, y radius=3.0mm] -- (1.732,1) arc [start angle=30, end angle=150, x radius=20.0mm, y radius=20.0mm];
			\draw[thick, dashed] (-1.732,1) arc [start angle=170, end angle=10, x radius=1.76, y radius=0.3];
			\node[right] at (1.732,1) {$\theta=\theta_0$};
			
		\end{tikzpicture}
	\end{center}
	\caption{Intersection of RT surface with the boundary at $\theta=\theta_0$. Polar cap $0\le \theta \le \theta_0$ is shown as the region shaded in gray.}
	\label{repfigsph}
\end{figure}
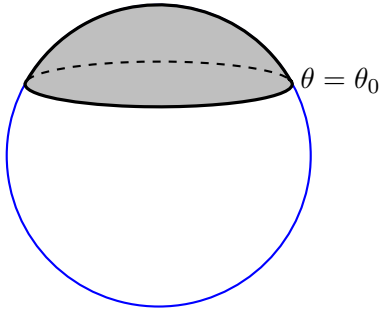

The theorem arises from an analysis of the asymptotic behaviour of the minimal area submanifold of $X\times T$. To gain some understanding for it, let us consider the  case where the spacetime is $AdS_{n+2} \times S^m$. For concreteness we can  work in Poincare coordinates for $AdS_{n+2}$ which has the metric 
\be
\label{metadsf}
ds_{AdS}^2=R^2 r^2 (-dt^2 + \sum^n_{i=1} (dx^i)^2) + R^2 \frac{dr^2}{r^2}.
\ee
The Sphere $S^m$ is taken to have radius $R \alpha$. Described in polar coordinates it has the  metric 
\be
\label{metsp}
ds^2= R^2 \alpha^2 (d\theta^2+ \sin^2(\theta) d\Omega_{m-1}^2),
\ee
where $d\Omega_{m-1}^2$ is the volume element for a unit $S_{m-1}$ and $\theta$ is the polar angle taking values $0\le \theta \le \pi$. 
For simplicity we consider a submanifold in the full space-time which lies at constant time, say at $t=t_0$, and in the $AdS_{n+2}$ wraps all the $x^i$ directions. On the $S^m$ the submanifold asymptotically, as $r\rightarrow \infty$,  wraps the surface $\theta=\theta_0$ for a fixed value of $\theta_0$ lying in the range $0< \theta_0\le \pi/2$. 
In other words what we are describing is   a minimal area  surface wrapping all the spatial directions on the boundary of the $AdS_{n+2}$ which is asymptotically pegged in the $S^m$ at the boundary of a spherical cap $0\le \theta\le \theta_0$. See Fig \ref{repfigsph}. 
Now  note also that the  hemisphere on the $S^m$ corresponds to taking $\theta_0=\pi/2$ and the boundary of this hemisphere (the analogue of a great circle for $S^2$) is then a minimal submanifold of $S^m$, as per our definition above,  since the  area of a spherical cap  is  proportional to $\sin(\theta)$ and its  first order variation is then proportional to $ \cos\theta$ and  vanishes at $\theta=\pi/2$.  
We will proceed to  show next that the minimal area requirement for the submanifold in $AdS_{n+2}\times S^m$ that we are considering will force its boundary on the $S^m$ asymptotically, as $r\rightarrow \infty$,  to   be the surface $\theta=\pi/2$. This, as we have just mentioned,  is the boundary of a hemisphere and  indeed   a minimal submanifold of $S^m$ -- in accordance with the GK theorem. 
As mentioned above, our analysis will involve the asymptotic region, $r\rightarrow \infty$ and it will be clear immediately that it would apply  equally well for $AdS_{n+2}$ in global coordinates. 
							
The minimal area surface  corresponds to a one dimensional locus in the $r,\theta$ coordinates, and will wrap all the $x^i$ coordinates and all the $m-1$ directions on the $S^m$ besides $\theta$. Its area functional is then given by 
\be
\label{af}	
A= \alpha^{m-1} R^{m+n}  V_B V_{S^{m-1}} \int_{\theta=0}^{\theta=\theta_0}   r^{n-1}  (\sin \theta)^{m-1} d\theta \sqrt{(r')^2+ r^2 \alpha^2   }.
\ee
Here $V_B$ is the total base space volume along the $x^i$ directions (which we have compactified),  $V_{S^{m-1}}$ is the volume of the unit $S^{m-1}$ and $r'={d r\over d\theta}$. 

We are considering surfaces which satisfy the condition that as $r\rightarrow \infty$, $ \theta\rightarrow \theta_0$, also $r$ attains its minimum value $r_0$, at $\theta=0$. These surfaces  close smoothly with the $S^{m-1}$ boundary on the $S^m$ shrinking to a point at the North pole,  at the turning point $r_0$.

%
Now the total range traversed along the $\theta$ direction is given by 
\be
\label{delt}
\Delta \theta=\int \left({d\theta \over d r} \right)dr,
\ee
and for this to remain finite we must have,
\begin{equation}
	\lim_{r \rightarrow \infty} r \dot{\theta} = 0. \label{assf}
\end{equation}
Here $\dot{\theta} = \frac{d \theta}{dr}$.
Given eq.\eqref{af} we can get Euler-Lagrange equation by varying $r(\theta)$ subject to the boundary conditions it satisfies. 
$r$ is fixed to take the value $r_{UV}\rightarrow \infty$, the UV cut-off, when $\theta=\theta_0$. At $\theta=0$, $r$ does not take a fixed value, however  since $r$ attains its minimum value there, $r'=0$. These mixed boundary conditions also lead to a well defined variational principle giving rise to the  Euler-Lagrange equation,
\begin{equation}
	\label{ELc}
	\frac{d}{d \theta} \left(\frac{r^{n-1} (\sin(\theta))^{m-1} r'}{ \sqrt{(r')^2+ \alpha^2 r^2}}\right) -  (\sin(\theta))^{m-1} \left( (n-1) r^{n-2} \sqrt{(r')^2+ \alpha^2 r^2} + \frac{\alpha^2 r^{n}}{\sqrt{(r')^2+ \alpha^2 r^2}}\right) =0 .
\end{equation}

In fact it is worth commenting that  the boundary term at $\theta\rightarrow 0$ which one obtains when varying eq.(\ref{af}) is given by 
\be
\label{bca}
{r^{n-1} (\sin(\theta))^{m-1} r' \delta r \over \sqrt{r'^2+r^2\alpha^2}}
\ee
and it  vanishes when $\theta\rightarrow 0$ both due to $r'$ vanishing and also because $\sin(\theta)$ vanishes. 

We should also note that had we been more cavalier and taking   $\theta$ to be a function of $r$ in eq.(\ref{af}), 
obtained an equation of motion by simply discarding any surface terms, we would have obtained, 

\begin{equation}
	\frac{d}{d r} \left(\alpha^2 \frac{r^{n+1} (\sin(\theta))^{m-1} \dot{\theta} }{ \sqrt{1+ \alpha^2 r^2 \dot{\theta}^2}}\right) - (m-1) \cos(\theta) (\sin(\theta))^{m-2} r^{n-1} \sqrt{1+ \alpha^2 r^2 \dot{\theta}^2}  =0.  \label{areaconstdil}
\end{equation}
One can show that eq.(\ref{ELc}) and eq.(\ref{areaconstdil}) are in fact the same away from the turning point, where $r'\rightarrow 0$. 

The rest of our analysis will be mostly in the asymptotic region, $r\rightarrow \infty$ and it will be more convenient to use
eq.(\ref{areaconstdil}) in our discussion below.

Let us  expand \eqref{areaconstdil} in the  large $r$ region. Then under the approximation eq.\eqref{assf},
\begin{equation}
	\sqrt{1+ \alpha^2 r^2 \dot{\theta}^2} \rightarrow 1,
\end{equation}
leading to,
\begin{equation*}
	\frac{d}{dr} (\alpha^2 r^{n+1} (\sin(\theta))^{m-1} \dot{\theta}) - (m-1) \cos(\theta) (\sin(\theta))^{m-2} r^{n-1}=0
\end{equation*}
Simplifying the above equation we get,
\begin{equation}
	\alpha^2 [(n+1) \sin(\theta_0) r \dot{\theta} + \sin(\theta_0) r^2 \Ddot{\theta} + \cos(\theta_0) r^2 \dot{\theta}^2]- (m-1) \cos(\theta_0)=0
\end{equation}
It is  easy to see that $r^2 \Ddot{\theta}$ must go to zero as $r \rightarrow \infty$ in accordance with the approximation eq.\eqref{assf}. Thus in the approximation eq.\eqref{assf}, the term in the square bracket vanishes and  the above equation is not satisfied unless $\theta_0$ meets the condition,
\begin{equation}
\cos{\theta_0} = 0 \implies \theta_0 = \frac{\pi}{2}.
\end{equation}

Thus we see that a necessary condition for the minimal area surface to exist is that it is pegged,  asymptotically as $r\rightarrow \infty$,   at the boundary  of a hemisphere, i.e. at   $\theta_0=\pi/2$. This is in agreement with the GK theorem as discussed above. 
							
It is worth summarising the key point of this analysis in a more physical manner. Since the sphere is compact the asymptotic velocity $\dot{\theta}$  can grow no faster than in eq.(\ref{assf}). But this requirement is inconsistent with the acceleration, being balanced by the force in the Euler Lagrange equation for the minimal area surface, unless the  minimal area surface  intersects the $S^m$ along a minimal submanifold. 
							
 \subsubsection{A more general analysis}
 \label{moregencomp}
Let us generalise the discussion above, to some extent, by  considering the product space $AdS_{n+2}$ $\cross$ $Y^m$ where $Y$ is not necessarily a compact space. The product space we consider has the following metric\footnote{This metric is related to metric in eq.\eqref{metadsf} , by the change of variables $r={R^2\over z}$},
\begin{equation}
\label{metxm}
	ds^2 = R^2 \left(\frac{1}{z^2} (-dt^2 + dz^2 + \sum_{i} dx_i ^2) + dy_1^2 + g^{\mu \nu} (y_1) dy_{\mu} dy_{\nu}\right).
\end{equation}	
Here $i$ runs from 1 to $n$ while $\mu, \nu$ run from 2 to $m$. Note that the coordinate $y_1$ is analogous to the polar angle $\theta$ above, but now, as mentioned above,  we do not require the internal manifold to be necessarily compact with a finite volume. 

We are interested in a surface which ends at the boundary,  $z\rightarrow 0$, on a co-dimension $1$ subspace of $Y$
that wraps all the  directions $y_\mu, \mu=2, \cdots m$, with $y_1$ taking a fixed value, $ y_{10}$. 
The area functional is given by,
\begin{equation}
\label{areafu}
	A = R^{n+m} \int \frac{V(y_1)}{z^{n+1}} \sqrt{dz^2 + z^2 dy_1 ^2}
\end{equation}
where $V(y_1)$ is the volume factor arising from the co-dimension $1$ submanifold of $Y$. For our analysis in the asymptotic region it is convenient to take $y_1$ to be a function of $z$. The EL equation then takes the form,
\begin{equation}
	\frac{d}{dz} \left(\frac{V(y_1) \dot{y_1}}{z^{n-1} \sqrt{1+ z^2 \dot{y_1 ^2}}}\right) - \frac{d V(y_1)}{dy_1} \frac{1}{z^{n+1}} \sqrt{1+ z^2 \dot{y_1 ^2}} =0 \label{ELY}
\end{equation}
where $\dot{y_1} = \frac{d y_1}{dz}$.

\noindent{\bf {Compact Internal Space}}

Now, if $y_1$ is compact then we have the following behaviour of $\dot{y_1}$,
\begin{equation}
	\lim_{z \rightarrow 0} z \dot{y_1} = 0, \label{assy1}
\end{equation}
and from the EL equation \eqref{ELY} we get,  as $z\rightarrow 0$,
\begin{equation}
\label{prgky}
	\frac{d}{dz} \left(\frac{V(y_1) z \dot{y_1}}{z^{n}}\right) - \frac{d V(y_1)}{dy_1} \frac{1}{z^{n+1}} =0.
\end{equation}						
Thus using eq.\eqref{assy1} we get, for $z\rightarrow 0$, 
\begin{equation}
	\frac{d V(y_1)}{dy_1} =0. \label{GKY}
\end{equation} 							
 So  we learn that the submanifold of $Y$ on which the RT surface ends asymptotically must be minimal, in agreement with the   GK theorem.

 Some of the steps above, in particular, eq.(\ref{assy1}) and eq.(\ref{GKY}) can be justified in a more careful manner as follows. 
Consider the asymptotic behavior of the solution as one approaches the AdS boundary $z \rightarrow 0$. 


First consider the case where the solution is asymptotically non-oscilating. More precisely, this means that there exists a point $z=z_0$ such that for $z<z_0$, we strictly have either $\dot{y}(z) \ge 0$ or $\dot{y}(z) \le 0$. We note that the asymptotic boundary conditions can be stated as
\be
\lim_{z \rightarrow 0}y_1(z)=y_{10}
\ee
where $y_{10}$ is the value for $y_1$ that we fix at the boundary. Since the internal space is compact, $y_{10}$ is finite. From this definition,
\be
y_1(z)=y_{10}+\int_0^z\dot{y}_1(z')dz'
\ee
This means

\be
\lim_{\epsilon \rightarrow o}\int_{0}^{\epsilon} \dot{y}_1(z)dz=0
\ee
Now using the mean value theorem we have
\be
\int_{0} ^{\epsilon} \dot{y}_1(z)dz=\epsilon \dot{y}_1(a(\epsilon) \times \epsilon)
\ee
where $a(\epsilon)$ is some function of $\epsilon$ that satisfies the condition
\be
0<a(\epsilon)<1
\ee
Putting everything together we get eq.(\ref{assy1},
provided that this limit exists. 


Also, obtaining eq.(\ref{GKY}) from  eq.(\ref{prgky}) assumes that one may set $ z \dot{y_1}$ in the limit of $z \rightarrow 0$ inside the derivative. This condition is not, however, necessary. To see this, integrate the two sides of the EL equation between $\epsilon$ and $2\epsilon$
\be
     \bigg[ \frac{\dot{y}_1 V(y_1)}{z^{n-1} \sqrt{ 1+ z^2 \dot{y}_1^2}} \bigg]_{\epsilon}^{2 \epsilon}= \int_{\epsilon}^{2 \epsilon} dz \frac{dV}{dy_1} \frac{\sqrt{ 1+ z^2 \dot{y}_1^2}}{z^{n+1}}
\ee
Using the mean value theorem we have
\be
      \epsilon \frac{\dot{y}_1(2 \epsilon) V(y_1(2 \epsilon))}{2 ^{n-1} \sqrt{ 1+ 4 \epsilon^2 \dot{y}_1(2 \epsilon)^2}}-\epsilon \frac{\dot{y}_1(\epsilon) V(y_1(\epsilon))}{ \sqrt{ 1+ \epsilon^2 \dot{y}_1(\epsilon)^2}}= \bigg[ \frac{\sqrt{ 1+ (a'(\epsilon) \times \epsilon)^2 \dot{y}_1 ^2}}{a'(\epsilon) ^{n+1}}\frac{dV}{dy_1}\bigg]_{z=a'(\epsilon) \times \epsilon} 
\ee    
where $a'(\epsilon)$ is this time some function of $\epsilon$ that strictly takes values between 1 and 2.
Taking the limit $\epsilon \rightarrow 0$ implies that we must have
\be
 \lim_{z \rightarrow o}\frac{dV}{dy_1}=0 \label{GHC}
\ee
which is the desired result. 

Next, consider  the class of asymptotically oscillating solutions. For these solutions there is no $z_0$ below which $\dot{y}_1(z)$ is monotonic or constant and the limit in eq.(\ref{assy1}) may not exist. But this necessarily means that there is an infinite number of points where $\dot{y}_1=0$ between any non-zero value of $z$ and $z=0$. We then define $\epsilon_1$ and $\epsilon_2$ as the two values of $z$ associated with a non-zero positive constant $\epsilon$, such that they are the biggest possible values satisfying the conditions
\be
\dot{y}_1(\epsilon_1)=\dot{y}_1(\epsilon_2)=0
\ee

\be
 \epsilon > \epsilon_1 > \epsilon_2
\ee
Evaluating the integral form of the EL equation once again we get
\be
     \bigg[ \frac{\dot{y}_1 V(y_1)}{z^{n-1} \sqrt{ 1+ z^2 \dot{y}_1^2}} \bigg]_{\epsilon_1}^{ \epsilon_2}= \int_{\epsilon_1}^{\epsilon_2} dz \frac{dV}{dy_1} \frac{\sqrt{ 1+ z^2 \dot{y}_1^2}}{z^{n+1}}
\ee
\be
 (\epsilon_1-\epsilon_2) \bigg[ \frac{\sqrt{ 1+ \sigma^2 \dot{y}_1 ^2}}{\sigma^{n+1}}\frac{dV}{dy_1}\bigg]_{z=\sigma }=0
\ee
where $\sigma$ is some value between $\epsilon_1$ and $\epsilon_2$. This then implies that

\be
 \bigg[\frac{dV}{dy_1}\bigg]_{z=\sigma }=0
\ee
which means that the solution must always be oscillating around a point satisfying this condition and taking the limit $\epsilon \rightarrow 0$ reproduces the condition of eq.(\ref{GHC}).
\\ 

\noindent{\bf{Non-Compact Internal Space}}
\\  \\
When the internal space is not compact eq.(\ref{assy1}) need not hold. Consider first
 \begin{equation}
 		\lim_{z \rightarrow 0} z \dot{y_1} = c. \label{assy2}
 	\end{equation}			
where $c$ is some nonzero constant.
Now $y_1$ diverges logarithmically as $z\rightarrow 0$, this could happen in a non-compact internal space $Y$. 
 

Then the EL equation eq.\eqref{ELY} reads, for $z\rightarrow 0$, 
\begin{equation}
 \frac{c}{\sqrt{1+c^2}}	\frac{d}{dz} \left(\frac{V(y_1)}{z^{n}}\right) - \frac{d V(y_1)}{dy_1} \frac{1}{z^{n+1}} \sqrt{1+c^2} =0.
\end{equation}			
leading to,
\begin{equation}
	\frac{c}{\sqrt{1+c^2}} \left[-n \frac{V(y_1)}{z^{n+1}} + \frac{d V(y_1)}{dz} \frac{1}{z^{n}}\right] - \frac{d V(y_1)}{dz} \frac{1}{c z^{n}} \sqrt{1+c^2} =0
\end{equation}
Further simplifying we get,
\begin{equation}
\label{fif}
	{d \log V\over d \log z}=- nc^2  \implies V(y_1) \sim z^{-n c^2}
\end{equation}					
Thus $V(y_1)$ has a power law divergence, as $z\rightarrow 0$.	This could happen when $Y$ is non-compact. 			
Finally consider the case when, as $z\rightarrow 0$,  		
  \begin{equation}
  	\lim_{z \rightarrow 0} z \dot{y_1} \rightarrow  \infty. \label{assy3}
  \end{equation}				
Then  eq.(\ref{fif}) is replaced by,
\begin{equation}
\label{fifa}
	{d \log V\over d \log z}=- n z^2{\dot y}^2.   
\end{equation}
We see that now  $V$ must diverge even more rapidly than in eq.(\ref{fif}).
E.g. if, as $z\rightarrow 0$
\begin{equation}
	z \dot{y_1} \rightarrow  \frac{1}{z^{\alpha}}, \alpha>0.
\end{equation}
 we get,
\begin{equation}
	V(y_1) \sim \exp(\frac{n}{2\alpha z^{2 \alpha}}).
\end{equation}
so that $V$  diverges exponentially fast. 					
					
The above analysis then reveals the following: when the internal manifold $Y$ is non-compact we can get extremal surfaces which end on the boundary on submanifolds of $Y$ which are of  non-minimal area, but in these cases the volume of the submanifold, $V$, diverges. 
\\ \\
\noindent{\bf{The Area of the Extremal Surface}}
\\ \\
Now consider parametrising the surface by the coordinate 
\be
\label{defx}
x=\log(z)
\ee
The EL equation, eq.(\ref{ELY}),  takes the form
\be
\label{neleq}
{d\over dx}
\bigl({V(y_1) y_1' e^{-nx} \over \sqrt{1+y_1'^2} }\bigr) 
-{dV \over dy_1} e^{-nx} \sqrt{1+y_1'^2}=0
\ee
where $y_1'={dy_1 \over dx}$.
At the boundary, where $z\rightarrow 0$, $x\rightarrow -\infty$. It is clear from eq.(\ref{neleq}) that if $y_1(x)$ is a solution meeting the boundary condition $y_1(-\infty)= y_{10}$ then so is $y_1(x+c)$ for any constant $c$. 
This shows that solutions come organized in sets of one parameter families of solutions, with solutions belonging to the same family all ending on the same asymptotic boundary conditions.  

Next let us introduce a cut-off in the radial direction at $z=\epsilon$ and take the shifted  $x$ coordinate $x\rightarrow -\log(\epsilon)+x=\log({z\over \epsilon})$. The boundary at $z=\epsilon$ corresponds now to $x=0$. $y_1$ takes the value
$y_{10}$ at the boundary. 
The Area, eq.(\ref{areafu}), in terms of this shifted coordinate takes the form
\be
\label{shifta}
A={R^{n+m}\over \epsilon^n} \int_0^{x_0} dx {V(y_1)\over e^{nx}} \sqrt{1+y_1'^2}
\ee
$x_0$, the turning point value for the $x$ coordinate, is determined completely in terms of the value of $y_{10}$ and  the EL equation eq.(\ref{neleq}). Thus the integral above is a function of $y_{10}$ alone. We denote 
\be
\label{defi}
I(y_{10})=\int_0^{x_0} dx {V(y_1)\over e^{nx}} \sqrt{1+y_1'^2}
\ee
leading to 
\be
\label{shittwo}
A={R^{n+m}\over \epsilon^n} I(y_{10})
\ee
which makes explicit the fact that $A$ diverges quite generally as ${1\over \epsilon^n}$. This scaling reflects the scaling symmetry of $AdS_{n+2}$, eq.(\ref{metxm}). 

Now if we take the partial derivative of the area with respect to $\epsilon$ while keeping $y_{10}$ fixed we get
\be
\frac{\partial A}{\partial\epsilon}={-nR^{n+m}\over \epsilon^{n+1}} I(y_{10})
\ee
This is essentially moving us from one solution in this family of solutions to another one that has the same $y_{10}$ at a different $\epsilon$. Similarly, we could have moved $y_{10}$ keeping $\epsilon$ fixed. In fact, if we do both variations simultaneously, we can easily tune them to keep moving along the same surface
\be
\frac{d A}{d \epsilon}= \Bigg[{-n\over \epsilon} A+\dot{y_{10}} \frac{\partial A}{\partial y_{10}}\Bigg]_{y_{10}=y_{10} (\epsilon)}.
\ee
Now changing things this way keeps us on the same surface and keeps the location of the turning point in our original coordinate $z$ fixed. Recalling that in terms of that coordinate the area is given by 

\be
A = R^{n+m} \int_\epsilon ^{z_*} dz \frac{V(y_1)}{z^{n+1}} \sqrt{1 + z^2 \dot{y_1} ^2}
\ee
where $z_*$ is the location of the turning point.

We easily see that the left hand side is nothing but negative the Lagrangian in the above expression while we have 
\be
\frac{\partial A}{\partial y_{10}}= -P_{y_1}
\ee
where $P_{y_1}$ is the momentum conjugate to $y_1$ at the cutoff $\epsilon$. This equation follows from the fact that we are varying an on shell solution and that first order variations of the area with respect to the turning point $z_*$ vanish. 

Putting everything together we get 
\be
\label{HJ}
A=\frac{\epsilon}{n}(\mathcal{L} - P_{y_1} \dot{y}_1)=-\frac{\epsilon}{n} \mathcal{H}
\ee

Where $\mathcal{H}$ is quite simply the "Hamiltonian" of the area functional, evaluated at the cutoff $\epsilon$.
To better understand this result, let us write eq.(\ref{HJ}) in a more explicit form
\be
\label{MR}
A=\frac{V(y_1)}{n\epsilon^n\sqrt{1+\epsilon^2\dot{y}_1^2}}
\ee
Now all one has to do to get the area of some surface ending on some cutoff $\epsilon$ is to plug the values for $y_1$ and $\dot{y}_1$ corresponding to that surface at $z=\epsilon$. \footnote{This expression evaluates the area of the surface from the cutoff to the turning point, with the turning point being defined as the point at which $\dot{y}_1$ diverges. If the surface doesn't terminate at that point, this wouldn't be the full area corresponding to the surface.}

In many cases, one is mostly interested in a specific one parameter family of solutions. For example, solutions that have a $y_1=0$ at the turning point. This forces a relation between $y_1$ and $\dot{y}_1$ at $z=\epsilon$.

In such cases, we can think $\dot{y}_1$ as a function of $y_1$ and the cutoff $\epsilon$. It can be then shown that this dependence takes the form
\be
\dot{y}_1= \left( \epsilon g(y_1) \right)^{-1}
\ee
with the function that we just introduced $g(y_1)$ satisfying the following differential equation 
\be
\frac { dg}{dy_1} =-\frac {g ^{2}+1}{V}   \left( g {\frac {dV}{dy_1}} +nV  \right) 
\ee
For the family of solutions with $y_1=0$ at the turning point, one only needs to solve this differential equation with the condition $g(0)=0$. Plugging back the result into eq.(\ref{MR}), we get the full expression for the area of any surface belonging to this family of solutions as a function of $y_1$ and $\epsilon$. \footnote{It should be noted that when applying this analysis, one should be careful in dealing with subtleties that may arise from the breakdown of parametrizations in terms of $y_1$ or $z$.}

We will discuss situations with a finite UV cutoff in more detail in section \ref{uvcutoff}. 

\subsection{Warped Product Spaces and a Way  Out}
\label{warp}
Here we will consider  warped product spaces, rather than direct products of the kind considered in the previous subsection. The transverse space will be compact. We will find that for warped products the conclusions of the GK theorem do not apply, and in such cases  the boundary of the minimal area surface in the transverse space need not    be minimal,

							
In particular we will consider  here the  case where the metric of the full spacetime is 
\be
\label{fullmet}
ds^2=g_{\alpha \beta} dX^\alpha dX^\beta + \Phi^2(X^\alpha) g_{mn}dy^mdy^n
\ee
with $X^\alpha$ being the coordinates of a space $X$ and $y^m$ of a compact transverse space $T$. $\Phi$ --  the dilaton -- depends on the coordinates $X^\alpha$, making this a warped product spacetime. 
	
	In fact, for concreteness we focus on  more specific cases where 
	eq.(\ref{fullmet}) is given by, 
	\begin{equation}
	\label{mgen}
	ds^2 = R^2 \left(-g_{00}(r)  dt^2 + \frac{dr^2}{f(r)} + g(r) \sum_{i} (dx^{i})^2 + \Phi(r)^2 d\Omega_m^2\right).
\end{equation}	
	We see that the transverse space $T$ is a sphere $S^m$ and the dilaton $\Phi$ only depends on a radial coordinate $r$. An example of such a space time is given by $AdS_{n+2}``\times" S^m$, 
\be
\label{fullaa}
ds^2= R^2 r^2 (-dt^2+\sum_i (dx^i)^2) +{R^2 \over r^2} dr^2 + R^2 \Phi(r)^2 (d\theta^2 + \sin^2(\theta) d\Omega_{m-1}^2).
\ee	
Other examples include $Dp$ brane geometries, and  metrics which are asymptotically flat.
						
Furthermore, in the metric eq.(\ref{mgen}),  we  consider as the minimal area surface a  submanifold,  also as in the previous subsection, which wraps the base space along  all the $x^i$ coordinates, at constant $t$, and which ends, at asymptotic $r\rightarrow \infty$, on an $S^{m-1}\in S^m$ given by $\theta=\theta_0$. 
The area functional  for  such a surface  given by 
\begin{equation}
	A = R^{m+n} V_n V_{s^{m-1}} \int g(r)^{\frac{n}{2}} \Phi^{m-1} (\sin(\theta))^{m-1} \sqrt{\frac{dr^2}{f(r)} + \Phi^2 d\theta^2}. \label{areaorg}
\end{equation}

							
We will find that the warping introduced due to the varying dilaton allows us to evade the conclusions of the GK theorem in these examples.
		
We will carry out an asymptotic analysis of these extremal surfaces, as  $r\rightarrow \infty$.

%

Two combinations of the metric components  will enter our  analysis,
\be
\label{defF}
\mathcal{F}(r) = \sqrt{\frac{1}{f(r)}} g(r)^{\frac{n}{2}} \Phi^{m-1},
\ee							
and 
\begin{equation}
	\label{defh}
	h(r) = f(r) \Phi^2.
\end{equation}						
In our analysis below we  assume that both $\mathcal{F}(r)$ and $h(r)$ grow like a power of $r$, for large $r$, more precisely in a power-law fashion, 
\be
{\cal F}(r)  =  c_F r^a(1+ O(1/ r^{a_1} )), \label{asF}
\ee
\be
h(r)  =  c_h r^b(1+ O({1\over r^{b_1}})), \label{ash}
\ee
with
\be
\label{condab}
a \ge 0, \ b>0,
\ee 						
and the exponents $a_1, b_1$ parametrise the corrections; these are positive, or  could vanish identically. 

						
						
						
The Euler-Lagrange equation obtained from eq.(\ref{areaorg}) -- see the related discussion in section \ref{GKtheorem} --  then gives,
\begin{equation}
	\label{assie}
	\frac{d}{dr} \left(\frac{\mathcal{F} (r) (\sin(\theta))^{m-1} h(r) \dot{\theta} }{\sqrt{1 + h(r) \dot{\theta}^2}}\right) - \mathcal{F} (r) (m-1)(\sin(\theta))^{m-2} \cos(\theta) \sqrt{1 + h(r) \dot{\theta}^2} =0
\end{equation}
where $h(r)$ is defined in eq.(\ref{defh}). 

 In the analysis below we also assume   $\theta$ to take the following asymptotic form, 
\begin{equation}
	\theta = \theta_0 - \frac{c}{\epsilon r^{\epsilon}},  \ \epsilon>0. \label{assf2}
\end{equation}

						

The resulting asymptotic analysis is a bit involved so we only state the main result here, and give more details in appendix \ref{Moredetana}. 

It turns out that a solution where $\theta$ has the form, eq.(\ref{assf2}) can only arise if  
\be
\label{condb}
b>2.
\ee
And in this solution 
\begin{equation}
	\epsilon = b-2 , c = \frac{(m-1) \cot(\theta_0)}{c_h (a+ 1)}. \label{Mastereqn}
\end{equation}						
						
We can now apply this result to a metric of the form eq.(\ref{fullaa}) with dilaton asymptotically taking the form, 
\be
\label{assdi}
\Phi\rightarrow c_\phi r^\delta, \delta >0.
\ee
Comparing eq.(\ref{fullaa}) with eq.(\ref{asF}), eq.(\ref{ash}), eq.(\ref{assdi})
we then get  that 
\begin{eqnarray}
	\label{valin}
	a & = & n-1+\delta (m-1), \label{valaa}\\
	b & = & 2+2\delta.\label{valba}
\end{eqnarray}						
The condition $a\ge 0$ in eq.(\ref{condab}) is non-trivial if $n=0$ and requires, 
\be
\label{condabc}
\delta (m-1) \ge 1.
\ee
We will assume eq.(\ref{condabc}) holds for the $n=0$ case. 
From eq.(\ref{valba}) it is clear that 
condition eq.(\ref{condb}) is indeed met, 
leading to 
\be
\label{valeps}
\epsilon=2\delta,
\ee
and 
\be
\label{valcc}
c={(m-1) \cot(\theta_0) \over c_{\phi}^2 (n+(m -1)\delta)}.
\ee						
						
We see then that for a metric and dilaton given asymptotically by eq.(\ref{fullaa}), eq.(\ref{assdi})  it is possible to evade the Graham Karch theorem. In contrast, we note that for a constant dilaton with no warping,  $\delta=0$.  As a result,  we see from eq.(\ref{valba})  that condition eq.(\ref{condb})  is not met and no non-trivial solution is allowed. This   is in agreement with the Graham Karch theorem.

To summarise, in this section we showed how in the presence of warping the GK theorem can be evaded. Our analysis was carried out for metrics of the form eq.(\ref{mgen}) which involves a warped  product with an $S^m$ factor. We considered simple situations involving a spherical cap with polar angle $\theta\le \theta_0$ on the $S^m$.
							We showed that if the conditions, eq.(\ref{asF}), eq.(\ref{ash}), eq.(\ref{condab}), eq.(\ref{condb}),  are met, then at least asymptotically the equations  allow  for the minimal area surface to end at the boundary of any such  cap, i.e. for any value of the polar angle, $\theta_0$. Thus the minimal area surface, asymptotically, does not have to be  a submanifold on the $S^m$ which is itself minimal.
							
One expects these arguments to generalise to other regions on the $S^m$ besides the  spherical caps we looked at and also to generalise for  other compact spaces besides $S^m$. We leave such an analysis for the future. Also, we assumed in our analysis that the asymptotic form of $\theta$ is of the power law form,  eq.\eqref{assf2}.
One can also explore relaxing this restriction and including  additional $\log$ corrections, we leave an investigation of this also for the  future. The   examples that we construct in the next subsection for warped cases agree with the asymptotic form  eq.\eqref{assf2}  and confirm the analysis done in this section.

\subsection{Some Explicit solutions with Warping}
\label{examples}
Here we will consider some examples  with a suitable warping  profiles etc. such that the conditions we had found above for the existence of an RT surface are indeed met, asymptotically. We will then solve the EL equation everywhere and construct the full RT surface. The full solution will be obtained numerically. 
							
Let us note at the outset that we have not investigated whether the metrics we study actually arise as solutions of Einstein equations in the presence of appropriate matter. 
							
We consider the case when the metric is of the form eq.(\ref{mgen}) with,
\begin{eqnarray}
	f(r) & = & r^2,\label{valf}\\
	g(r) & = & r^2.\label{valg}
\end{eqnarray}
As a result, eq.(\ref{defF}), eq.(\ref{defh}), 
\begin{eqnarray}
	{\cal F}(r) & = & r^{n-1} \Phi^{m-1},\label{valcurlf} \\
	h & = & r^2 \Phi ^2.\label{valh}
\end{eqnarray}							
							
Let us investigate the behaviour near the turning point in the interior. We are interested in surfaces of ``disk-like" topology, rather than cylinderical topology, i.e. in which the spherical cap shrinks to zero at the turning point. For such surfaces, $\theta\rightarrow 0$ at the turning point $r_0$. 
	Near the turning point we take 
\be
\label{funr}
r(\theta)=r_0 + {1\over 2} k \theta^2,
\ee
so that $r'={dr\over d\theta}$ vanishes at $\theta=0$.						 
							
Given eq.\eqref{areaorg}, we can take $r$ to be a function of $\theta$. The EL equation then gives, 
\begin{eqnarray}
	&& \frac{d}{d \theta} \left(\frac{ r^{n-1}  \Phi^{(m-1)} (\sin(\theta))^{m-1}  r'}{ \sqrt{(r')^2+  r^2 \Phi^2}}\right) \nonumber \\
	& -&  (\sin(\theta))^{m-1} r^{n-2}  \Phi^{(m-2)} \left( ((n-1) \Phi + r (m-1) \partial_r \Phi)  \sqrt{(r')^2+  r^2 \Phi^2} + \frac{ r^2 \Phi^2 ( \Phi+ r\partial_r \Phi)}{\sqrt{(r')^2+ r^2 \Phi^2}}\right) =0. \nonumber
\end{eqnarray}
This gives to leading order behaviour
\begin{eqnarray}
	k \frac{d}{d \theta} (r_0 ^{n-2}  \Phi_0 ^{(m-2)} \theta ^{m}) - \theta^{m-1} r_0^{n-2}  \Phi_0^{(m-2)} \left( ((n-1) \Phi_0 + r_0 (m-1) \partial_r \Phi)  r_0 \Phi_0 + r_0 \Phi_0 (\Phi_0 + r_0 \partial_r \Phi)\right) = 0, \nonumber
\end{eqnarray}
where
\begin{equation}
	\Phi_0 = \Phi(r_0).
\end{equation}
Equating coefficients gives, 
\be
\label{eqc}
k =\frac{1}{m} r_0 \Phi_0 ( n \Phi_0 + r_0 m \partial_r \Phi|_{r=r_0}).
\ee

Now to be explicit let us further take the dilaton to be 
\be
\label{dilp}
\Phi= \alpha + \beta r^{\delta}.
\ee							
							
The resulting extremal surfaces are then shown in Fig \ref{fig:alphasanddeltas}. We have taken $m=n=2$, and considered different values of $\alpha$, $\beta$ and $\delta$. Instead of varying $\theta_0$ we have varied the turning point $r_0$ and found that $\theta_0$ changes monotonically, increasing as $r_0$ is decreased, for any given values  of $\alpha,\beta,\delta$.
							
(Note for ease of integration we actually start the numerical integration very close to the turning point, at  $\theta_* = \frac{\pi}{10^5}$. And take $r_*$ to be given by eq.(\ref{funr}). We have also verified that asymptotically $\dot{\theta}$ is of the form eq.(\ref{assf2}) with $\epsilon$ and $c$ agreeing with their values, (\ref{Mastereqn})).

							%
							%
							
{\begin{figure}
		\centering
		\subfigure[]{\includegraphics[width=0.32\textwidth]{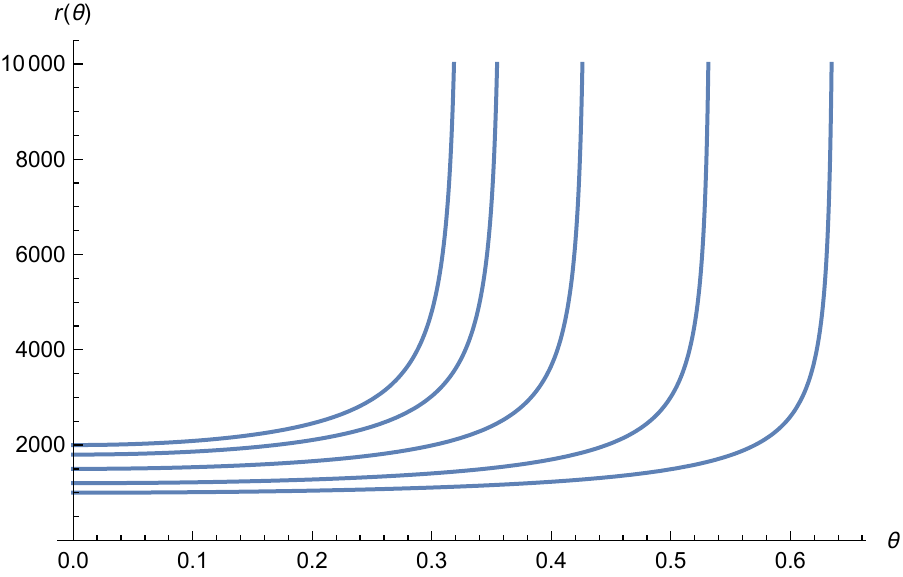}}
		\subfigure[]{\includegraphics[width=0.33\textwidth]{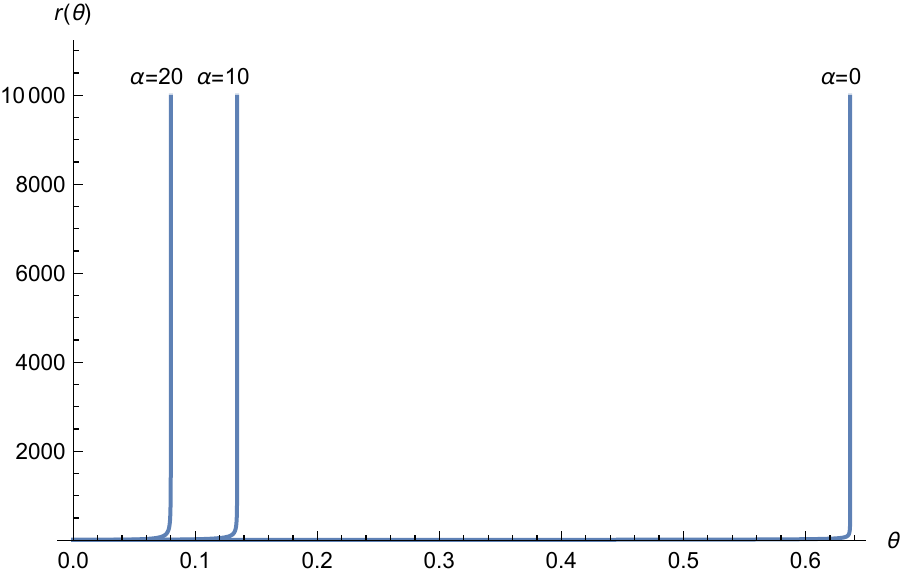}} 
		\subfigure[]{\includegraphics[width=0.32\textwidth]{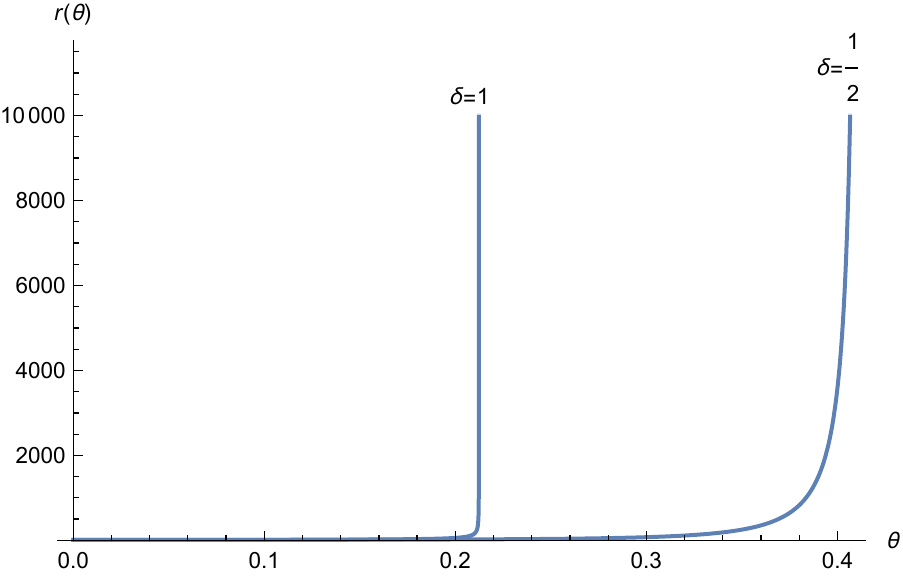}} 
		\caption{(a) $\beta=10^{-3},\alpha=0,\delta=1.$ We see that  higher values of $\theta_0$ lead to smaller values of $r_0$. (b) $\beta=10^{-1},r_0=10, \delta=1$. We see that higher values of $\alpha$ lead to lower values of $\theta_0$, for the same $r_0$. (c) $\beta=10^{-1},r_0=10,\alpha=5$. Here we see that for the same $r_0$ and $\alpha$ a smaller value  of $\delta$ leads to a  higher  $\theta_0$.  }
		\label{fig:alphasanddeltas}
\end{figure}}
							
\subsubsection{Comparison with Another Surface }
Here we will again consider a metric eq.\eqref{mgen}. We will work with a finite and large cut-off $r_{UV}\gg 1$ in the radial direction, at which the boundary is located, and would like to compare the area of the surface we have obtained above with another one which does not venture radially inward but ``drapes" over the transverse sphere. The surface we have studied in the previous few subsections has the area, eq.\eqref{areaorg}, 
\begin{eqnarray}
	A_1 &=& R^{m+n} V_n V_{s^{m-1}} \int g(r)^{\frac{n}{2}} \Phi^{m-1} (\sin(\theta))^{m-1} \sqrt{\frac{dr^2}{f(r)} + \Phi^2 d\theta^2}  \nonumber \\
	&=& R^{m+n} V_n V_{s^{m-1}} \int_{r=r_0}^{r=r_{UV}} \mathcal{F} (r) (\sin(\theta))^{m-1} \sqrt{1 + h(r) \dot{\theta}^2} dr,
\end{eqnarray}
where ${\cal F}$ and $h$ are given in eq.(\ref{valcurlf}), eq.(\ref{valh}).							 
							
The second surface we would like to compare it with, extends over the spherical cap,  $\theta\le \theta_0$ of the $S^m$, and as we mentioned above, is located at the boundary $r_{UV}$. Its area is 
\begin{align}
A_2 =& V_{S^{m-1}} R^{n+2} V_n \Phi_{UV}^{m} (g(r_{UV}))^{\frac{n}{2}} \int_{0}^{\theta_0} d\theta  (\sin(\theta))^{m-1} \nonumber \\
=& V_{S^{m-1}} R^{n+2} V_n \mathcal{F} (r_{UV}) \sqrt{h(r_{UV})} \int_{0}^{\theta_0} d\theta  (\sin(\theta))^{m-1}.
\end{align}
Simplifying we get,
\begin{equation}
A_2 =  V_{S^{m-1}} R^{n+2} V_n r_{UV}^{a+\frac{b}{2}} \int_{0}^{\theta_0} d\theta  (\sin(\theta))^{m-1}.
\end{equation}
Near the UV boundary   with ${\cal F}$, $h$ asymptotically going like eq.(\ref{asF}), eq.(\ref{ash}), the behaviour of $A_1$ can be estimated from the asymptotic behaviour discussed in subsection \ref{warp}. 
From eq.(\ref{Mastereqn}), eq.(\ref{assf})  it then follows  that as $r\rightarrow r_{UV}$
\begin{equation}
	h(r) \dot{\theta}^2 \sim r_{UV}^b \left(\frac{(m-1) \cot{\theta_0}} {a+1}\right)^2 \frac{1}{r_{UV} ^{2b-2}} \sim \frac{1}{r_{UV} ^{b-2}} \ll 1.
\end{equation}
So we can approximate the $r_{UV}$ dependence of  $A_1$ to be 
\begin{equation}
	A _1= V_{S^{m-1}} R^{n+2} V_n \frac{1}{a+1} r_{UV} ^{a+1} (\sin(\theta_0))^{m-1}\sim r_{UV}^{a+1}.
\end{equation}							
This is clearly smaller than $A_2$ since $b$ satisfies eq.(\ref{condb}). Thus we see that when the surface $A_1$ exists it has lower area in the limit when the UV cut-off $r_{UV}\rightarrow \infty$, or when it  is finite but sufficiently big. 
							
\subsection{Asymptotically $AdS$, Flat and $Dp$ brane geometries}
\label{other}
The analysis above was carried out for a general metric of the form, eq.(\ref{mgen}) and can be applied to an extremal RN geometry in asymptotically $AdS$ or flat space and for  the metric for $Dp$ branes. 
							
Let's start with the eRN black hole in asymptotically $AdS_{d+1}$ space. We consider global coordinate. The metric is given by 
\be
\label{metassfe}
ds^2=-f(r) dt^2 +{1\over f(r)} dr^2 + r^2 d\Omega_{d-1}^2,
\ee							
where $f(r)\rightarrow {(r-r_h)^2\over R_2^2}$ near the horizon $r_h$, and $f(r)\rightarrow {r^2\over R_{d+1}^2}$ as $r\rightarrow \infty$. 
							
Comparing with eq.(\ref{mgen}) we see that $f(r)$ in eq.(\ref{mgen}) is in fact the same as the emblackening factor  in eq.(\ref{metassfe}). Also, $n=0$, $m=d-1$, and $\Phi=r$. From eq.(\ref{defF}), eq.(\ref{defh}) we see that ${\cal F}={r^{m-1}\over \sqrt{f}}$, $h=fr^2$, and therefore asymptotically as $r\rightarrow \infty$, ${\cal F} \rightarrow r^{m-2}, h\rightarrow r^4$. The conditions, eq.(\ref{condab}) are then met for $m\ge 2$, i.e., $d\ge 3$, and we also see that since $b=4$, eq.(\ref{condb}) is met. As a result all requirements for a non-trivial RT surface of the kind we considered above are met. Of course for this case, this is only to be expected since from the point of the UV $AdS_{d+1}$ the RT surface we  are considering is the conventional one with the sphere $S^m$ being the spatial directions of the  base space in the UV boundary. 

A detailed analysis of the resulting RT surfaces as a function of the polar angle $\theta_0$ and varying values of $r_h\over R_{d+1}$ is left for the future. It is worth noting that in this case, unlike the planar case studied in \ref{ads2}, the presence of the black hole leads to an important ``homology" constraint, which needs to be taken into account in the analysis. 


						
Next, let us look at an eRN geometry in asymptotically flat space. We  consider the  $3+1$ dimensional case for concreteness, similar conclusions apply for higher dimensions as well. The metric is given by,
\be
ds^2  =  -(1-{Q\over r})^2  dt^2 + {dr^2\over (1-{Q\over r})^2} + r^2 d\Omega_2^2.\label{mext} 
\ee						
Comparing with eq.(\ref{mgen}) we see that $n=0$, $m=2$, $R=1$, $f(r)=(1-{Q\over r})^2$ and $\Phi=r$. It then follows that  $h(r)=(1-Q/r)^2 r^2$, ${\cal F}={r \over (1-Q/r)}$, eq.(\ref{defh}), eq.(\ref{defF}).   and asymptotically as $r\rightarrow \infty$, $h(r) \rightarrow r^2, {\cal F} \rightarrow r$. As a result, eq.(\ref{condab}) is met. However we note that  now $b=2$, and therefore condition eq.(\ref{condb}) is not met. We therefore conclude that the RT surface considered above does not exist in this case. 
						
Next let us consider the $Dp$ brane geometry in the near-horizon  region, \cite{Itzhaki:1998dd}, with Einstein frame metric,
\be
\label{dpmet}
ds^2=e^{(-{\phi\over 2})}[H^{-1/2} (-dt^2+\sum_{i=1}^p (dx^i)^2) +H^{1/2}(dr^2+r^2d\Omega_m^2)].
\ee						
Here $e^{(-{\phi\over 2})}$ is the factor involving the dilaton of string theory needed  convert the metric to Einstein frame with   the metric within the square brackets being the string frame metric, and $H$ is the harmonic function, with  
\begin{eqnarray}
	\label{convf}
	H & = & {R^{7-p}\over r^{7-p}},  R^{m-1} =   ( 4\pi)^{m-3\over2} \Gamma({m-1\over 2}) l_s^{m-1} g_sN,\\
	e^{(-{\phi\over 2})}  & = & H^{(p-3)\over 8}.
\end{eqnarray}						
And  $n=p$ and $m=8-p$.  
We alert the reader that $\phi$ is different from the warping factor $\Phi$ in eq.(\ref{mgen}), and from eq.(\ref{dpmet}) note that $\Phi^2= e^{(-{ \phi\over 2})} H^{1/2}r^2$. 
						
						
Comparing with eq.(\ref{mgen}) we see that now $f(r)=e^{({ \phi\over 2})} H^{-1/2}$ and therefore ${\cal F}, h(r)$, are 
\begin{eqnarray}
	{\cal F} & = & H^{1/2} r^{m-1}=(R r )^{({7-p\over 2})}, \label{valcfh}\\
	h(r) = & f \Phi^2 & = r^2. \label{valdhh}
\end{eqnarray}						
Asymptotically, as $r\rightarrow \infty$, it follows that ${\cal F} \rightarrow r^{({7-p\over 2})}$. Thus for $p\le 7$ eq.(\ref{condab}) is valid and our analysis above holds. However we see from eq.(\ref{valdhh}) that $b=2$ and therefore we see that condition eq.(\ref{condb}) is not met and the RT surface does not exist.  
						
Note in particular that our conclusions above also apply to  the  $p=0$ case corresponding to $D0$ branes. Later in the paper we will consider the $D0$ brane geometry in the presence of a UV cut-off in the radial direction.

\section{Geometries with a UV cut-off}						
\label{uvcutoff}
\subsection{Product Space geometries with a UV cut-off}	
\label{Productspacegeo}
In section \ref{GKtheorem} we reviewed the GK theorem and showed how  in some specific cases it arises for asymptotically $AdS$ spaces, when the boundary of $AdS$ is taken to  infinity, i.e. when $r\rightarrow \infty$, where $r$ is  the radial coordinate in eq.(\ref{metadsf}). In this section we  consider situations where there is a compact transverse space with a boundary which  is at a finite but large $UV$ cutoff, i.e. where the boundary value of the radial coordinate, $r_{UV}$,   is finite. More specifically we will consider spaces of the form $AdS_{n+2} ``\times" S^m$, with a radial cut off $r_{UV}$ and consider surfaces of the kind we considered in section \ref{GKtheorem} which end on the boundary on an  $S^{m-1}$ spherical cap.
To begin in this subsection we in fact consider the direct product $AdS_{n+2}\times S^m$ and  then in section \ref{Warpedspacegeo} consider examples with warping.


For the  product space $AdS_{n+2}$ $\cross$ $S^m$, with metric  given in eq.\eqref{metadsf} and eq.\eqref{metsp}, the area functional for the surface ending on a spherical cap of $\theta= \theta_0$ was given in eq.\eqref{af}. We reproduce it here for easy reference (we have set $\alpha=1$),
\begin{equation}
	A=R^{m+n}  V_B V_{S^{m-1}} \int   r^{n-1}  (\sin \theta)^{m-1} \sqrt{(dr)^2+ r^2  d \theta^2 }. \label{af2}
\end{equation}
The Euler-Lagrange equation then gives, with $\theta$ as a function of $r$,  
\begin{equation}
	\frac{d}{d r} \left( \frac{r^{n+1} (\sin(\theta))^{m-1} \dot{\theta} }{ \sqrt{1+ r^2 \dot{\theta}^2}}\right) - (m-1) \cos(\theta) (\sin(\theta))^{m-2} r^{n-1} \sqrt{1+ r^2 \dot{\theta}^2}  =0. \label{producteom}
\end{equation}
Here ${\dot \theta}= {d\theta\over dr}$, more generally we follow the notation that a ``dot" superscript indicates derivatives with respect to $r$.

{\it $AdS_{n+2} \times S^n$}:
Before proceeding let us discuss the case with  $n=m$, i.e.,  $AdS_{n+2}\times S^n$. In this case eq.(\ref{producteom})  admits an exact solution,
\begin{equation}
\label{exactsol}
	{r\over r_0}= \sec (\theta) 
\end{equation}
where $r_0$ is the turning point. This solution satisfies  the initial condition: as $r \rightarrow r_0$, $\theta \rightarrow 0$. The turning point $r_0$ is determined by demanding that: as $r \rightarrow r_{UV}$, $\theta \rightarrow \theta_0$.  Also note that this solution is in accordance with the GK theorem; since as $r \rightarrow \infty$, $\theta \rightarrow \frac{\pi}{2}$. 
 For a finite value of $r_{UV}$, $r_0$ is determined by the relation, 
 \be
 \label{rera}
 r_0=r_{UV} \cos (\theta_0)
 \ee
 and we see that as $\theta_0\rightarrow \pi/2$ $r_0\rightarrow 0$, i.e. goes to the horizon. 
 Thus we see that in this case RT surfaces can go deep into the bulk, till the horizon, as $\theta_0$ is varied. 
 The area can also be obtained analytically and is given by 
 \be
 \label{anaxx}
 A=R^{m+n}  V_B V_{S^{m-1}} {r_{UV}^n\over n} \sin(\theta_0)^n.
 \ee
 
 More generally, we cannot solve eq.(\ref{producteom}) exactly and carry out instead a linearised analysis here. We know that every solution to eq.(\ref{producteom}) must eventually end up at the equator of the transverse sphere once the UV cut-off is taken to infinity in accordance with the GK theorem. When $r_{UV}$ is large, for investigating the behaviour  near $r_{UV}$ we expand $\theta$ as $\theta = \frac{\pi}{2} + \epsilon$. Eq.(\ref{producteom}) at linear order in $\epsilon$  then takes the form, 
\begin{equation}
	r^2 \Ddot{\epsilon} + (n+1) r \dot{\epsilon} + (m-1) \epsilon=0. \label{osceom}
\end{equation}
This is the equation for a damped harmonic oscillator (as can be made explicit by changing variables from $r$ to $e^x$),
with solution
\be
\label{sole}
\epsilon=  A r^{(-{n \over 2}+{\sqrt{n^2+4-4m}\over 2}) } + B r^{(-{n \over 2}-{\sqrt{n^2+4-4m}\over 2} )}
\ee
where $A,B$ are integration constants.

For $n^2 < 4 (m-1)$ we get underdamped oscillation while for $n^2 > 4 (m-1)$ we get over damped oscillation. 
In both cases, for $m>1$, the solution decays so that $\theta\rightarrow {\pi\over 2}$, and in the underdamped case this decay is accompanied by oscillations. 
Note, as is easy to see,  that the above solution in eq.(\ref{sole})  is in agreement with the exact solution eq.\eqref{exactsol}
with $B=0$. 


We end with some  comments pertaining to the underdamped case. 
Since $\theta$ oscillates one can achieve a given value $\theta_0$ at the boundary by considering surfaces which have different turning points $r_0$ in the interior. In fact this can happen in two ways, either there are multiple values of $r_0$ for which we reach the same value of $\theta_0< \pi/2$- this corresponds to a polar cap including the north pole that  closes at $r_0$.  Or in some cases, we  can overshoot the value $\pi/2$, in this case the polar angle can  reach a value  $\pi-\theta_0$ at the boundary- which corresponds to a polar cap now including the south pole that  closes at $r_0$.  One would like to know in these cases which of these surface has the lowest area? 

Consider for concreteness $AdS_5\times S^5$ (with $n=3$ and $m=5$). Eq.(\ref{sole}) gives
\begin{equation}
	\epsilon = r^{-\frac{3}{2}} \left(A \cos(\frac{\sqrt{7}}{2} \log(r)) + B \sin(\frac{\sqrt{7}}{2} \log(r))\right).
\end{equation}
We have found numerically that the lowest area surface arises (for a given $r_{UV}$) when the turning point $r_0$ takes the biggest value possible, i.e. when $r_0\over r_{UV}$ takes the biggest value, among all the surfaces. 
This is shown in Fig \ref{D3branearea} (b) where the Area $A$, in suitably normalised units,  is plotted as a function of $r_0$. We see in general that $A$ is monotonically decreasing  with increasing $r_0$. This also implies that among all surfaces which reach the same value of the polar angle (either $\theta_0$, or $\pi-\theta_0$, as mentioned above)  in the plot, the surface with the largest value of $r_0$ will have the lowest area. Our numerics also show that  the extent of the overshoot beyond $\pi/2$ is in fact small. This is shown in  Fig \ref{D3branearea} (a) where we  plot $\theta-\pi/2$, and find that its maximum value is $\sim 2 \times 10^{-2}$. 
We leave a more general investigation beyond $AdS_5\times S^5$, for the future.

{\begin{figure}
		\centering
		\subfigure[]{\includegraphics[width=0.45\textwidth]{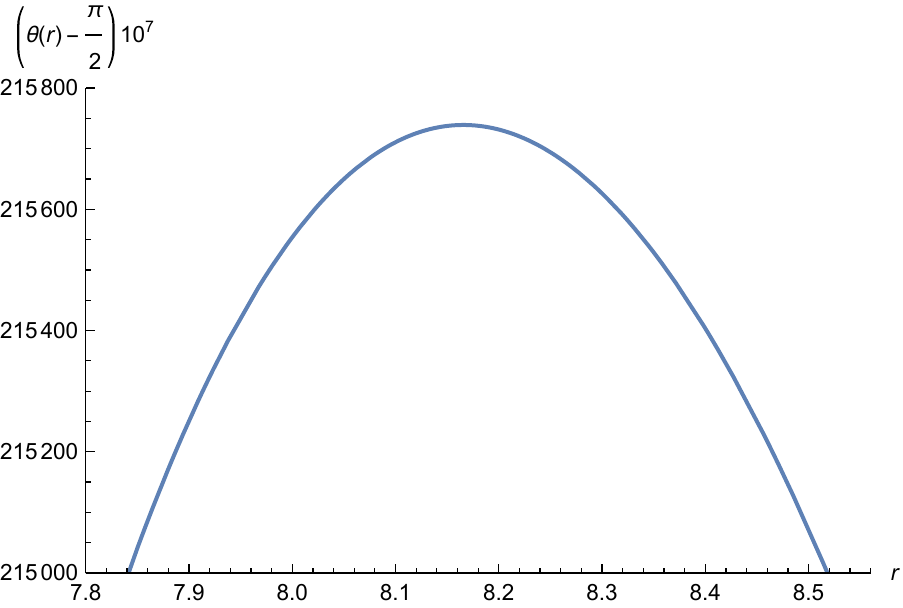}}\hfill
		\subfigure[]{\includegraphics[width=0.45\textwidth]{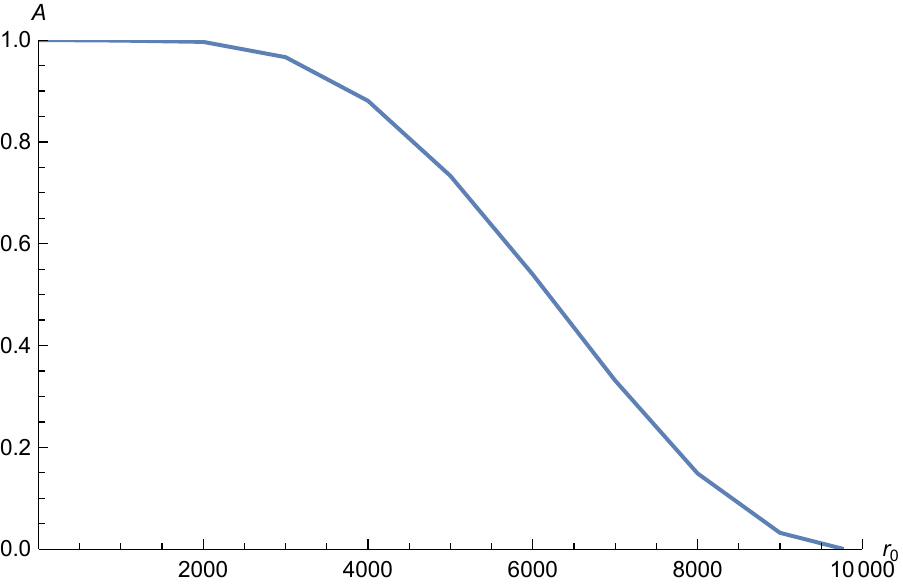}} 
		\caption{ (a) We have plotted the difference between the value of $\theta$ and $\frac{\pi}{2}$ as a function of $r$ after enhancing it by $10^7$. As is clear from the plot the maximum occurs around 8.16. (b) We have plotted the area as a function of $r$. We have divided the area eq.\eqref{af2} by the value it gives at $r_0 =5$. The plot then clearly represents a function that is monotonically decreasing from 1 to 0 as $r$ increases from $r_0=5$ to $r_{UV} =10^4$.   }
		\label{D3branearea}
\end{figure}}

\subsubsection{Warped product case}
\label{Warpedspacegeo}
Next we turn to warped  product cases and examine the behaviour of extremal surfaces when there is a boundary at a finite cut off $r_{UV}$ in the geometry. In particular, we will consider spherical ``caps" of the kind considered above.  We are interested in knowing if there are  oscillations in $\theta$ around $\frac{\pi}{2}$ as one approaches the boundary.

To  start, the area is given by   the following area functional,
\begin{equation}
	A=R^{m+n}  V_B V_{S^{m-1}} \int   r^{n-1}  (\sin \theta)^{m-1} \Phi^{m-1} \sqrt{(dr)^2+ r^2  \Phi^2 d \theta^2 } \label{afwp}
\end{equation}
As before we assume $\theta$ to be a function of $r$, then the EL equation is,
\begin{equation}
	\frac{d}{d r} \left( \frac{r^{n+1} \Phi^{m+1} (\sin(\theta))^{m-1} \dot{\theta} }{ \sqrt{1+ r^2 \Phi^2 \dot{\theta}^2}}\right) - (m-1) \cos(\theta) (\sin(\theta))^{m-2} r^{n-1} \Phi^{m-1} \sqrt{1+ r^2 \Phi^2 \dot{\theta}^2}  =0. \label{warpedproducteom}
\end{equation}
We assume the following form for $\Phi$ for concreteness,
\begin{equation}
	\Phi = \alpha + \beta r^{\delta}.
\end{equation}
Taking $\theta = \frac{\pi}{2}+\epsilon$ and expanding the above equation to $\order{\epsilon}$, gives
\begin{equation}
\label{eqwt}
	(m-1)\epsilon + r (\alpha + \beta r^{\delta}) \left( (\alpha (n+1) + \beta r^{\delta} (n+1 + \delta(m+1))) \dot{\epsilon} + r (\alpha + \beta r^{\delta}) \Ddot{\epsilon} \right) =0
\end{equation}
In the limit $\delta,\beta \rightarrow 0, \alpha\rightarrow 1$, we recover eq.\eqref{osceom}. 

Next, we  take $\delta >0$, so that the dilaton grows towards the boundary. In the large $r$ limit we get from eq.(\ref{eqwt}),
\begin{equation}
  (n+1 + \delta(m+1)) r \dot{\epsilon} + r^2 \Ddot{\epsilon}=0
\end{equation}
Solution to the above equation is,
\begin{equation}
	\epsilon = c_1 - c_2 \frac{1}{n + \delta(m+1)} r^{-(n+\delta(m+1))}.
\end{equation}
Hence we see that one  only gets a power law decay and not an oscillating solution as in the direct  product case above. 
The fact that as $r\rightarrow \infty$, $\epsilon\rightarrow c_1$, i.e. to a constant, is in agreement with our analysis in section \ref{warp} where we showed that the GK theorem can be avoided in the presence of warping. 
\subsection{The $Dp$ brane  geometry with a UV cut-off}
\label{Dpbranegeo}						
We saw in section \ref{other} that non-trivial RT surfaces cannot exist in the $D0$ brane geometry if the UV cut-off is taken to infinity. However the supergravity approximation breaks down in this case as $r\rightarrow \infty$ so it is reasonable to consider what happens if we impose a UV cut-off at $r_{UV}$ where the curvature becomes of order the string scale. Our conclusion above  for the $r_{UV}\rightarrow \infty$ case will then not apply and RT surfaces will exist. However, one might expect that the turning point of such a surface will lie ``quite close"  to $r_{UV}$ and not sufficiently in the interior. 
We will investigate this issue below by considering the more general case of the near horizon $Dp, p< 3$,  brane geometries.

						%
						
The near horizon metric of $N$ coincident extremal $Dp$ branes with $p< 3$ in the string frame is of the form,
\begin{equation}
	ds^2 =  H(r) ^{-\frac{1}{2}} (-dt^2 + \sum_{i=1} ^{p} dy_i ^2) + H(r) ^{\frac{1}{2}} (dr^2 + r^2 d \theta ^2 + r^2 \sin^2{\theta} d \Omega_{n}), \label{DOhemisphmet}
\end{equation}
\begin{equation}
	H(r) = \left(\frac{R}{r}\right)^{n}, \hspace{1cm} R^n = (4 \pi)^{\frac{n-2}{2}} \Gamma\left(\frac{n}{2}\right) l_s ^n (g_s N).
\end{equation}						
where $n=7-p$ , $l_s$ is the string length and $g_s$ is the string coupling. The Ricci Scalar $\mathcal{R}$ for the above metric is,
	\begin{equation}
	\label{rc}
	\mathcal{R} \propto \frac{r^{\frac{3-p}{2}}}{R^{\frac{7-p}{2}}}.
\end{equation}
To go to Einstein frame we make the change,
\begin{equation}
	ds_E ^2 = e^{-\frac{\phi}{2}} ds^2
\end{equation}
where,
\begin{equation}
	e^{\phi} = H(r) ^{\frac{3-p}{4}}.
\end{equation}					
Supergravity is valid, with $\alpha'$ corrections being small,  as long as, $\mathcal{R} \sim \frac{1}{l_s ^2}$, this gives, eq.(\ref{rc}), 
\begin{equation}
	\label{valgsn}
	r \sim R (g_s N)^\frac{4}{(7-p)(3-p)}.
\end{equation}						
This provides the upper cutoff for $r$ which we denote by $r_{UV}$,  while the lower cutoff is obtained by considering the limit where dilaton becomes too large. This happens for,
\begin{equation}
	r \sim R.
\end{equation}						
Thus our analysis for the extremal area surfaces is valid in the following regime.
\begin{equation}
	1 \ll \frac{r}{R} \ll (g_s N)^\frac{4}{(7-p)(3-p)}.
\end{equation}						
\subsubsection{Minimal Surface}
\label{extsurDp}
We consider a surface whose boundary ends on a spherical cap at $\theta = \theta_0$. Then the area in the Einstein frame is given by,
\begin{eqnarray}
	A &=& V_{S^n} V_p \int H(r) ^{\frac{n+1-p}{4}}  r^n e^{-\frac{n+1+p}{4} \phi} (\sin(\theta))^n \sqrt{dr^2 + r^2 d\theta^2} \nonumber \\
	&=& V_{S^n} V_p \int H(r) ^{\frac{4-p}{2}}  r^n e^{-2 \phi} (\sin(\theta))^n \sqrt{dr^2 + r^2 d\theta^2}. 
\end{eqnarray}
Simplifying further we get,
\begin{equation}
	A = V_{S^n} V_p \int r^n H(r) ^{\frac{1}{2}} (\sin(\theta))^n \sqrt{dr^2 + r^2 d\theta^2} = V_{S^n} V_p R^{\frac{n}{2}} \int r ^{\frac{n}{2}} (\sin(\theta))^n \sqrt{dr^2 + r^2 d\theta^2}. \label{Dpbranearea}
\end{equation}						
We will construct extremal area surfaces numerically  which end at $r_{UV}$ as discussed above. 
						
We are interested in extremal surfaces which start at different values of the polar angle $\theta_0$ at $r_{UV}$. But it is easier numerically to construct the surfaces starting from the turning point $r_0$ where $\theta\rightarrow 0$ and then integrating out to larger values of $r$. As $r_0$ is varied the final polar angle $\theta_0$ also changes. 
						
	Some more details are as follows. 					
From \eqref{Dpbranearea} we get the following Lagrangian.
\begin{equation}
	\mathcal{L} =  r ^{\frac{n}{2}} (\sin(\theta))^n \sqrt{r'^2 + r^2},
\end{equation}
where,
\begin{equation}
	r' = \frac{d r}{d \theta}, n=7-p.
\end{equation}
The Euler Lagrange equation for the above Lagrangian is,
\begin{equation}
	\label{eoldp}
	\frac{d}{d\theta} \left(\frac{ r^{\frac{n}{2}} (\sin(\theta))^n r'}{\sqrt{r'^2 + r^2}}\right) - r^{\frac{n-2}{2}} (\sin(\theta))^n \left(\frac{n}{2}  \sqrt{r'^2 + r^2} + \frac{r^2}{\sqrt{r'^2 + r^2}} \right) =0.
\end{equation}
We assume that near the turning point,
\begin{equation}
\label{ntr}
	r'= k \theta^{\gamma} \implies r= r_0 + \frac{k}{\gamma +1} \theta^{\gamma+1}
\end{equation}						
where $k$ and $\gamma$ are positive constants. Then from eq.(\ref{eoldp}) we get 
\begin{equation}
	\gamma=1, \hspace{1cm} k = \frac{n+2}{2 (n+1)} r_0 \ = \frac{9-p}{2(8-p)} r_0.
\end{equation}						
						
Now we use the above starting point near $r_0$ and integrate to larger values of $r$ to obtain the surface. 
In the numerical plots  we take $R=1$ and set $r_{UV}=10^4$. From  eq.(\ref{valgsn}) we see that this fixes the value of $(g_s N)$. 

Actually,  away from the turning point it is more convenient to solve for $\theta$ as a function of $r$; this is because close to the boundary one can have oscillations of the type we discussed in section \ref{Productspacegeo}, which makes $r$ a multi-valued function of $\theta$. Eq.(\ref{eoldp}) is equivalent to 
\begin{equation}
	\label{eomdp}
	\frac{d}{dr} \left(\frac{ r^{\frac{n}{2} +2} (\sin(\theta))^n \dot{\theta}}{\sqrt{1 + r^2 \dot{\theta}^2}}\right) - n \sin(\theta))^{n-1} \cos(\theta) r^{\frac{n}{2}} \sqrt{1 + r^2 \dot{\theta}^2} =0.
\end{equation}
From eq.(\ref{ntr}) we obtain near the turning point, 
${d \theta\over dr }={1\over k} {1\over \theta^\gamma}$. Inputing  this in eq.(\ref{eomdp}) we can solve for $\theta(r)$. 
The resulting  plots are shown in figures \ref{fig:D0braneplot}, \ref{fig:Dpbraneplot} for $p=0,1,2$, and for varying values of $r_0$. 

Note also that eq.(\ref{eomdp}) is invariant under the scaling $\theta\rightarrow \theta$, $r\rightarrow \lambda r$. 
This means  $\theta$   depends on $\theta_0$, its value at $r_{UV}$ and  the ratio $r\over r_{UV}$.
		
In Fig \ref{fig:D0braneplot}  	we see that as $r_0$ is decreased  one reaches a bigger value of $\theta_0$, with $\theta_0=\pi/2$ being reached when $r_0=1435.7$, i.e. for $r_0/r_{UV}=0.1436$. For purposes of further discussion we refer to this value as 
\be
\label{mintpd0}
\xi_{min}=( r_0/r_{UV})_{min}=0.1436.
\ee
 Decreasing $r_0$ further beyond this value leads to oscillations; the amplitude of these oscillations is very small so they cannot be seen in Fig \ref{fig:D0braneplot}. 
Similar behaviour is also observed for $p=1,2$ Fig \ref{fig:Dpbraneplot} and as discussed in section \ref{Productspacegeo}, for the $p=3$ case. 

Due to oscillations, \cite{Anous:2019rqb},  there can be several surfaces which reach the same value of $\theta_0$ on the boundary, as was also discussed for $p=3$ in section \ref{Productspacegeo}. We have checked that for $p=0,1,2,$ as in the $p=3$ case,  the minimum area is obtained by the surface with the largest value of $r_0$, i.e. when the turning point is closest to the boundary. Thus the RT surfaces, which minimise the area, do not go beyond $\xi_{min}$, eq.(\ref{mintpd0}). We see from  Fig \ref{fig:D0braneplot} and eq.(\ref{mintpd0}) that this turning point does not go ``deep" into the interior, in any parametric sense. Since we choose $r_{UV}$ to be located where supergravity breaks down, this teaches us that the surface of minimum area lies everywhere in the  strong curvature region and as a result its area cannot be computed reliably in the supergravity approximation. 


%

{\begin{figure}[]
	\centering
	\includegraphics[scale=0.8]{{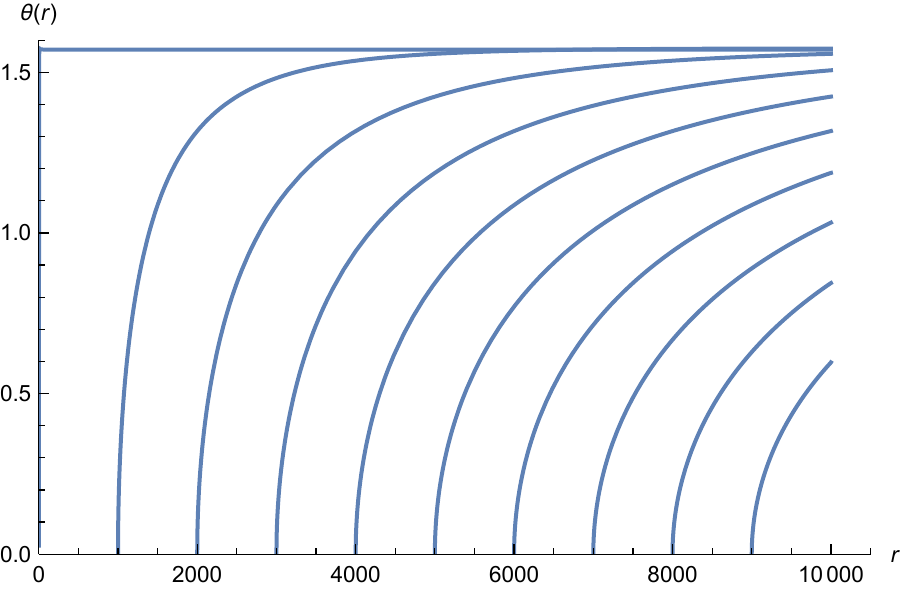}}
	\caption{Plot of $\theta(r)$ as a function of r, for various values of $\theta_0$,  in the case of the $D0$ brane, with $R=1,  r_{UV}=10^4$. As is clear from the plot, as  $\theta_0$ increases,  $r_0$ decreases. Also, as  $r\rightarrow \infty$, $\theta$ asymptotes to $\pi/2$.}
	\label{fig:D0braneplot}
\end{figure}}
	
{\begin{figure}
		\centering
		\subfigure[]{\includegraphics[width=0.45\textwidth]{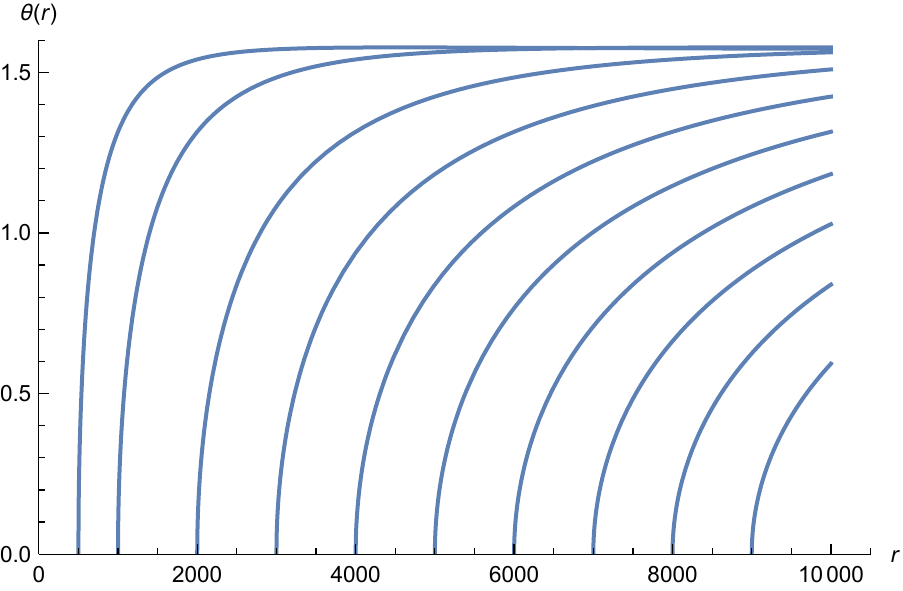}}\hfill
		\subfigure[]{\includegraphics[width=0.45\textwidth]{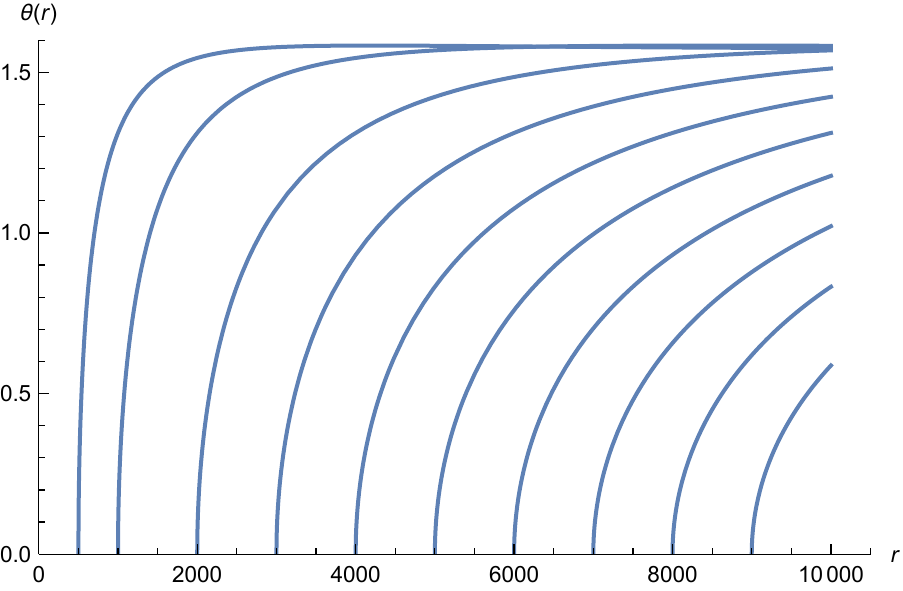}} 
		\caption{Plot of  $\theta(r)$ as a function of $r$,  for surfaces with various values of $\theta_0$ in the case of $Dp$ branes. (a) $p=1$, (b) $p=2$. With $R=1,  r_{UV}=10^4$.}
		\label{fig:Dpbraneplot}
\end{figure}}

To get a rough estimate of  the area of the RT surface it is convenient to define the variable
\be
\label{defxxi}
\xi={r\over r_{UV}}.
\ee
Eq.(\ref{Dpbranearea}) then becomes, 
\be
\label{sdpba}
A =  V_{S^n} V_p R^{\frac{n}{2}} r_{UV}^{{n\over 2}+1} \int^1_{\xi_0} d \xi \xi ^{\frac{n}{2}} (\sin(\theta))^n \sqrt{1 + 
\xi^2 ({d\theta\over d\xi})^2}.
\ee
 The turning point $\xi_0$ is determined by $\theta_0$ -- the value of $\theta$ at $r_{UV}$. 
 Thus we see that the Area only depends on $r_{UV}$ and on $\theta_0$. 
 For fixed  $\theta_0$,  
 \be
 \label{behavx}
 A\sim    r_{UV}^{ {n\over 2}+1} f(\theta_0)
 \ee
 where $f(\theta_0)$ is given by the integral in eq.(\ref{sdpba})
 \be
 \label{ftwo}
 f(\theta_0)=\int^1_{\xi_0} d \xi \xi ^{\frac{n}{2}} (\sin(\theta))^n \sqrt{1 + 
\xi^2 ({d\theta\over d\xi})^2}
\ee
 We see that $A$ is strongly dependent on the UV cut-off.
 E.g. setting $n=7-p$ we find for $D0$ branes that  it goes like 
 \be
 \label{behadz}
 A\sim r_{UV}^{{9\over 2}} f(\theta_0)
 \ee
 This is similar to eq.\eqref{shittwo} obtained in section \ref{moregencomp}. This shows that any entanglement entropy $A$ corresponds to, must be strongly dependent on the UV. We also learn from eq.(\ref{behadz}) that no subtraction can get rid of the UV divergence.

We end this section with a comment. The oscillations in $\theta$ mentioned above,  which occur as one goes towards the boundary can be understood as follows.  
Writing $\theta=\pi/2+\epsilon$, eq.(\ref{eomdp}) to linear order in $\epsilon$ gives, 
\begin{equation}
	 r^2\Ddot{\epsilon} + \frac{n+4}{2} r \dot{\epsilon} + n \epsilon =0
\end{equation}
The solution to the above equation is,
\begin{equation}
\label{eocc}
	\epsilon = C_1  r^{\frac{-n-2 + \sqrt{n^2 -12 n + 4}}{4}}+ C_2r^{\frac{-n-2 -\sqrt{n^2 -12 n + 4}}{4}}
\end{equation}
where, $C_{1,2} $ are constants. 
The   square root term in the exponent is negative for  $n\ge 1$, i.e. $p \leq 6$. Hence for the $Dp$ branes we are mainly considering here, $p\le 3$, this term is negative leading to  damped oscillations in $\epsilon$. 


\section{Discussion}
\label{disc}

In most known examples of gauge-gravity duality, internal symmetries of a field theory get geometrized into internal spaces which enter as factors in  direct product or warped product space-times. The main motivation behind this paper is to try and gain a better understanding of the role that these internal spaces, and the associated degrees of freedom in the boundary hologram, may play in the emergence of a smooth bulk spacetime. Towards this goal, we have investigated here, Ryu Takayanagi (RT) surfaces which are  associated with a subregion $R$ of the internal space, and end or are  ``anchored", on the boundary  of  $R$. Such surfaces, if they exist, could   teach us about some aspects of the dual field theory, and perhaps could also  play a role in understanding   how the bulk is reconstructed from the boundary theory. 
Above, we have referred to such surfaces as internal RT surfaces. 

It is worth emphasising that these surfaces, anchored on the boundary of a subregion of the internal space, are novel and not of the conventional type usually studied. For example, for 
product space times of the form  $Y^{AdS}_{n+2} \times K_m$, where $Y^{AdS}_{n+2}$ is an asymptotically $AdS$ space time and $K_m$ is a compact space, the conventional RT surfaces  wrap $K_m$ and are associated with a subregion of the $Y^{AdS}_{n+2}$ boundary (the base space of the dual theory). Instead, the internal RT  surfaces are associated with  a subregion $R$ of the internal space $K_m$, and wrap the spatial boundary of $Y^{AdS}_{n+2}$. Despite this difference,  the mathematical question of the existence of internal RT surfaces  is also well posed and one might hope that something interesting can be learnt from them\footnote{More generally, one can also consider extremal  surfaces which are associated with  a region which is a union of subregions of the internal and base spaces. We  leave  an investigation of such surfaces for the future.}.

The Graham Karch theorem, as we reviewed in section \ref{compact}, states that for spacetimes which  asymptotically are direct products of the form $AdS_{n+2}\times K,$ where $K$ is compact, internal RT surfaces associated with a region $R$ of $K$  can exist only if $\partial R$ is a minimal surface of $K$ \footnote{As explained earlier, minimal in this context 
 means surfaces whose area is unchanged under perturbations, to first order.}. 
One of our key results is that the Graham Karch theorem can be evaded if we consider warped product spacetime instead of direct products. As asymptotic analysis including warping was carried out in section \ref{compact},  where some explicit examples were also constructed. Actually, the simplest example is of a warped product $AdS_2``\times" S^m$, with which we began our study in section \ref{ads2}.  

The GK theorem can also be avoided if there is a boundary  in  space-time corresponding to a UV cut-off in the dual theory. 
 
 With these observations in mind we have carried out an investigation of several types of internal RT surfaces in the paper. 
 There are roughly three kinds of different behaviours we found which we  illustrate in the following  examples. 
 
 \subsection{Some Examples:}
 
 I ) The example of  $AdS_2``\times" K_m$, with $K_m=R^m, $ or $ S^m$,   was studied in section \ref{ads2}. Here we state the results in a somewhat general manner, more precise formulas can be found in section \ref{ads2}. We found   that the RT surfaces go deep into the bulk,  closer and closer to  the horizon, as the size of the region $R$ on $K_m$ increases. When $K_m=S^m$ increasing the size of the region sufficiently, so that the surface goes into the bulk, far from   the boundary in a parametric manner,  requires that the radius of 
 $S^m$, and of $AdS_2$,  satisfy the condition, $R_{S^m}\gg R_{AdS^2}$, i.e., that  the radius of the sphere is much bigger than  $R_{AdS^2}$. This condition is met by a big  extremal RN black hole (eRN)  in $AdS_{2+m}$,  which arises when a big  chemical potential ($R_{AdS} \mu \gg 1$) has been  turned on, but not by an extremal black holes in asymptotically flat space, where $R_{AdS^2}\sim R_{S^m}$. 
 The condition for the  size of the region  $l$ to be big is $l \gg R_{AdS^2}$, eq.(\ref{statcd}). 
 For such big regions the leading contribution to the area is eq.(\ref{fpads2})
 \be
 \label{areaxx}
 A= \mu^{m} (l)^m, 
 \ee
 and this is independent of the UV cut-off, to leading order. $\mu$ is the chemical potential for the eRN case, more generally it is the horizon value  of the dilaton, eq.(\ref{vala}), in suitable units.   
 
 We see that the entanglement scales like the volume of the region $R$ on the $T^m$. The analysis reveals also that  the entanglement,  can be calculated reliably in the IR 
 $AdS_2``\times"  K_m$ geometry with little sensitivity to the UV cut-off \footnote{In the eRN UV completion,  eq.(\ref{areaxx}) is the  leading behaviour of the renormalised area, after we remove the usual UV divergent term.}.
 Since RT surfaces go arbitrary deep  into the geometry, this also suggests that the  full bulk geometry should be reconstructable from the boundary theory, in terms of the IR quantum mechanics dual to the near-horizon $AdS_2$ theory, using  suitable degrees of freedom. The  entanglement entropy of these degrees  is calculated by the internal RT surface, and given by eq.(\ref{areaxx}). 
 
 One more comment is worth making, see section \ref{mgads2}. Note that while the  area, eq.(\ref{areaxx}), only depends on the horizon value of the dilaton, the existence of the RT surface, with its behaviour of going deep into the bulk, depends crucially on the radial variation of  the dilaton which acquires  its minimum value at the horizon. In contrast  the  deviation in the $2$ dimensional metric along the $r,t$ directions, from its leading order $AdS_2$ form,  while of the same order as the dilaton's radial evolution,  are unimportant. This fact is a reflection, in the entanglement calculation, of what is also known about  near-$AdS_2$ systems from low-energy scattering, etc. and is the reason why quite universally the near horizon region can be described by a Schwarzian theory, \cite{nayak2018dynamics, moitra2019extremal, maldacena2016remarks, iliesiu2021statistical, Heydeman:2020hhw}.
 
II) The $AdS_{n+2} ``\times" R^m$, with $n>0$, case was studied in section \ref{adsn}, this is a higher dimensional analogue of I). 
We found that the behaviour is quite different. 
As long as the dilaton, i.e. the warp factor, reaches a non-vanishing value at the horizon, the RT surfaces reveal the existence of a  gap, i.e. of a finite  correlation length,  along the internal space directions. Once the size, $l$,  in length units,   of the region $R$ becomes bigger than a critical size $l_{crit}$, given by eq.(\ref{lcrit}), the RT surface breaks up into two disconnected  components, each of which goes all the way into the horizon. And the area then grows like the perimeter
of the boundary region -- indicative of a correlation length of order $l_{crit}$. The area of the RT surface,  it turns out,   is strongly dependent on the UV cut off, which is denoted by $r_{UV}$ in section \ref{adsn}, see eq.(\ref{calx}), eq.(\ref{a3g}). 
This means the entanglement entropy is strongly dominated by short distances correlations of order the UV cutoff 
 and cannot be calculated reliably in the IR theory. 

These conclusions are supported by a known example, \cite{d2009magnetic}, section \ref{consdil}, where an example was studied in which a geometry which is $AdS_5$ in the UV flows to a $AdS_3\times R^2$ geometry in the IR. In the boundary theory this is due to a magnetic field having been turned on. Both scales $l_{crit}$ and $r_{UV}$ are of the same order in this case and set by the magnetic field  being turned on in the UV theory.  More generally our analysis above shows that 
any dimension changing RG flow, where a higher dimensional $AdS_{n+2+m}$  flows to a lower dimensional $AdS_{n+2}``\times" R^m$ can arise only if the system develops a mass gap. 

It is also worth noting that when a small temperature is turned on  the change in the area of the RT surface from its zero temperature value is finite and temperature dependent, see appendix \ref{fta}, eq.(\ref{finitearea1T}), eq.(\ref{rea5}), and can be reliably calculated in the IR geometry. 
This means that the renormalised area  obtained after subtracting a state independent  UV cutoff dependent term, is determined by the IR geometry alone and can be used to distinguish between different states in the IR theory, reflecting change   in the entanglement of the corresponding boundary states.  

%

Let us also note that   the fact that the RT surfaces goes in all the way to the horizon in these examples, suggests that a  reconstruction of bulk supergravity operators should be possible for the full bulk spacetime in terms of a suitable subalgebra of observables  in the IR theory. 
 
III) Finally we consider the case of the near horizon $D0$ brane geometry, discussed, along with other $Dp$ brane geometries, in section \ref{Dpbranegeo}. In this case we find that the RT surface in the presence of a UV cut-off $r_{UV}$, which we can take to  be located at a radial location where $\alpha'$ corrections become important, does not go deep inside, with the minimum value $r_{min}/r_{UV}\sim 0.14$. Thus the entanglement wedge, to the extent that  our analysis is  reliable, only covers the region of spacetime which is strongly curved and not the region where the geometry is smooth and  described by supergravity. Also the area of the surface depends strongly on $r_{UV}$,  as in case II) above, going here like $r_{UV}^{9/2}$ (see discussion after eq.(\ref{sdpba})).
This means that degrees of freedom whose entanglement is being calculated by  the RT surface  are strongly correlated at the scale of the cut-off. 

To end,  let us also briefly mention two more cases. For $p=3$ case, section \ref{Productspacegeo}, we are describing $AdS_5\times S^5$ with a cutoff.  In this case when we take $r_{UV}\rightarrow \infty$ all surfaces end up at  an equator of the boundary $S^5$ -- as follows from the GK theorem. Thus if a subalgebra of observables  exists in the cutoff theory, whose entropy is related to the RT surface, it must become under RG flow,  a universal type of subalgebra in the UV complete ${\cal N}=4$ theory, which accounts for half the degrees of freedom on the $S^5$. 

The second case is   $AdS_{n+2}\times S^n$ with a cutoff, studied in section \ref{Productspacegeo}.  Here the single connected component RT surfaces go all the way to the horizon in the  bulk, covering all of it. The area goes like $r_{UV}^n$, eq.(\ref{anaxx})  and is again  strongly UV dependent. However, as in example  II above, we expect that a UV subtraction which is state independent will give finite results for changes in the geometry, for example when a small temperature is turned on.

The bottom line, as  the examples above illustrate,  is  that  internal  RT surfaces  are quite interesting  and can teach us   in fact quite a bit  about the boundary theory. One can also hope to learn something about bulk reconstruction from them as mentioned above; we turn to this shortly below.

\subsection{ Subalgebras and Bulk Reconstruction:}

Let us end this discussion with some  comments on subalgebras and bulk reconstruction. Our  comments are  mostly preliminary and  meant more as suggestions for future  directions of work. 

\begin{figure}[h]
	\begin{minipage}{0.5\linewidth}
		\centering
		\begin{tikzpicture}
			\draw[thick] (0,0) circle [radius=2cm];
			\draw[very thick, blue] (0.84,1.812) .. controls (0.4,0.5) and (-0.4,0.5) .. (-0.84,1.812);
			\node[above right] at (0,2) {B};
			\draw[very thick, red] (-0.347,1.97) .. controls (0.0408,0.64) and (-0.614,0.18) .. (-1.732,1);
			\node[above left] at (-1.147,1.683) {A};
		\end{tikzpicture}
	\end{minipage}
	\hfill
	\begin{minipage}{0.5\linewidth}
		\centering
		\begin{tikzpicture}
			\draw[thick] (0,0) circle [radius=2cm];
			\draw[very thick, teal] (0.84,1.812) .. controls (0.4,0.5) and (-0.4,0.5) .. (-0.84,1.812);
			\node[above right] at (0,2) {B};
			\draw[very thick, teal] (-0.347,1.97) .. controls (0.0408,0.64) and (-0.614,0.18) .. (-1.732,1);
			\node[above left] at (-1.147,1.683) {A};
			
			\begin{scope}
				\clip (0.84,1.812) .. controls (0.4,0.5) and (-0.4,0.5) .. (-0.84,1.812) -- (-0.84,1.812) arc (115:65:2cm);
				\draw[very thick, orange] (-0.347,1.97) .. controls (0.0408,0.64) and (-0.614,0.18) .. (-1.732,1);
			\end{scope}
			
			\begin{scope}
				\clip (-0.347,1.97) .. controls (0.0408,0.64) and (-0.614,0.18) .. (-1.732,1) -- (-1.732,1) arc (150:100:2cm);
				\draw[very thick, orange] (0.84,1.812) .. controls (0.4,0.5) and (-0.4,0.5) .. (-0.84,1.812);
			\end{scope}
		\end{tikzpicture}
	\end{minipage}
	\caption{Strong subadditivity: since $S_{A\cup B} \le S_{green}$ and $S_{A\cap B} \le S_{orange} \implies S_{A\cup B} + S_{A\cap B} \le S_A + S_B$.}
	\label{repfigsph2}
\end{figure}
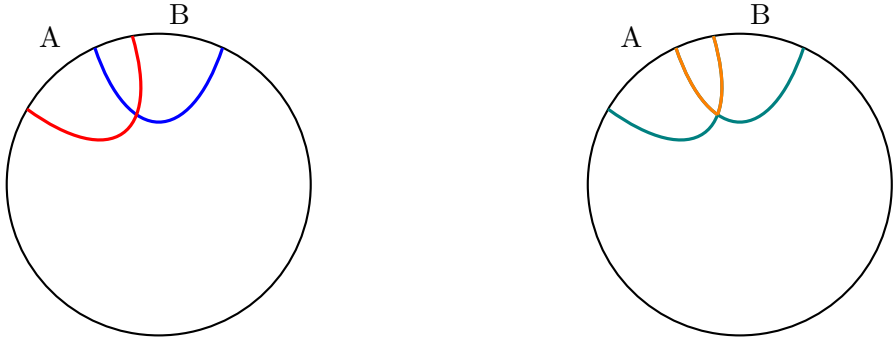

{\it Strong Subadditivity}: Before proceeding let us note that internal RT surfaces also satisfy the constraint which comes  from strong subadditivity.  For two  regions $A,B$ in the  internal space, and the corresponding RT surfaces which we also denote by $A,B,$ one can easily see that 
\be
\label{ssa}
{\rm Area}_{(A\cup B)}+{\rm Area}_{(A\cap B)} \le {\rm Area}_{(A)}+{\rm Area}_{(B)}.
\ee
This follows from the same geometric argument that applies to the conventional base space RT surfaces, as is indicated in Fig \ref{repfigsph2}, where the boundary now refers to the internal space (the  base space directions which the internal RT surface completely wraps are not shown in this figure).  For examples, like I)  above, where the IR internal space arises from the base space in the UV, this is of course only to be expected, but we see that it is more generally true. 
This property of internal RT surfaces  is consistent with  them   being related to the  entanglement entropy of some subalgebra.

%
We now turn to a discussion of subalgebras and bulk reconstruction. 

Conventional RT surfaces, associated with a subregion $R$ of the spatial boundary of $AdS$ space, are related to the subalgebra of operators in the dual  field theory,   localised to  also lie in  $R$. The entropy associated with this subalgebra is  the area of the RT surface and bulk operators within the entanglement wedge of the RT surface can be reconstructed in terms of this subalgebra, in a manner which protects them against erasure errors in the complement of $R$ \cite{harlow2017ryu, harlow2018tasi}. On general grounds one would expect many of these properties to also hold for internal RT surfaces. In particular, one expects that internal RT surfaces should also be associated with a suitable subalgebra of physical observables in the boundary theory. 

For cases where the internal space, when extended in the UV, arises from part of the base space, as in the $AdS_{2+m}\rightarrow AdS_2``\times"S^m$ example discussed earlier, this is automatically follows  from the correspondence between base space RT surfaces and the subalgebra ${\hat A}$ of operators localised in corresponding region of base space, mentioned above. Roughly speaking then,  the operators in ${\hat A}$, after RG flow to the IR, give rise to subalgebra corresponding to the internal RT surface in the IR theory. It is easy to see that this subalgebra is of the second type, eq.(\ref{0-6}),  (\ref{0-7}), discussed in section \ref{sec:intro} -- since ${\hat A}$ in the UV theory is of this type to begin with (we are starting with gauge invariant operators and localising them in $R$ to obtain ${\hat A}$). 

However, this provides a description of the subalgebra relevant for the IR region in terms of the UV boundary theory which is $m+1$ dimensional. From the IR perspective, we expect that the near $AdS_2$ geometry is dual to a $1$ dimensional Quantum Mechanics (QM) -- indeed recent developments on the  ${\rm near} \ AdS_2/ {\rm near} \  CFT$ correspondence, provide considerable evidence for this, \cite{nayak2018dynamics, moitra2019extremal, maldacena2016remarks, iliesiu2021statistical, moitra2019jackiw}. This means that there should be a subalgebra in the QM, associated with the corresponding part of the internal space, $K_n$, which reconstructs supergravity operators in the IR region of the entanglement wedge. And the IR part of the RT surface calculates the entanglement entropy associated with this subalgebra. A better understanding of such  subalgebras in the IR QM theory would be worth obtaining.



More generally, the various examples  we have studied  suggest that  an  internal RT surface  computes the entanglement of some  subalgebra  ${\cal A}$ of observables in the dual theory. When ${\cal A}$ can  reconstruct supergavity modes deep in the bulk the corresponding RT surfaces penetrate far inside. But in situations where the RT surface ``hangs" close to the UV boundary this  subalgebra  cannot carry out  a reconstruction of the  region sufficiently deep inside the bulk. 

In some cases though, it might  be reasonable to believe that  bulk reconstruction is still  possible, even for  the interior bulk region  where no  internal RT surfaces  can reach. For example, consider the near-horizon $D0$ geometry (similar comments apply for all $Dp$,  $p<3$ brane geometries). We know, from gauge-gravity  duality, \cite{maldacena1999large},\cite{aharony2000large}, that  this theory is dual to the corresponding $D0$ brane gauge theory. Our results in this paper show that RT surfaces do not go deep into the interior where Sugra is a good approximation, in this case. However one would expect that an HKLL type of reconstruction for bulk supergravity operators localised in the some regions deep in the interior, where Sugra is trustworthy, should still be possible in terms of a suitable sub-algebra in the boundary QM\footnote{The detailed map between bulk fields and operators in the QM has been discussed in \cite{Sekino:1999av}}. For example, there are Rindler observers in the bulk and the region in the bulk lying within their horizon can be associated with the causal wedge of a  region  of the boundary $S^8$. This suggests that there is another subalgebra for this region in the $D0$ brane case, and more generally in similar cases, which allows for bulk reconstruction. But the  entropy of this subalgebra would not be  given by an  RT surface. It is possible that such a subalgebra could be of the first type considered in section \ref{sec:intro}, eq.(\ref{0-1}), (\ref{0-3}).

Finally in terms of open questions,  to turn things around, our investigation also suggests, from a boundary theory perspective, that it might be possible  to calculate the entropy, for some subalgebras associated with target space constraints,  by means of a replica trick calculation. The gravity dual of such a calculation would then map to the RT surface anchored on the corresponding region of the internal space. Investigating this question  in field theory itself would  quite be interesting.

\section{Acknowledgments} We thank Abhijit Gadde, Apratim Kaviraj, Shiraz Minwalla, Onkar Parrikar  and Sunil Sake   for discussions.  A.K.,  K. K. N., G. M. and S. P. T. acknowledge   the support of the Govt. Of India, Department of Atomic Energy, under Project No. 12-R\&D-TFR-5.02-0200 and support from the Quantum Space-Time Endowment of the Infosys Science Foundation. The work of S.R.D. and M.H. Radwan is partially supported by National Science Foundation grants NSF-PHY/1818878, NSF-PHY/2111673 and by a Jack and Linda Gill Chair Professorship. S.R.D. would like to thank T.I.F.R for hospitality during the completion of this paper.
						
\appendix						
\section{Finite temperature extension}
\label{fta}
\subsection{Near extremal RN branes calculation}
Here we consider near extremal black branes which flow from $AdS_4$ in the UV to $AdS_2\times R^2$ in the IR. The discussion here is a continuation of section \ref{flowads4to2} which discussed a similar flow for extremal black branes. We will be interested in extremal surfaces in the near extremal RN geometry here. The notation below for metric components etc. is taken from section \ref{flowads4to2}.  
						
Near-extremal black branes have a small temperature ${\mathcal T}$ compared to the chemical potential $\mu$, i.e. ${\mathcal T} \ll \mu$. 
						
Like section \ref{flowads4to2}  we   consider a strip extending in the $x$ direction. The main difference now is that in the near horizon limit ,
\begin{equation}
	f(r) = \frac{(r-r_h) (r-r_+)}{R_2 ^2}.
\end{equation}						
Here $r_+$ is the outer horizon of the brane which in the small temperature limit is given by,
\begin{equation*}
	\label{shifth}
	r_+ = r_h + T.
\end{equation*}						
Note that $f(r)$ has a single zero at the horizon as $r\rightarrow r_+$. $T$ in eq.(\ref{shifth})  is the  re-scaled temperature given in terms of the   physical temperature $\mathcal{T}$ by,
\begin{equation}
	T = \frac{2 \pi}{3} R_4 ^2 \mathcal{T}.
\end{equation}
When ${\mathcal T}  \ll \mu$, 
\be
\label{condtT}
T\ll r_h.
\ee
						
In the far horizon region the answer eq.\eqref{farho} for $\Delta x$ does not change much. 
\begin{equation}
	\Delta x_{FH} \sim \frac{R_4}{r_+}.
\end{equation}					
So we focus on the near horizon here. Integrating both sides of eq.\eqref{deltax}  in the near horizon limit we get,
\begin{align}
	\Delta x &= \int_{r_0} ^{r_B} \sqrt{\frac{r_0 ^4}{f(r) (r^6 - r_0 ^4 r^2) }} dr  \nonumber\\
	& = \int_{r_0} ^{r_B} \sqrt{R_2 ^2 \frac{r_0 ^4}{(r-r_h) (r-r_+) (r^6 - r_0 ^4 r^2) }} dr  = \int_{r_0} ^{r_B} \frac{R_2 r_0 ^2}{ r \sqrt{(r-r_h) (r-r_+)(r^4 - r_0 ^4)}} dr \nonumber \\
	& = \int_{r_0} ^{r_B} \frac{R_2 r_0 ^2}{ r_0 \sqrt{4 r_0 ^3 (r-r_h) (r-r_+) (r - r_0)}} dr  = \int_{r_0} ^{r_B} \frac{R_2}{2 r_0 ^{\frac{1}{2}}\sqrt{(r-r_h) (r-r_+)(r - r_0)}} dr \nonumber \\
	& = \int_{0} ^{y_B} \frac{R_2}{2 r_0 ^{\frac{1}{2}} \sqrt{(y+r_0-r_h) (y +r_0-r_+)y}} dy.  \hspace{1cm} (\text{$y=r-r_0$})
\end{align}						
We assume that $r_0$ will be very close to $r_+$ for a large enough $\Delta x$. In fact we will assume that 
\be
\label{condaax}
{r_0-r_+\over r_+-r_h}\equiv \epsilon \ll 1.
\ee 						
This will be proved self consistently below. 
						
We now rewrite the above integral after rescaling ${\tilde y}={ y \over r_+-r_h}$, as 
\be
\label{newdelxa}
\Delta x= {R_2\over 2 \sqrt{r_0 (r_+-r_h) }} \int_0^{\tilde y_B} 
{d{\tilde y} \over \sqrt{  ({\tilde y} + 1 + \epsilon ) ({\tilde y} + \epsilon) {\tilde y} }}
\ee
where ${\tilde y_B}= {y_B\over r_+-r_h}= {y_B \over T}\gg 1$.						 
						
By examining the behaviour of this integral near lower limit ${\tilde y}=0$ it is easy to see since $\epsilon \ll 1$ that to leading order 
\be
\label{valdelx}
\Delta x \simeq {R_2\over 2 \sqrt{r_0 (r_+-r_h)}}\ln({1\over \epsilon}).
\ee
This gives,
\begin{equation}
	\label{hclose}
	(r_0 - r_+) \simeq T \exp{\left( - 2 \Delta x \frac{\sqrt{r_+ T}}{R_2}\right)}.
\end{equation}						
We see that when the term in the exponent is sufficiently big, i.e., 
\be
\label{condx}
\Delta x \frac{\sqrt{r_+ T}}{R_2}\simeq \Delta x \frac{\sqrt{r_h T}}{R_2}\gg 1,
\ee
the condition eq.(\ref{condaax})  is indeed met.							 
							
							Thus we learn that for small temperatures, when the  strip size  $\Delta x$ is very large the entangling surface goes very close to the horizon. 
The result for the area now becomes,
\begin{equation}
	A \sim \Delta x \Delta y r_+ ^2= \Delta x \Delta y (r_h+T)^2\simeq \Delta x \Delta y( r_h^2 + 2 r_h T).
\end{equation}		
Thus the leading temperature dependence is linear in $T$.							 
							
We also note that for intermediate values of  $\Delta x$,
	\be
\label{condbbx}
{R_2 \over r_h} \ll \Delta x \ll {R_2 \over \sqrt{r_h T}},
\ee						
the surface will go into the near horizon $AdS_2$ region but will not lie very close to the horizon. Finally, for even smaller strip sizes $\Delta x \le {R_2 \over r_h}$ the effect of the temperature will be small and as in the zero $T$ case the surface will not enter the near-horizon $AdS_2$ region. 
							
							\subsection{Finite Temperature $AdS_{n+2} \times R^m$}
							\label{ftadsn}
The discussion in this subsection is a continuation of section \ref{consdil} and our notation is borrowed mostly from that section. In particular we will be interested in extremal surfaces in a geometry where the $AdS_{n+2}$ is replaced by a finite temperature black brane. The extremal surface we will study, like in section \ref{consdil}, at the boundary of $AdS_{n+2}$ ends in turn on the boundary of a strip in the $y_1$ direction, and wraps all the other $y^\mu$ directions. 							
							
						
The metric for the finite temperature case goes as,
\begin{equation}
	ds^2 = R^2 \left(- r^2 f(r) dt^2 + \frac{dr^2}{r^2 f(r)} + r^2 \sum_{i} (d x_{i})^2\right) + R^2 \alpha^2 \sum_{\mu} (dy_\mu)^2, \label{finiteTmet}
\end{equation}
where, $R$ is the AdS radius and,
\begin{equation}
	f(r) = 1 -\left(\frac{r_h}{r}\right)^{n+1}.
\end{equation}
The temperature is given by,
\begin{equation}
	T = \frac{n+1}{4 \pi} r_h. \label{temp}
\end{equation}
We are interested in the regime
\be
\label{regimest}
T \ll r_{UV}\implies r_h\ll r_{UV}.
\ee						
						
	The area functional becomes,
\begin{equation}
	\label{areafunc}
	A = 2 R^{m+n} \alpha^{m-1} V_{m-1} V_{n} \int_{r_0}^{r_{UV}} r^n \sqrt{\frac{dr^2}{r^2 f(r)} + \alpha^2 dy_1^2}.
\end{equation}
So the function we have to extremize is,
\begin{equation}
	\mathcal{L} = r^n \sqrt{\frac{1}{r^2 f(r)} + \alpha^2 \Dot{y_1}^2}.
\end{equation}
The conjugate momentum to $y_1$
\begin{equation}
	P_y = \frac{r^n \alpha^2 \Dot{y_1}}{\sqrt{\frac{1}{r^2 f(r)} + \alpha^2 \Dot{y_1} ^2}},
\end{equation}
has the value
\begin{equation}
	\alpha r_0 ^n.
\end{equation}
Rearranging the equation we get ,
\begin{equation}
	\alpha  \Dot{y_1} = \frac{r_0 ^n}{\sqrt{r^2 f(r)(r^{2n} - r_0 ^{2n})}}.
\end{equation}
Integrating both sides then gives,
\begin{equation}
	\alpha \Delta y_1 = \int_{r_0} ^{r_{UV}} \frac{r_0 ^n}{\sqrt{r^2 f(r)(r^{2n} - r_0 ^{2n})}} dr. \label{finiteTyint}
\end{equation}

Computing it in the large $r$ limit, where $f(r) \rightarrow 1$ we get,
\begin{equation}
	\alpha \Delta y_1 = C  - \frac{r_0 ^n}{n r_{UV} ^n},
\end{equation}
where $C$ is a constant. We see that the RHS has a well defined limit for $r_{UV} \rightarrow \infty$. The temperature dependence will arise from the behaviour of the surface in the interior, away from the UV cut-off. 						
						
In general the area functional depends on three dimensionful parameters, $r_0, r_h, r_{UV}$, and thus two dimensionless ratios. In parallel with  the discussion for $T=0$ there are three kinds of surfaces to consider. The first surface goes in from the boundary and moves in both $r, y_1$. The second ``hangs" close to the boundary, without going radially inward. Finally the third goes in straight radially without moving in the $y_1$ direction.
						
						Let us first consider the regime where 
						
\be
\label{condvc}
{r_0-r_h\over r_h} \ll 1 .
\ee						
						
And start with the first surface. When eq.(\ref{condvc}) is met we expect  the surface to correspond to a strip size $\Delta y_1$ which is sufficiently big. We will see that this expectation is borne out self consistently in our analysis. 
						
With the approximation eq.(\ref{condvc})  we get, to leading order,  that the contribution close to the horizon  satisfies the condition,
\begin{eqnarray}
	\alpha \Delta y_1 &=& \int_{r_0} ^{r_B} \frac{r_0 ^n}{\sqrt{r^2 (r^{2n} - r_0 ^{2n})}} \sqrt{\frac{r^{n+1}}{r^{n+1} - r_h ^{n+1} }} dr \nonumber \\
	& \simeq & \int_{r_0} ^{r_B} dr \frac{1}{\sqrt{(r-r_0) (r-r_h)}} \frac{1}{\sqrt{2n (n+1)}} \nonumber \\
	& \simeq & \int_{0} ^{z_B} \frac{2}{\sqrt{2n (n+1)}}dz \frac{1}{\sqrt{(z^2 + \epsilon ^2)}},
\end{eqnarray}						
where $z = \sqrt{r-r_0}, \epsilon = \sqrt{r_0 - r_h}$. Here $r_B$ is  the cutoff in the near-horizon region. Integrating we obtain,
\begin{equation}
	\alpha \Delta y_1 \simeq \frac{1}{\sqrt{2n (n+1)}} \log{\frac{r_B - r_0}{r_0 - r_h}},
\end{equation}
leading to						 
\begin{equation}
r_0 - r_h \simeq r_h \exp( - \alpha \sqrt{2n (n+1)} \Delta y_1). \label{diffdelyT}
\end{equation}
Here we have assumed $r_B \sim \order{r_h}$. We see, as expected, that when $\Delta y_1$ is sufficiently big meeting the condition, 
\be
\label{condyy}
\alpha \Delta y \gg 1,
\ee						
the turning point will come close to the horizon meeting the condition, eq.(\ref{condvc}). 
						
The area in this case is given by, 
\begin{equation}
	A_1 = 2 R^{m+n} \alpha^{m-1} V_{m-1} V_{n} \int_{r_0} ^{r_{UV}} dr \frac{r^{2n-1}}{\sqrt{f(r) (r^{2n} - r_0 ^{2n})}}. \label{A1finiteT}
\end{equation}						
This is divergent as $r_{UV}\rightarrow \infty$, with the contribution near $r=r_{UV}$ going like
\be
\label{divc}
\Delta A_{UV}=2 R^{m+n} \alpha^{m-1} V_{m-1} V_{n} ( {r_{UV}^n\over n} + \cdots),
\ee						
where the ellipses indicate contributions which vanish as $r_{UV}\rightarrow \infty$. The main temperature dependent contribution comes from close to the turning point and is given by 
\begin{equation}
	\Delta A \simeq 2 R^{m+n}  V_{m-1} V_{n} r_h ^n \alpha^{m}  \Delta y_1 \sim 2 R^{m+n}  (V_n T^n) (V_{m-1}   \Delta y_1 \alpha^{m}),
	\label{finitearea1T}
\end{equation}
with,
\be
\label{suma1}
A_1=\Delta A_{UV}+ \Delta A.
\ee						
We see that the temperature dependent  contribution $\Delta A$ is extensive in both the volume $V_n$ along the spatial direction at the boundary of $AdS_{n+2}$, with a scale set by the temperature $T$, and  also  in the volume along the $R^m$ directions, $V_{m-1} \Delta y_1$,  with the scale set by $\alpha$. We reiterate that eq.\eqref{finitearea1T} is only valid in the limit of eq.\eqref{condyy} where turning point $r_0$ is very close to the horizon $r_h$ such that eq.\eqref{condvc} is met.

The second surface  which ``hangs" near the boundary, it is easy to see has an area,
\begin{equation}
	A_2 =2   R^{m+n} ( V_{n} r_{UV} ^n) (V_{m-1}   \Delta y_1 \alpha^{m}). \label{finitearea2}
\end{equation}						

Finally, one might want to consider a  third surface, which is analogous to the one that goes straight in along the radial direction, without moving along $y_1$, at $T=0$. However we note now, eq.(\ref{finiteTmet}),  that the metric component along the $(dx_\alpha)^2$ directions does not go to zero at the horizon, since $r_h>0$, unlike the zero temperature case. 
Thus to get a surface which is homologous to the  region of interest at the boundary we must add an extra segment to the surface which traverses the required interval in the $y_1$ direction. This gives the area:
\be
\label{a32}
A_3= 2 R^{m+n} \alpha^{m-1} V_n V_{m-1} \int_{r_h}^{r_{UV}} r^{n -1} {dr \over \sqrt{f(r)}} + 2 R^{m+n} \alpha^{m-1} V_n V_{m-1} r_h^n \alpha \Delta y_1
\ee
where the first term comes from two  segments that go straight along the radial direction till the horizon, and the second due to a segment at $r_h$ that moves along $y_1$. 
The resulting surface does not satisfy the minimal area condition though and we expect that its area will be bigger than $A_1$, as we will verify  shortly below. 
  
We can write the integral on the RHS of eq.(\ref{a32})   as
\be
\label{rea3}
\int_{r_h}^{r_{UV}} r^{n -1} {dr \over \sqrt{f(r)}} =  \left({r_{UV}^n\over n}- {r_{h}^n\over n}+r_h^n \int_1^{\xi_{UV}} \xi^{n-1} d\xi \left[{1\over \sqrt{1-{1\over \xi^{n+1}} }}-1\right]\right).
\ee
In the last integral $\xi_{UV}={r_{UV}\over r_h}\gg 1$. Writing this last integral as	
\begin{eqnarray}
	I &=& \int_1^{\xi_{UV}} \xi^{n-1} d\xi \left[{1\over \sqrt{1-{1\over \xi^{n+1}} }}-1\right] \nonumber \\
	&=& \int_1^\infty \xi^{n-1}d\xi \left[{1\over \sqrt{1-{1\over \xi^{n+1}} }}-1\right] -\int_{\xi_{UV}}^\infty 
	\xi^{n-1} d\xi \left[{1\over \sqrt{1-{1\over \xi^{n+1}} }}-1\right] \label{lastint}
\end{eqnarray}
we get that 
\be
\label{lastint2}
I= {\hat C} + \frac{1}{n} - {r_h\over 2 r_{UV}} + \order{{r_h^2\over r_{UV}^2}},
\ee
where 
\be
\label{valch}
{\hat C}= {-\sqrt{\pi} \Gamma({1\over 1+n})\over n \Gamma(-{1\over 2}+{1\over 1+n})} \geq 0 \hspace{0.5cm} \text{for} \hspace{0.5cm} n \geq 1 ,
\ee
and we see that this integral converges when we take ${r_{UV}\over r_h} \rightarrow \infty$, 
leading to 
\be
\label{rea4}
A_3=2  R^{m+n} \alpha^{m-1} V_n V_{m-1} \bigl({r_{UV}^n\over n}+r_h^n({\hat C}  + \cdots)\bigr) + 2 R^{m+n} \alpha^{m-1} V_n V_{m-1} r_h^n \alpha \Delta y_1
\ee						
Comparing $A_3$ with $A_1$ above we see that in both the UV divergent terms, scaling like $r_{UV}^n$, agree. The comparison of the temperature dependent terms between both the areas  is in general not so straightforward. However in the limit eq.\eqref{condyy} we see that $A_3$ in eq.\eqref{rea4} has an extra  temperature dependent finite term (reproduced below in eq.\eqref{rea5}) that is not in $A_1$. Thus in the limit where eq.\eqref{condvc} is met, such that eq.\eqref{condyy} is valid, $A_1 < A_3$. 
\be
\label{rea5}
\Delta A_3=2 R^{m+n} \alpha^{m-n} V_n V_{m-1} r_h^n ({\hat C}).
\ee						

\begin{figure}[]
\centering
\includegraphics[scale=0.8]{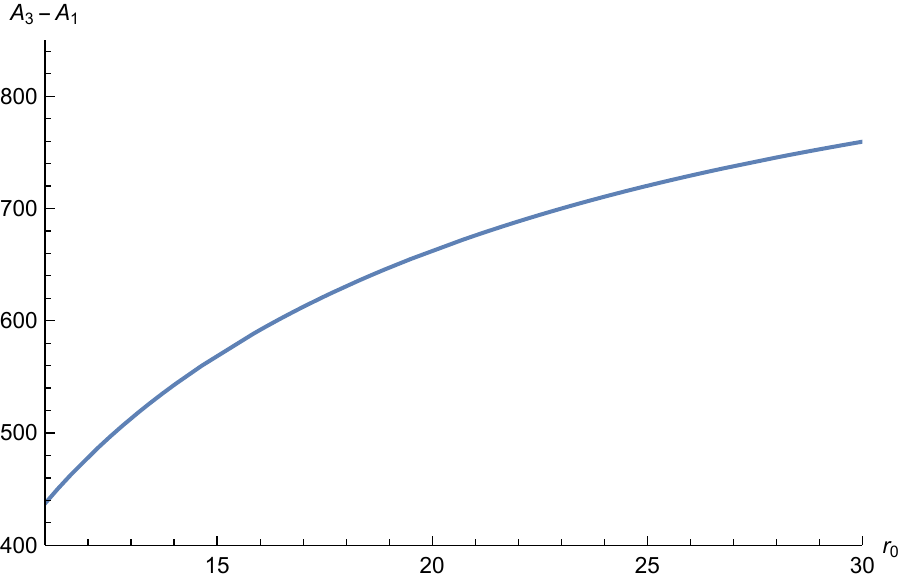}
\caption{Here we have plotted the difference between $A_3$, given in eq.\eqref{a32}, and $A_1$, given in eq.\eqref{A1finiteT}, as a function of the turning point $r_0$, for $r_{UV} = 10^4, n=3,r_h=10$.}
\label{finiteTareaplot}
\end{figure}

More generally, when ${(r_0-r_h)\over r_h} \sim >\order{1}$ we find numerically that $A_1$ continues to  have the lower value., see Fig \ref{finiteTareaplot} where we have plotted $A_3 - A_1$ for representative values of $r_{UV},n$.

The reader might object that eq.\eqref{rea5} vanishes for $n=1$ and hence the preceding arguments don't hold in the case of $AdS_3$. While it is true that eq.\eqref{rea5} does vanish for $n=1$ but in that case one can exactly integrate eq.\eqref{rea4} and eq.\eqref{A1finiteT} without any approximation and show that $A_3 > A_1$.

The picture which then emerges is that for any finite temperatures, $T< r_{UV}$, unlike in the zero temperature case, the exchange of dominance doesn't happen. For all values of $\alpha \Delta y_1$, the surface corresponding to $A_1$ continues to exist and has the lowest area. The resulting temperature dependence of its area is  extensive in both the volume $V_n$ along the spatial directions at the boundary of $AdS_{n+2}$, and also in the volume $V_{m-1} \Delta y_1$ along the $R^m$ directions with a temperature dependence going like $T^n$. 

\section{$AdS_{n+2} \times R^m$ in Global coordinates}
\label{globaladsn}
This section examines a situation related to the one considered in section \ref{consdil}.  We analyse extremal surfaces here  of the type considered in section \ref{consdil} after replacing the Poincare patch of $AdS_{n+2}$ by hyperbolic space in global coordinates. 
The metric for the $AdS_{n+2}$ $\cross$ $R^m$ in the global coordinates is,
\begin{equation}
	ds^2 = R^2 \left(- (1+r^2) dt^2 + \frac{dr^2}{1+r^2} + r^2 d\Omega_n ^2\right) + R^2 \alpha^2 \sum_{i} (dy_i)^2. \label{globaladsmet}
\end{equation} 
We consider the following strip in the $y_1$ direction.
\begin{equation}
	0 \leq y_1 \leq 2\Delta y_1.
\end{equation}
To find the entanglement entropy of this strip we minimise the Area functional,
\begin{equation}
	A_1 = R^{m+n} \alpha^{m-1} V_{m-1} V_{S^n} \int r^n \sqrt{\frac{dr^2}{1+r^2} + \alpha^2 dy_1^2}. \label{areaglobal}
\end{equation}						
$V_{m-1}$ is the volume of the rest of $m-1$ $y$ directions while $V_{S^n}$ is the volume of $S^n$. So the function we have to minimize is,
\begin{equation}
	\mathcal{L} = r^n \sqrt{\frac{1}{1+r^2} + \alpha^2 \Dot{y_1} ^2}.
\end{equation}
The conjugate momenta to $y_1$
\begin{equation}
	P_y = \frac{r^n \alpha^2 \Dot{y_1}}{\sqrt{\frac{1}{1+r^2} + \alpha^2 \Dot{y_1} ^2}} \label{pyglobal}
\end{equation}
is conserved. And it's value is
\begin{equation}
	\alpha r_0 ^n,
\end{equation}						
where $r_0$ is the turning point. Then rearranging the equation we get,
\begin{equation}
	\alpha \Dot{y_1} = \frac{r_0 ^n}{\sqrt{(1+r^2)(r^{2n} - r_0 ^{2n})}}.
\end{equation}						
Now integrating both sides and changing coordinates to $\xi = \frac{r}{r_0}$, we get,
\begin{equation}
	\alpha \Delta y_1 = \int_{1}^{\frac{r_{UV}}{r_0}} \frac{1}{\sqrt{(\frac{1}{r_0 ^2} + \xi^2)(\xi^{2n} - 1)}} d \xi.\label{globalintdely}
\end{equation}						
						
We take the $UV$ cut-off $r_{UV}\gg 1$. We now assume, self-consistently as we will see, that the turning point satisfies the condition, 
\begin{equation}
	\label{condee}
	1 \ll r_0 \ll r_{UV}.
\end{equation}						
We can then expand the first factor in the square root in the integral in eq.(\ref{globalintdely}) and we get,
\begin{equation}
	\alpha \Delta y_1 = \int_{1}^{\frac{r_{UV}}{r_0}}d \xi \frac{1}{\xi \sqrt{(\xi^{2n} - 1)}} - \int_{1}^{\frac{r_{UV}}{r_0}} d \xi \frac{1}{2 r_0^2}\frac{1}{\xi^3\sqrt{(\xi^{2n} - 1)}}.
\end{equation}						
In the second integral above we can now take, to good approximation,  the upper limit to infinity, giving,
\begin{equation}
	\alpha \Delta y_1 = \frac{\pi}{2n} - \frac{r_0 ^n}{n r_{UV} ^n} - \frac{1}{2 r_0^2} \frac{\sqrt{\pi} \Gamma(\frac{1}{n} + \frac{3}{2})}{ (n+2) \Gamma(\frac{1}{n} + 1)}+ \order{\left(\frac{r_0}{r_{UV}}\right)^{n+2}}. \label{globaldely}
\end{equation}						
From this expression we can find the maximum value for $\Delta y_1$ as a function of $r_0$. Denoting the corresponding value of the turning point as $r_0 ^{min}$ we get 
\begin{equation}
	(r_0 ^{min}) ^{n+2} = \frac{\sqrt{\pi} \Gamma(\frac{1}{n} + \frac{3}{2})}{ (n+2) \Gamma(\frac{1}{n} + 1)} r_{UV} ^n. \label{turningpointeq}
\end{equation}
The maximum value of $\alpha \Delta y_1$ is then,
\begin{equation}
	\alpha \Delta y^{max}_1 = \frac{\pi}{2n} - \frac{\gamma}{ 2 n (r_0 ^{min})
		 ^2} + \order{\frac{1}{r_{UV}^2}}, \label{maxyglobal}
\end{equation}
where,
\begin{equation}
	\gamma = \frac{\sqrt{\pi} \Gamma(\frac{1}{n} + \frac{3}{2})}{\Gamma(\frac{1}{n} + 1)}.
\end{equation}						
						
Representative figures for $\alpha \Delta y_1$ as a function of $r_0$, with $n=2, r_{UV}=10^5$, and $r_{UV}=10^6$  is given in Fig \ref{globaldelyplot}. We see that as $r_0$ decreases from very large values to $r_{0}^{min}$, $\Delta y_1$ increases attaining its maximum value at $\Delta y_1^{max}$. Decreasing $r_0$ further leads to $\Delta y_1$ decreasing from $\Delta y_1^{max}$. As a result there are generically two values of $r_0$ corresponding to  a given value of $\Delta y_1$. 
						
The area for the surface with turning point $r_0$  is given by,
\begin{equation}
	A_1 =2 R^{m+n} \alpha^{m-1}  V_{m-1} V_{S^n}\int_{r_0} ^{r_{UV}}  \frac{r^{2n}}{\sqrt{(1+r^2)(r^{2n} - r_0 ^{2n})}} dr
\end{equation}
which can be written in the variable $\xi$ as follows,
\begin{equation}
	A_1 = 2 R^{m+n} \alpha^{m-1}  V_{m-1} V_{S^n} r_0 ^n \int_{1}^{\frac{r_{UV}}{r_0}} d \xi \frac{\xi^{2n}}{\sqrt{(\frac{1}{r_0 ^2} + \xi^2)(\xi^{2n} - 1)}}.
\end{equation}						
Again expanding the square root and integrating we get to leading order,
\begin{equation}
	A_1 =2 \frac{1}{n} R^{m+n} \alpha^{m-1}  V_{m-1} V_{S^n} \sqrt{r_{UV} ^{2n} - r_0 ^{2n}}. \label{areaeq1}
\end{equation}						
Let us denote the two values of $r_0$ for a given value of $\Delta y_1$ by   $r_1$ and $r_2$, with  $r_2$ bigger than $r_0^{min}$ and  $r_1 <r_0^{min}$. To know which is the actual turning point we need to find which leads to the lower area. Eq. (\ref{areaeq1}) suggests  that the larger value of $r_0$, i.e. $r_2$ would lead to a lower area. This can be checked to be true numerically as well. Thus the value of $r_0> r_0^{min}$ is always the true turning point. It also follows that for the true turning point eq.(\ref{condee}) is  met. 
						
Finally we note from eq.(\ref{turningpointeq}) that as the UV cut-off $r_{UV}\rightarrow \infty$,  $r_0 ^{min} \rightarrow \infty$,  and thus all the minimal area surfaces go off to infinity. 
						
The surface that hangs near the boundary has the area,
\begin{equation}
	A_2 =2 R^{m+n} \alpha^{m-1}  V_{m-1} V_{S^n} r_{UV} ^n \alpha \Delta y_1. \label{globalarea2}
\end{equation}
The surface that does not have a turning point has area,
\begin{equation}
	A_3 = 2R^{m+n} \alpha^{m-1}  V_{m-1} V_{S^n} \int_{0}^{r_{UV}} \frac{r^n}{\sqrt{1+r^2}} dr .\label{globalarea3}
\end{equation}						
 Fig \ref{globaldelyplot}  confirms the above analytical results.  In  Fig \ref{globalarea} we have plotted $\frac{A_1}{A_2}$ and $\frac{A_1}{A_3}$ showing that $A_1$ gives the  minimum area. 
						
						
{\begin{figure}
		\centering
		\subfigure[]{\includegraphics[width=0.49\textwidth]{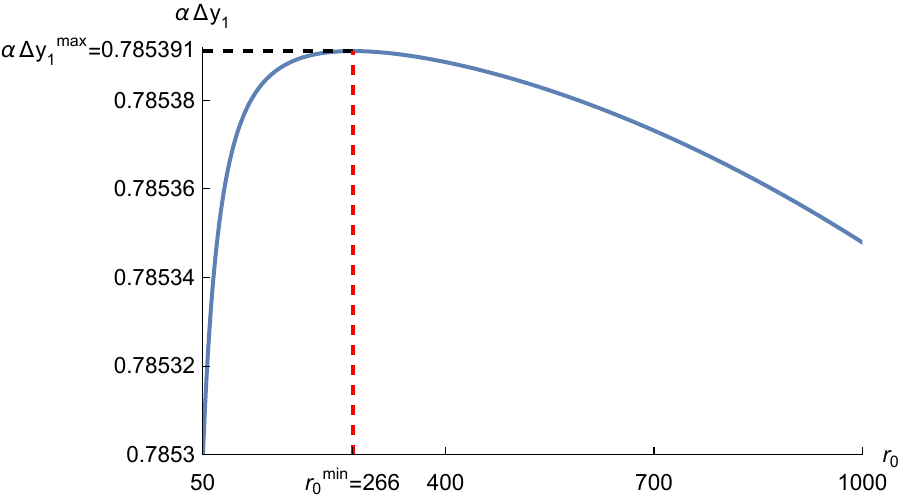}} \hfill
		\subfigure[]{\includegraphics[width=0.49\textwidth]{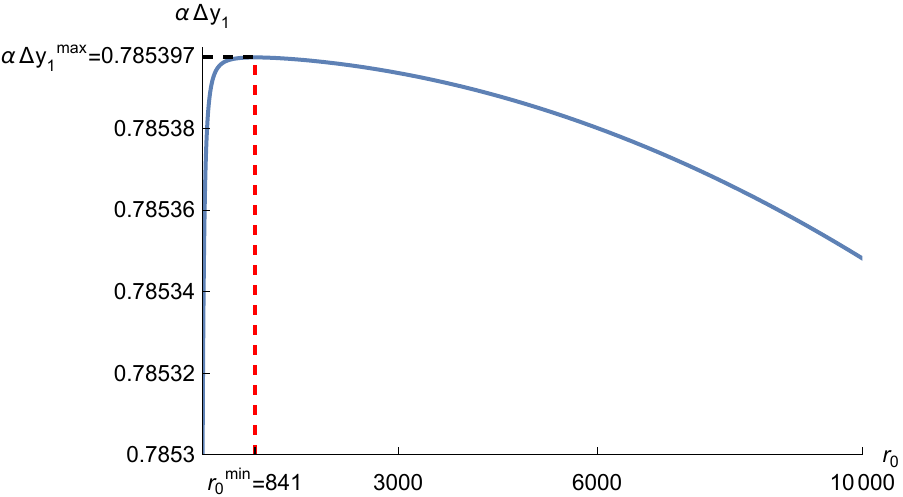}} 
		\caption{ Plot of $\alpha \Delta y_1$ vs $r_0^{min}$. (a) For $r_{UV}=10^5, n=2$, with  $r_0^{min}\simeq 266$. (b) For $r_{UV}=10^6, n=2$, with $r_0^{min}\simeq 841$. The plot also shows the two values of $\alpha \Delta y^{max}_1$,  given by eq.\eqref{maxyglobal}.}
		\label{globaldelyplot}
\end{figure}}
{\begin{figure}
		\centering
		\subfigure[]{\includegraphics[width=0.45\textwidth]{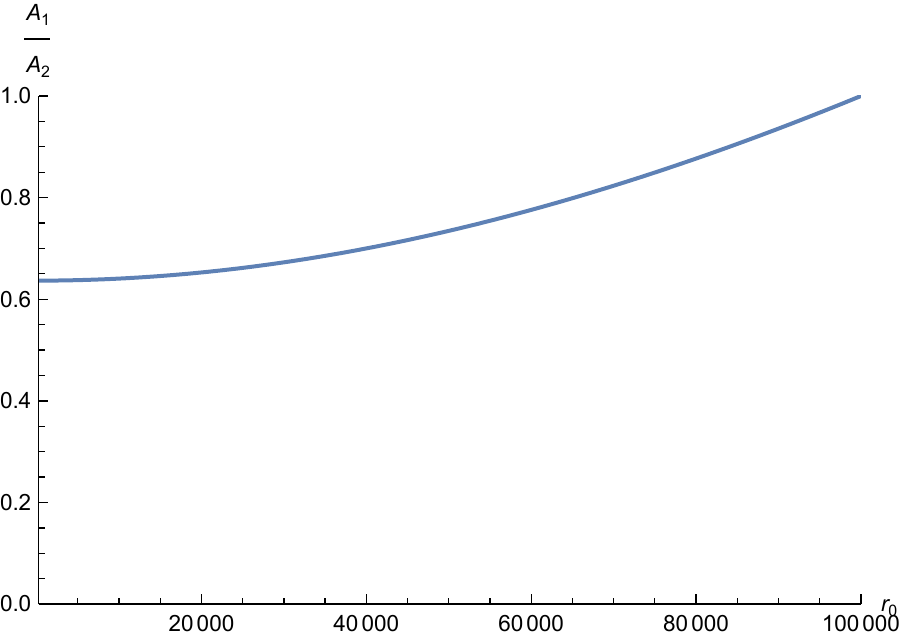}}\hfill
		\subfigure[]{\includegraphics[width=0.45\textwidth]{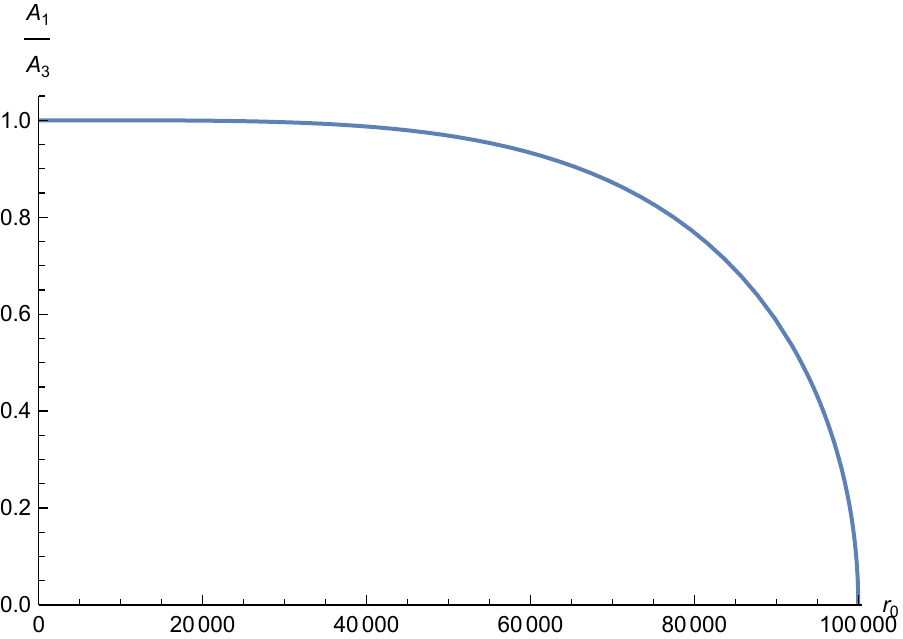}} 
		\caption{ Plot of ${A_1\over A_2}$ and ${A_1\over A_3}$ for $r_{UV} = 10^5, n=2$. In both  plots $r_0 ^{min}\simeq 266$.}
		\label{globalarea}
\end{figure}}
						
						
\section{More Detailed Analysis for Warped Compact Cases}
\label{Moredetana}
Here we  give further details about solutions to eq.(\ref{assie}). 
						
There are three cases to consider:
						
\textbf{Case-1:} $h(r) \dot{\theta}^2 \rightarrow \infty$
						
In this case we get,
\begin{equation*}
	\frac{d}{dr} \left(\mathcal{F} (r) (\sin(\theta))^{m-1} \sqrt{h(r)}\right) - (m-1)(\sin(\theta))^{m-2} \cos(\theta) \mathcal{F} (r) \sqrt{h(r)} \dot{\theta} =0.
\end{equation*}						
Assuming a power law behaviour for ${\cal F}, h(r)$, as we are doing, the two terms scale with $r$ like, 
\begin{equation}
	\frac{\mathcal{F} (r) \sqrt{h(r)}}{r}, \hspace{1cm} \mathcal{F} (r) \sqrt{h(r)} \dot{\theta}
\end{equation}						
respectively. Because of eq.\eqref{assf2}, the first term is dominant over the second term and if, $\mathcal{F},h$ grow with $r$, this term cannot be zero and we can have no consistent solution.

\textbf{Case-2:}  In this case 
\be
\label{condff}
h(r) \dot{\theta}^2 \rightarrow  d+ \order{1/r^\gamma}, \gamma>0
\ee
From the above assumption and eq.\eqref{defh} we get,
\begin{equation}
	\dot{\theta} \sim r^{-\frac{b}{2}} \label{newth}
\end{equation}
So we get,
\begin{equation*}
	\frac{d}{dr} \left(\frac{\mathcal{F} (r) (\sin(\theta))^{m-1} h(r) \dot{\theta} }{\sqrt{1 + d}}\right) - \mathcal{F} (r) (m-1)(\sin(\theta))^{m-2} \cos(\theta) \sqrt{1 +d}=0.
\end{equation*}						
Then using the ansatz eq.\eqref{asF} and eq.\eqref{ash} and the result eq.\eqref{newth} the two terms scale with $r$ as,
\begin{equation}
	r^{a-1+\frac{b}{2}}, \hspace{1cm} r^a
\end{equation}
respectively. Then for both the terms to be comparable we must have,
\begin{equation}
	b=2.
\end{equation}
Then putting this value into eq.\eqref{newth} shows that, eq.\eqref{newth} is in direct contradiction with eq.\eqref{assf2}. Thus we don't have a consistent solution.

\textbf{Case-3:} Finally we assume $h(r) \dot{\theta}^2 \rightarrow 0$, this requires, 
\be
\label{assre}
b <2(1+\epsilon).
\ee
Then we have using the assumption,
\begin{equation}
	\frac{d}{dr} (\mathcal{F} (r) (\sin(\theta))^{m-1} h(r) \dot{\theta}) - \mathcal{F} (r) (m-1)(\sin(\theta))^{m-2} \cos(\theta) =0
\end{equation}
Simplifying we get,
\begin{equation}
	(\mathcal{F} (r)' h(r) + \mathcal{F} (r) h(r)') (\sin(\theta))^{m-1} \dot{\theta} + \mathcal{F} (r) h(r) (\sin(\theta))^{m-1} \Ddot{\theta} - \mathcal{F} (r) (m-1)(\sin(\theta))^{m-2} \cos(\theta)=0
\end{equation}		
This has a solution only if all the terms are comparable. From eq.(\ref{asF}, \ref{ash}), eq.(\ref{assf2}), we get eq.(\ref{valaa}), eq.(\ref{valba}).
For this to be consistent with eq.(\ref{assre}) we need eq.(\ref{condb}). 

\section{CFT interpretation of the AdS$_4 \to$ AdS$_2$}
\label{CFT}

The discussion here is related to that in  Section \ref{ads2}, although it presents a simpler treatment, allowing for a simple derivation of the subleading term in the extremal area.

As in Section \ref{ads2}, we will take the metric as
\[
ds^2 = -dt^2 f(r) + f(r)^2 dr^2 + \Phi^2(r) (dx^2 + dy^2)
\]
where, for the RN BH in AdS$_4$
\begin{align}
	&\Phi(r) = r,
	\nonumber \\
	& f(r) = \frac{(r-r_h)^2}{r^2}   \frac{r^2 + 2 r r_h + 3 r_h^2}{R_4^2}
	= \frac{z^2}{(1+z)^2} (1 + \fr23 z + \fr16 z^2) \fr{r_h^2}{R_2^2}, \quad r=: r_h(1+z) 
	\label{rn-bh}
\end{align}
Note that this $z$ is different from the one used in section \ref{ads2}. For the analysis below, we will focus on the near horizon region
$r-r_h \ll r_h$ where $z\ll 1$. We will also consider the 4D geometry to be
more general, with the following near-horizon behaviour of $f(r)$ and $\Phi(r)$ 
\begin{align}
	& \Phi(r) =\Phi(r_h) + \alpha (r-r_h)^p = \Phi_h (1+ \tilde\alpha z^p), \; \tilde\alpha= \alpha r_h^p/\Phi_h^p \nonumber\\
	& f(r)= \fr{\gamma^2}{4} z^2 (1+ az + \order{z^2})
	\label{gen-near-hor}
\end{align}
For the RN BH metric \eq{rn-bh} we have
\begin{align}
	\Phi_h = r_h, \; \alpha = 1, \; p=1, \; \gamma = \fr{2 r_h}{R_2}, \; a = -\fr43.
	\label{rn-bh-parameters}
\end{align}
In Section \ref{ads2}, we considered the entanglement entropy of the infinite strip, ${\cal A}= \{ x\in \left(-x_m,x_m \right), \ y\in (0,L\rightarrow\infty)\}$ in the boundary theory, where $x_m = l/2$. The corresponding RT surface was anchored at this strip at $r = r_m \to \infty$. For finite $r_m$, the area of the RT surface computes the entanglement entropy of the field theory at a UV cutoff $\Lambda = r_m/R_4^2$. The area of the strip in the boundary is $V_2 = R_4^2 \Delta x \Delta y$, where $\Delta x= 2 x_m = l,\; \Delta y= L$. 

In this Appendix we will consider a sliding cut-off in the spirit of holographic RG. Thus, we will bring the cut-off surface $r=r_m$ into the near-horizon region $r_m= r_h (1+ z_m)$, $z_m \ll 1$. The sliding cut-off will be identified with
\begin{align}
	\Lambda= r_m/R_4^2 = \fr{r_h}{R_4^2}(1+ z_m) = \fr{\mu}{\sqrt 3} (1+ z_m)
	\label{sliding-cutoff}
\end{align}
In the above, we have used the definition of the chemical potential \eq{chemp}.
In the IR limit
\begin{align}
	r_m \to r_h, z_m \to 0, \Lambda \to \Lambda_{IR} \equiv \fr{\mu}{\sqrt 3}.
	\label{IR}
\end{align}
The area functional for the RT surface is (using the $x\to -x$ symmetry)
\[
A = 2 L\int_0^{x_m} dx\ \Phi(r) \sqrt{\frac{{r'}^2}{f(r)} + \Phi^2(r)}
\]
where $r$ is thought of as function of $x$. The extremization of the area is performed with the Dirichlet-Neumann boundary conditions $r(x_m)=r_m$ and $r'(0)=0$.  Since the Lagrangian does not have any explicit dependence on the ``time'' $x$ as ``time'', the  Hamiltonian is a constant (independent of $x$):
\begin{align}
	\mathcal{H} \equiv P_r r' - \mathcal{L} = -\frac{\Phi^3}{\sqrt{\frac{r'^2}{f(r)} +\Phi^2(r)}}= {\rm constant} =-\Phi^2_0.
	\label{ham}
\end{align}
In the fourth step we evaluate the constant at $x=0$ where $r'(0)=0$, yielding $-\Phi^2(x=0)$. In the following we will use the notation $r(0) \equiv r_0$. 
From \eq{ham} we immediately get (i) a differential equation for $r(x)$
\begin{align}
	r'^2 = \Phi^2(r) f(r) \left(\frac{\Phi^4(r) - \Phi_0^4}{\Phi_0^4}\right)
	\label{r-x-eq}
\end{align}
and (ii) a simple exact expression for the extremal area of the RT surface
\begin{align}
	A = 2 L \int_{0}^{x_m} dx \frac{\Phi^4}{\Phi_0^2}
	\label{exact-extr}
\end{align}
It is straightforward to evaluate the above expression by using the {\it ansatze} \eq{gen-near-hor}, once one solves for $r(x)$ (equivalently $z(x)$) from \eq{r-x-eq}. Carrying this out for the RN BH parameters \eq{rn-bh-parameters}, we get
\[
z' = \gamma z \sqrt{z -z_0}\left[1 + \fr{z}{3} - 2 z_0 +\order{z^2}\right],
\quad \gamma= \fr{2r_h}{R_2},\; z_0=z(x=0)= z(r_0)
\]
At leading order in $z$, we can solve this equation exactly
\begin{equation}
	z' = \gamma z\sqrt{z-z_0}(1+ \order{z,z_0}) \Rightarrow z=z_0 \sec^2\left(\frac{\sqrt{z_0} \gamma}{2} x \right), 
	\label{z-diff}
\end{equation}
The constant $z_0$, corresponding to the turning point, can be determined by imposing the boundary condition $z(x_m)= z_m$, yielding
\begin{equation}
	z_m =z_0 \sec^2\left(\frac{\sqrt{z_0} \gamma}{2} x_m \right), 
	\label{z0-zm}
\end{equation}
Using this solution, and the RN BH parameters \eq{rn-bh-parameters} it is easy to evaluate the extremal area \eq{exact-extr}
\begin{align}
	A &= 2 L r_h^2 \int_{0}^{x_m} (1 - 2z_0 + 4 z) dx  \nonumber \\
	&=  2 L r_h^2\left[ x_m \left(1 - 2 z_0 \right) +  8 \fr{\sqrt{z_0}}{\gamma}  \tan\left(\fr\gamma{2} x_m \sqrt{z_0}\right) \right]  \nonumber \\
	&= \Delta x \Delta y r_h^2 \left(1 + 2z_0 \right) \nonumber \\
	&= \fr{V R_4^2 \mu^2}{3} \left(1 + 2z_m \right) 
\end{align}
In the first line we have used $\Phi=r = r_h(1+z)$, $\Phi_0= r_0=
r_h(1+z_0)$ and used the near-horizon assumption $z, z_0 \ll 1$. In
the second line we have changed the integration variable from $x$ to
$z$ using the expression for $z'(x)$ from \eq{z-diff}.  In the third
line we have used $2 x_m= \Delta x, \ L= \Delta y$, and the near
horizon approximation $z_m, z_0 \ll 1$ which lets us use the linear
approximation to the $\tan$ function.  In the last line we have used
the physical 2D volume of the strip $V = R_4^2 \Delta x \Delta y$
and $r_h = R_4^2 \mu/\sqrt 3$.

Using the holographic EE formula
\begin{align}
	S &= \fr{\rm Area}{4 G_N}= c\ V N^2 \mu^2 \left(1 + 2 z_m\right)
	\nonumber\\
	&= c\ V N^2 \mu^2 \left(1 + 2 \fr{\Lambda - \Lambda_{IR}}{\Lambda_{IR}}\right)
	\label{RG}
\end{align}
Here $c$ is a numerical constant. The leading answer is the same as found
in the main text (Section \ref{ads2}), but we have kept the first
correction. In the second line, we have used \eq{sliding-cutoff} and \eq{IR}, which gives an RG interpretation for the entanglement entropy as obtained from the holographic calculation.

\newpage
\bibliographystyle{JHEP}
\bibliography{refs}
					
\end{document}